\documentclass[twocolumn,tigten,letteredappendix,appendixfloats,twocolappendix,10pt]{aastex63}

\hypersetup{linkcolor=blue,citecolor=blue,filecolor=blue,urlcolor=blue,
	pageanchor=false,hidelinks,colorlinks}

\PassOptionsToPackage{unicode}{hyperref}
\PassOptionsToPackage{naturalnames}{hyperref}	

\usepackage{amsmath,amstext}
\usepackage{apjfonts}
\usepackage{comment}
\usepackage{footnote}
\usepackage{booktabs}
\usepackage{float}

\usepackage{url}
\usepackage[toc,page]{appendix}

\usepackage{graphicx,booktabs,array}
\usepackage{natbib}
\usepackage{hyperref}

\usepackage[hyphenbreaks]{breakurl}
\usepackage{soul}

\usepackage{multirow}

\providecommand{\url}[1]{\href{#1}{#1}}
\providecommand{\dodoi}[1]{doi:~\href{http://doi.org/#1}{\nolinkurl{#1}}}

\submitjournal{ApJ}
\shorttitle{Manuscript}
\shortauthors{Ling \& Yan}

\begin{document}
        
        \title{Morphological Evolution of the Hosts of Far-infrared/Submillimeter Galaxies}
        \author{Chenxiaoji Ling}
        \affiliation{Department of Physics and Astronomy, University of Missouri-Columbia, USA}
        \email{clvdb@mail.missouri.edu}
        
        \author{Haojing Yan}
        \affiliation{Department of Physics and Astronomy, University of Missouri-Columbia, USA}
        \email{yanha@missouri.edu}

        \begin{abstract}
                
               We present a host morphological study of 1266 far-infrared galaxies (FIRGs) and submillimeter galaxies (SMGs) in the Cosmic Evolution Survey field using the F160W and F814W images obtained by the Hubble Space Telescope. The FIRGs and SMGs are selected from the Herschel Multi-tiered Extragalactic Survey and the SCUBA-2 Cosmology Legacy Survey, respectively. Their precise locations are based on the interferometry data from the Atacama Large Millimeter/submillimeter Array and the Very Large Array. These objects are mostly at  $0.1\lesssim z\lesssim 3$.  The SMGs can be regarded as the population at the high-redshift tail of the FIRGs. Most of our FIRGs/SMGs have a total infrared luminosity ($L_{\rm IR}$) in the regimes of luminous and ultraluminous infrared galaxies  (LIRGs, $L_{\rm IR} = 10^{11-12}L_\odot$; ULIRGs, $L_{\rm IR}>10^{12}L_\odot$). The hosts of the SMG ULIRGs, FIRG ULIRGs, and FIRG LIRGs are of sufficient numbers to allow for detailed analysis, and they are only modestly different in their stellar masses. Their morphological types are predominantly disk galaxies (type D) and irregular/interacting systems (type Irr/Int). There is a morphological transition at $z\approx 1.25$ for the FIRG ULIRG hosts, above which the Irr/Int galaxies dominate and below which the D and Irr/Int galaxies have nearly the same contributions. The SMG ULIRG hosts seem to experience a similar transition. This suggests a shift in the relative importance of galaxy mergers/interactions versus secular gas accretions in  ``normal'' disk galaxies as the possible triggering mechanisms of ULIRGs. The FIRG LIRG hosts are predominantly D galaxies over $z=$ 0.25--1.25, where they are of sufficient statistics.
                
        \end{abstract}

        
         \section{Introduction} \label{sec:intro}
        
        Ultraluminous infrared galaxies (ULIRGs) were discovered as the brightest objects in the infrared sky \citep[][]{Aaronson1984,Houck1984, Houck1985}, with an extremely high total infrared luminosity (integrated over rest-frame 8--1000~$\mu$m) of $L_{\rm IR}>10^{12} L_\odot$. While some of them harbor active galactic nuclei (AGNs), it is generally believed that their extreme IR emission is mainly due to a starburst enshrouded by dust, which reprocesses the strong UV photons from  young stars and emits them in the IR-to-millimeter regime \citep[see][for a  review]{Lonsdale2006}. Their analogs at a lower luminosity level, luminous  infrared galaxies (LIRGs), which have $L_{IR}=10^{11-12} L_\odot$, are believed  to be of similar origin. While both ULIRGs and LIRGs (hereafter ``(U)LIRGs'') are rare objects in the nearby universe \citep[see, e.g.,][]{Soifer1991}, they are more common at high redshifts \citep[e.g.,][]{LeFloch2005, Magnelli2013}; therefore, understanding them has important cosmological implications.  
        
        A key question about (U)LIRGs is the cause of their active star formation, which is still under debate today. In the local universe, nearly all ULIRGs and a large fraction of LIRGs are mergers \citep[][]{Veilleux2002}, and this has led to the picture that (U)LIRGs are mostly merger-driven. Over the past decade, however, it has been called into question whether this picture holds at high redshifts, especially at $z\approx 2$, when the cosmic star formation has reached its peak. Generally speaking, star-forming galaxies have increasing star formation rates (SFRs) when looking back from  today to higher redshifts, with a factor of $\sim$20 increase from  $z\approx 0$ to 2 \citep[e.g.,][]{Elbaz2011, Karim2011}. For this reason, it has been argued that the SFRs of LIRGs ($>$10~${ M_\odot}$~yr$^{-1}$) and  ULIRGs ($>$100~${ M_\odot}$~yr$^{-1}$) are not so extreme at $z\approx 2$ because a ``normal'' star-forming galaxy at the high-mass end of the so-called ``galaxy main sequence'' would naturally have such a high SFR. Some numerical simulations do suggest that in situ, secular gas accretion through disk instability could sustain such a high SFR in a normal disk galaxy \citep[e.g.,][]{Dekel2009,Dave2010}; putting in a sufficient amount of dust, such a galaxy would appear (U)LIRG-like. Along with this argument, there has  also been a trend to use the term ``dusty star-forming galaxies''  \citep[DSFGs; see][and references therein]{Casey2014} to replace  ``(U)LIRGs'' at high-$z$ because the significance that the latter term implies would (arguably) no longer be justified at high-$z$. 
        
        Observationally, this becomes the question of whether (U)LIRG phenomena  are observed at high-$z$ in normal-looking galaxies, and if so, whether mergers become insignificant in triggering (U)LIRGs at such redshifts. To answer this question, a morphological study of (U)LIRG hosts at high-$z$ is the most direct approach. For historical reasons, submillimeter galaxies (SMGs; mostly selected in the $\sim$850~$\mu$m window) are often viewed as the representatives of high-$z$ ULIRGs, and they are shown to lie at $z\approx 2$--3 \citep[see][and references therein]{Chapman2005}. The far-IR (FIR; 100--500~$\mu$m) surveys done by the Herschel Space Observatory from 2009 to 2013 detected orders of magnitude more sources than the SMG  surveys, and most of these FIR galaxies (FIRGs) are (U)LIRGs at  $z\approx 0.8$--2 \citep[][]{Casey2012a, Casey2012b}. It has become possible to study the morphologies of (U)LIRG hosts up to $z\approx 3$ using large  samples. This, however, needs to overcome two major difficulties.  
        
        The first is how to properly identify (U)LIRG counterparts in optical/near-IR images where morphological studies can be carried out. This has been a long-standing problem because of the coarse spatial  resolutions of the single-dish instruments through which SMGs and FIRGs are discovered. For example, the Submillimetre Common-User Bolometer Array 2 \citep[SCUBA2;][]{Holland2013} at the 15 m James Clerk Maxwell Telescope (JCMT) has an FWHM beam size of 14\arcsec.9 at 850~$\mu$m. The most sensitive band at the 3.5 m Herschel was the 250~$\mu$m channel of the Spectral and Photometric Imaging Receiver \citep[SPIRE;][]{Griffin2010}, which has an FWHM beam size of 18\arcsec.1.  To combat this problem, the best approach to date is to use  submillimeter/millimeter interferometry. The most powerful instrument is the Atacama Large Millimeter/submillimeter Array (ALMA), which can achieve subarcsecond resolution. The next in line is radio interferometry; for example, the observations with the Very Large Array (VLA) in the extended configurations can achieve similar resolution. The connection between radio and FIR/submillimeter is through the tight FIR--radio relation in star-forming  galaxies, which spans decades in luminosities from ULIRGs to subgalactic star-forming regions \citep[e.g.,][]{Helou1985, Condon1992}. Lacking either interferometry, the most commonly used alternative nowadays is to utilize mid-IR data of better (although still coarse) resolution as the ``ladder'' to  make a connection to optical/near-IR images through a deblending process. In  most cases, the mid-IR data are the images taken by the Multiband  Imaging Photometer for Spitzer \citep[MIPS;][]{Rieke2004}, which surveyed a number of well-studied extragalactic fields to significant depths at 24~$\mu$m (6\arcsec \ resolution). Other methods include using  the statistical ``likelihood ratio'' method \citep[][]{Sutherland1992} and the ``major contributor'' method that directly identifies the major  contributor(s) of a FIR source by using optical/near-IR images for the position priors \citep[][]{Yan2014}. 
        
        The second major difficulty is how to deal with the possible change of morphology at different rest-frame wavelengths, or the so-called  ``morphological $k$-correction.'' At $z\gtrsim 1$, the optical bands in the observer's frame would sample the targets in their rest-frame UV. As young stellar populations emit strongly in the UV, a star-forming galaxy with segregated star-forming regions would show knotty features in the  rest-frame UV, which would then bias the morphological classification to favor the irregular category. The most appropriate approach is to use bands of progressively longer wavelengths for increasing redshifts to ensure that the classification is always done in the rest-frame optical, which samples the matured, dominant stellar populations of the galaxy. In other words, observations in near-IR bands would be needed for $z\gtrsim 1$. In  addition, such near-IR observations need to be of high resolution to  discern the details at $z\gtrsim 1$, which means that one would have to rely on the near-IR imaging with either adaptive optics or the instruments at the Hubble Space Telescope (HST). 
        
        Despite all the difficulties, significant progress has been made over the past two decades on the morphological study of high-$z$ (U)LIRG hosts.   {\citet[][]{Chapman2003} and \citet[][]{Conselice2003} studied about a dozen SMGs whose positions were precisely determined through radio and/or millimeter interferometry. They analyzed the high-resolution optical images obtained by the Space Telescope Imaging Spectrograph on board the HST and concluded that $61\% \pm21$\% of their sources were active mergers. A high merger rate among SMGs was supported by \citet[][]{Smail2004}, who obtained optical images from the HST Advanced Camera for Surveys (ACS) for 20 SMGs with accurate radio positions and found that they were all mergers or interacting systems. \citet[][]{Swinbank2010} obtained both optical and near-IR images of 25 SMGs using the HST ACS and the Near Infrared Camera and Multi-Object Spectrometer, respectively, and they concluded that the SMG host morphologies differ more between the rest-frame UV and optical bands than typical star-forming galaxies. They interpreted this as evidence for structured dust obscuration, rather than merger.} \citet[][]{Targett2013} studied 24 SMG hosts in the HST Wide Field  Camera 3 (WFC3) images of the Cosmic Assembly Near-IR Deep Extragalactic  Legacy Survey \citep[CANDELS;][]{Grogin2011, Koekemoer2011}, whose counterpart identification was mostly done through the VLA imaging. They concluded that most of their sources are disklike in the WFC3 F160W (hereafter $H_{160}$) images, and therefore SMGs should be considered as part of the normal star-forming galaxies in this epoch. However, their results included many cases where analytic profiles were forced to fit galaxies with obvious irregularities.  \citet[][]{Wiklind2014} studied 10 SMGs using the same CANDELS WFC3 images, with the difference that the source counterparts were located by the ALMA identifications from the larger ALESS sample of \citet[][]{Hodge2013}. They found a diverse range of host morphologies, although no definite conclusion could be  made due to the small sample size. This was extended by \citet[][]{Chen2015}, who used 45 SMGs from the ALESS sample and found that $\sim$82\% of their hosts have disturbed morphologies in $H_{160}$ indicative of mergers. When  forcing a fit of the analytic profiles to these galaxies, they also obtained  disklike fits. The ALMA imaging of the SCUBA-2 Cosmology Legacy Survey \citep[S2CLS;][]{Geach2017} in the UltraDeep Survey (UDS) field has resulted in a large SMG sample (708 sources detected at 870~$\mu$m) with precise  locations \citep[AS2UDS;][]{Simpson2017, Stach2018, Stach2019}. Among them, 47 sources are within the coverage of the CANDELS $H_{160}$ image; one of the conclusions of \citet[][]{Stach2019} is that $50$\% $\pm 10$\% of these sources show clear merger morphologies or have likely companions.  A similarly large fraction of mergers are also observed in the hosts of the Herschel FIRGs.  \citet[][]{Kartaltepe2012} studied the CANDELS WFC3 $H_{160}$ morphologies of 52 ULIRGs at $z\approx 2$ that were selected using the Herschel 100  and 160~$\mu$m data in the southern field of the Great Observatories Origins  Deep Survey \citep[GOODS;][]{Giavalisco2004}. They found that galaxies with irregular features and/or  in interactions comprise $\sim$72\% of this sample. The $H_{160}$ counterparts  were identified by using the MIPS 24~$\mu$m data as the ``ladder.'' \citet[][]{Hung2014} carried out a morphological study of 246 Herschel 100--500~$\mu$m sources selected in the Cosmic Evolution Survey (COSMOS) field \citep[][]{Scoville2007}, which are (U)LIRGs spanning $z\approx 0.5$--1.5. Also using MIPS 24~$\mu$m data as the  intermediate, they identified the host galaxies in the ACS F814W (hereafter $I_{\rm 814}$) images that sample the rest-frame optical. They compared to a sample of local (U)LIRG hosts  ($z<0.3$) and concluded that the merger fraction of (U)LIRGs decreases by at most $\sim$20\% from $z\approx 0$ to 1.  
        
        In this work, we carry out a systematic morphological study of (U)LIRG hosts over $0.5<z<3$ in the rest-frame optical. All data used are from the public archives. Our sample includes 1266 FIRGs and SMGs in the COSMOS field that are selected by Herschel/SPIRE and JCMT/SCUBA2. The source locations are  precisely determined by the ALMA and/or VLA interferometric data in this field, and the host morphologies are determined based on the HST WFC3 $H_{160}$ and ACS $I_{814}$ images. This is the largest high-$z$ (U)LIRG morphological sample to date, which allows us to study the morphological evolution over $0.5<z<3$. We focus on the central question of whether (U)LIRGs could happen in normal disk galaxies and the relative importance of secular gas accretion versus merger in triggering (U)LIRGs. 
        
        This paper is organized as follows. We describe the data sets in Section 2 and the source selection in Section 3. The morphological analysis is detailed in Section 4. The implications are discussed in Section 5, and we conclude with a summary in Section 6. We adopt the standard $\rm \Lambda$CDM cosmology with $\Omega_M$ = $0.27$, $\Omega_\Lambda$ = $0.73$, and $H_0$ = $71$~km~s$^{-1}$~Mpc$^{-1}$. All magnitudes are in the AB system.
        
         \section{Data Description}
        
        Our samples are based on the FIR and submillimeter data in the field of COSMOS \citep[][]{Capak2007,Scoville2007}. The FIRGs were selected using the data from the Herschel Multi-tiered Extragalactic Survey \citep[HerMES;][]{Oliver2012}, while the SMGs were selected using the data from the S2CLS \citep[][]{Geach2017} and the SCUBA2 COSMOS Survey \citep[S2COSMOS;][]{Simpson2019}.  We obtained their morphological information mainly based on the HST WFC3 $H_{160}$ images from the COSMOS--Drift And SHift (COSMOS-DASH) program \citep[][]{Mowla2019}, and in some cases, we also used the ACS $I_{814}$ data from the original COSMOS program. To identify the counterparts of the FIRGs/SMGs on the HST  images, we used the high-resolution interferometry data from the Automated Mining of the ALMA Archive in the COSMOS Field \citep[A$^3$COSMOS;][]{Liu2019a}, as well as the deep 3~GHz radio data from the COSMOS-VLA program \citep[]{Smolcic2017a}. In a few cases, the identification was also aided by the $K_s$ data from the UltraVISTA program \citep[][]{McCracken2012}. These data are briefly described below.
        
        	 \subsection{ FIR Data from HerMES}
                
                The entire COSMOS field has been mapped by the HerMES program using SPIRE in 250, 350, and 500~$\mu$m. The FWHM beam sizes are 18\farcs1, 24\farcs9, and 36\farcs6 in these three bands, respectively. We used the maps and source catalog included in the fourth data release (DR4) of the program.\footnote{Released in 2016 July; see \url{https://hedam.lam.fr/HerMES/index/dr4}} The maps in the COSMOS field cover $\sim$4.76 deg$^2$. The three-band source catalog was built using extractions based on the ``blind'' 250~$\mu$m positions (referred to as “xID250”), which was done in the same way as in the earlier releases \citep[][]{Wang2014}. The 250 $\mu m$ map has the typical instrumental noise of 0.8 and confusion noise of 3.3 mJy beam$^{-1}$.

             \subsection{Submillimeter Data from S2CLS and S2COSMOS}

                The S2CLS program used the SCUBA2 instrument at the JCMT to survey a total of $\sim$5 deg$^2$ in 850~$\mu$m over eight different fields. Due to the immature termination of the program, however, the observations in the $\sim$2.2 deg$^2$ COSMOS field reached the designed depth only in half of the area. The average 1$\sigma$ depth of this half is 1.6 mJy beam$^{-1}$ and the image point-spread function (PSF) FWHM is 14\farcs8. We used the 850~$\mu$m map from the first data release (DR1) of the program.\footnote{Released in 2016 July; see \url{http://dx.doi.org/10.5281/zenodo.57792}}
                
                The S2COSMOS program was the extension of the S2CLS in the COSMOS field to complete what the S2CLS did not finish. By combining the data from both, the final S2COSMOS 850~$\mu$m map has reached the 1$\sigma$ depth of 1.2 mJy~beam$^{-1}$ over the ``main coverage'' (MAIN) of 1.6~deg$^2$ and 1.7~mJy~beam$^{-1}$ over an additional ``supplementary'' (SUPP) area of 1~deg$^2$. While the S2COSMOS source catalog has been made public, the new 850~$\mu$m map has not. Fortunately, the S2CLS DR1 map in the COSMOS field has incorporated partial data from S2COSMOS so that it also covers the other half of the field, although it does not reach the desired depth in this half. Therefore, we selected SMGs from the S2COSMOS catalog and used the map of the S2CLS whenever necessary.

             \subsection{HST Near-IR Data from COSMOS-DASH}
                
                The COSMOS-DASH program imaged the COSMOS field using the WFC3 in $H_{160}$. Its own observations consisted of 456 pointings covering a total of 0.49~deg$^2$. In addition, it incorporated the $H_{160}$ data from the CANDELS program and 10 other HST \, General Observer  (GO) programs. The program made a mosaic of all of these data using a uniform scale of 0\farcs1~pixel$^{-1}$, which covers 0.66~deg$^2$ in total. The PSF FWHM is 0\farcs 21 \citep[][]{Momcheva2017}.
                
                We used the v1.2.10 data release of the program.\footnote{Released in 2019 July; see \url{ https://archive.stsci.edu/hlsp/cosmos-dash}} In general, the mosaic has reached the 5$\sigma$ sensitivity of $H_{160}$ = $25.1$~mag for point sources (within an aperture of 0\farcs3 diameter), but in some regions, it is significantly deeper. The data release does not include the source catalog; therefore, we carried out the source extraction on our own using SExtractor \citep[][]{Bertin1996}. We weighted the science image by the included weight map and tuned the parameters to optimize the detection of faint sources.  Specifically, we used a 2 pixel FWHM, $5\times 5$ Gaussian convolution kernel for filtering; set the detection threshold to a factor of 0.7 above the background fluctuation; and required that a source must have at least 4 connected pixels above the threshold. In addition, we set the minimum contrast between flux peaks for deblending to 0.0001. To minimize the number of fake sources, we only retained the detections at $H_{160} \leqslant 27.0$~mag. Our final source catalog contains 499,093 objects.

             \subsection{HST Optical Data from COSMOS}
                
                The original COSMOS program obtained the ACS $I_{814}$ images over 1.7~deg$^2$. The image mosaic has a scale of 0\farcs 03~pixel$^{-1}$ and reaches the 5$\sigma$ sensitivity of $I_{814}=27.2$~mag. We used the products retrieved from the team's version 2.0 data release.\footnote{Released in 2010 Feb, see \url{https://irsa.ipac.caltech.edu/data/COSMOS/images/acs_mosaic_2.0/tiles/}}
                
             \subsection{$\rm K_{s}$-band Data from UltraVISTA}
                
                UltraVISTA was a deep near-IR survey in the COSMOS field and was carried out at the European Southern Observatory’s 4 m Visible and Infrared Survey Telescope for Astronomy (VISTA). We used DR4 of the program,\footnote{Released in 2019 Aug, see \url{https://irsa.ipac.caltech.edu/data/COSMOS/images/Ultra-Vista/mosaics/}} in particular the $K_s$-band mosaic over 1.5 $\times$ 1.2 deg$^2$ that reached the 5$\sigma$ depth of 24.5~mag (within an aperture of 2\farcs1 diameter). The PSF FWHM in $K_s$ is 0\farcs78.

             \subsection{Submillimeter Data from A$^3$COSMOS}
        
                The most ideal localization of FIRG/SMGs galaxies discovered by single-dish facilities is through submillimeter/millimeter interferometry. In the COSMOS field, there have been a significant number of such observations obtained by ALMA at various pointings. The A$^3$COSMOS program is an ongoing data mining project that takes advantage of these data. It processes the continuum data from the individual programs, detects  sources in the processed data, and provides source catalogs with photometry and other ancillary information. In this work, we used the processed images and the blind source catalog from its version 20180801 data release,\footnote{Released in 2018 July; see \url{https://sites.google.com/view/a3cosmos/data/dataset_v20180801}} which incorporates 1908 continuum images that cover a total of $\sim$280~arcmin$^2$ \citep[][]{Liu2019a}. The latest release is version 20200310; however, we found that it would only add a few objects to our final sample. As the images corresponding to this release are not yet publicly available, we chose not to use this version but to use the previous version instead.
                
                As the individual ALMA programs had a wide range of instrumental settings, the processed images vary significantly in their characteristics. First of all, these data were obtained at different wavelengths over ALMA Bands 3, 4, 6, 7, 8, and 9 (corresponding to a wide wavelength range from 3.4 to 0.4 mm). Second, the images have very different spatial resolutions, with the major beam size ranging from 0\farcs0175 to 4\farcs6. Third, their sensitivities also vary greatly. For example, the 1$\sigma$ sensitivity of the Band 6 ($\sim$1.1 mm) images varies from 5.6 to 131.8~$\mu$Jy~beam$^{-1}$. All of this creates difficulty in constructing our sample, and we will discuss this further in Section 3.2.
                
             \subsection{Radio Interferometry Data from the VLA-COSMOS 3 GHz Survey}
                
                As the A$^3$COSMOS data only cover a small fraction of the COSMOS field, we used the radio interferometry data from the VLA-COSMOS program, which imaged 2.6~deg$^2$ at 3 GHz. We used the images and associated catalog that were released in 2017 March.\footnote{Released in 2017 March; see \url{https://irsa.ipac.caltech.edu/data/COSMOS/tables/vla/} and \url{https://irsa.ipac.caltech.edu/data/COSMOS/images/vla/}} The 1$\sigma$ sensitivity of the map is 2.3~$\mu$Jy~beam$^{-1}$, on average, and the image FHWM is 0\farcs 75. The catalog consists of 10,830 3~GHz sources to 5$\sigma$ \citep[][]{Smolcic2017a}. A subset of 9161 sources \citep[][]{Smolcic2017b} have been matched with optical/near-IR counterparts using the catalogs of \citet[][]{Laigle2016}, \citet[][]{Ilbert2009}, and \citet[][]{Sanders2007}. For better consistency, however, we did not use this result but rather performed our own matching of the counterparts for our analysis, which will be detailed in Section 3.7.

         \section{Source selection}
                
        Here we describe the selection of the FIRGs and SMGs for this study and the use of the ALMA and VLA data for the counterpart identification.
                
             \subsection{Sources from the HerMES/S2COSMOS Catalogs}
                
                We selected the FIRGs from the HerMES DR4 catalog and the SMGs from the S2COSMOS catalog. For quality assurance, we aimed to study only the highly reliable sources and therefore applied the signal-to-noise ratio (S/N) cut at $\mathrm{S/N}\geqslant 5$ for both. For the FIRGs, their S/N values were calculated using $\mathrm{S/N}=f_{250}/et_{250}$, where $f_{250}$ and $et_{250}$ are the flux density and associated error (confusion noise included) in 250~$\mu$m, respectively. For the SMGs, we directly used the S/N values listed under the \texttt{snr} column. The requirement of $\mathrm{S/N}\geqslant 5$ resulted in 10,738 FIRGs and 574 SMGs.
                        
                We cross-matched these two samples to check how they overlap. The matching radius was chosen based on the positional uncertainties in both. Following Equations (1) and (2) in \citet[][]{Ma2015} (see also \citet[][]{Ivison2007b}),
                \begin{equation}
                \label{eqn:equ1}
                 \sigma_{\rm pos}=\frac{0.6}{S/N} \sqrt{\theta^{2}_{a}+\theta^{2}_{b}} =\frac{0.6\times 1.414 \times \theta}{S/N} ,
                \end{equation}
                
                where $\theta_{a}$ and $\theta_{b}$ are the beam size along the major and the minor axes, respectively, and $\theta$ is the total beam size when the beam is symmetric. For the FIRGs, their quoted positions are based on 250~$\mu$m; therefore, we adopted $\theta=18$\farcs1 because this is the beam size of 250~$\mu$m. For the SMGs, we adopted $\theta=14$\farcs9, which is the PSF FWHM as measured in the S2COSMOS map \citep[][]{Simpson2019} \footnote{The beam size of SCUBA2 at 850 $\mu m$ is 13\farcs1; however we used the measured PSF FWHM on the S2COSMOS map to be more generous in matching.}. Here we applied Equation 1 at the fixed $\mathrm{S/N}=5$. Putting this in for both 250 and 850~$\mu$m and adding the two terms in quadrature, we obtained a total positional matching error of $\mathrm{ \sigma_{pos}} = $ 3\farcs98. For simplicity, we used 4\farcs0 as the matching radius between the FIRGs and the SMGs. We obtained 280 common objects. 
                
                There are 10,458 and 294 unmatched FIRGs and SMGs, respectively. About 45\% of the unmatched FIRGs are outside of the S2COSMOS coverage. Within the S2COSMOS area, the vast majority of the unmatched SMGs are rather due to their having S/N $<5$ than being genuine nondetections in 250~$\mu$m. On the other hand, the vast majority of the unmatched FIRGs are genuine nondetections in 850~$\mu$m, which can be explained by the source redshift distribution (see  Section 5.2). Some of the unmatched sources are due to the blending problem. Generally speaking, the SMGs make up a subpopulation of the FIRGs, although our sample of SMGs does not constitute a complete subset of the FIRGs due to the choice of S/N threshold. For the ease of discussion, however, we still refer to our sources as FIRGs or SMGs in this paper.
                
             \subsection{Sources from the A$^3$COSMOS Blind Source Catalog}

                The blind catalog of A$^3$COSMOS contains 1144 records, and the same ALMA source can have multiple records if it was observed multiple times (e.g., in different bands or epochs and/or with different settings). Therefore, we must first consolidate the multiple records of the same ALMA source before deciding on which record to use for the FIRG/SMG counterpart identification.
                
                For this purpose, we used the Python package \texttt{pyfof},\footnote{https://github.com/simongibbons/pyfof} which is based on the friends-of-friends algorithm, to group the ALMA positions that are within 1\arcsec. This resulted in 859 unique sources. Some of them, however, are not suitable for the counterpart identification because of their low S/N. We calculated the source S/N based on the \texttt{Total\_flux\_pbcor} and \texttt{E\_Total\_flux\_pbcor} columns. For the sources that have multiple records, the highest S/N was chosen as the source S/N. We decided that we would only retain the ones that have $\mathrm{S/N}\geqslant 3$, which amount to 819 unique sources. We examined all of the images of these sources and removed six of them that are likely spurious. In the end, we obtained 813 unique ALMA sources, among which 652 and 161 have single and multiple records, respectively.
                
                For those that have multiple records, we used the one that has the smallest $\mathrm{ \sigma_{pos}}$ (according to Equation \ref{eqn:equ1}) for counterpart identification. The values for $\theta_a$ and $\theta_b$ were taken from the \texttt{Maj\_beam} and \texttt{Min\_beam} columns in the catalog, respectively.
                
             \subsection{Sources from the VLA 3~GHz Catalog}
                
                We used all of the 10,830 sources in the catalog of \citet[][]{Smolcic2017a}, which retains only the sources detected at S/N $\geqslant$ 5. A very small fraction of these sources are flagged in the catalog as being made of multiple components, but these multiple components (67 in total) do not enter the catalog individually; instead, the catalog includes them only as merged, unique objects and quotes only the averaged positions and the added flux densities from the individual components. In most cases, merging such individual radio components is justified because the components are parts of  single galaxies. However, we found a few cases where the merging would not be  appropriate because the radio components are clearly on different galaxies. In addition, there are a few more similar cases (beyond the reported 67) where a single entry in the catalog actually corresponds to two very close but still clearly separated sources on the VLA map; while these ``sources'' are not flagged in the catalog as being made of multiple components, we would have to separate the components because they correspond to different galaxies. We  will discuss such special cases in more detail in Section 3.7 if they enter our final morphological sample.
                
             \subsection{A$^3$COSMOS Counterparts of FIRGs/SMGs}

                The A$^3$COSMOS images are discrete and in most cases not centered on our targets. Due to the small field of view of ALMA, it is not uncommon that an A$^3$COSMOS image is near an FIRG/SMG but only covers a small fraction of the area 18\farcs1/14\farcs9 around the target center, which is not sufficient for the counterpart identification. Therefore, we should determine whether an FIRG/SMG is effectively covered by the A$^3$COSMOS data. To this end, we have decided that an FIRG/SMG is effectively within the A$^3$COSMOS coverage if the area within 3\arcsec \ in radius around its center is covered by an A$^3$COSMOS image.  Using this criterion, there are 879 FIRGs and 258 SMGs (146 in common) effectively covered by A$^3$COSMOS.
                
                In determining the matching radius to the A$^3$COSMOS sources, we considered the positional errors due to the uncertainties in both HerMES/S2COSMOS and A$^3$COSMOS, which we denote as $\mathrm{ \sigma_{pos}^{FIR}}$ or $\mathrm{ \sigma_{pos}^{SMG}}$, and $\mathrm{ \sigma_{pos}^{A}}$, respectively. These terms were calculated based on Equation (\ref{eqn:equ1}) and then were added in quadrature to obtain the total error, $\mathrm{ \sigma_{pos} =  \sqrt{( \sigma_{pos}^{FIRGs/SMGs})^2 + ( \sigma_{pos}^{A})^2}}$. In most cases, the former is the dominant term. For the FIRGs and the SMGs, we use $\mathrm{\theta = 18\farcs1}$ and 14\farcs9 in Equation 1, respectively. For reference, these correspond to $\mathrm{ \sigma^{FIRGs}_{pos} = 3\farcs07}$ and $\mathrm{ \sigma^{SMGs}_{pos} = 2\farcs53}$, respectively, at $\mathrm{S/N = 5}$ and are smaller at higher S/N. To simplify the matching process, we followed the approach in \citet[][]{Yan2020}. We first searched for A$^3$COSMOS sources within the radius of $\theta$ for a given FIRG/SMG. Then we calculated the positional offsets, $\mathrm{\Delta_{pos}}$, between the center of the FIRG/SMG and the centers of all of the A$^3$COSMOS counterpart candidates and determined the ratios $\mathrm{\Delta/ \sigma}$ between $\mathrm{\Delta_{pos}}$ and $\mathrm{ \sigma_{pos}}$. The matches with either $\mathrm{\Delta_{pos}}\leqslant$~3\arcsec \ or $\Delta/ \sigma\leqslant 3$ were deemed to be the corrected counterparts. These criteria were slightly different from those were used in \citet[][]{Yan2020}, but what we adopted here were more appropriate for this current analysis.
                
                We applied additional treatment for some special cases. First of all, the common objects among the FIRG and the SMG samples have two possible choices of positions, namely, the  HerMES positions and the S2COSMOS positions. Therefore, for each of them, we matched its HerMES and S2COSMOS positions with A$^3$COSMOS sources individually and then combined all of the matched A$^3$COSMOS sources together.
                
                In total, we found 413 FIRGs and 223 SMGs (132 sources in common) that were matched by 544 A$^3$COSMOS counterparts. Among these 504 unique FIRGs/SMGs, 444 and 60 have single and multiple A$^3$COSMOS counterparts, respectively.
                
                We also extended the identification to those whose ALMA coverages are smaller than the effective coverage as defined above. This was done following the same procedures above, as the identification method itself does not depend on the size of the coverage. We identified four more FIRGs and two more SMGs (one in common), which correspond to five A$^3$COSMOS counterparts.  Combining both, we obtained 509 unique FIRGs/SMGs (417 FIRGs, 225 SMGs, and 133 in common) with 549 A$^3$COSMOS counterparts. 
                
                There are 466 FIRGs and 35 SMGs (14 in common) that have effective A$^3$COSMOS coverage but lack identification. The simplest explanation is that the current ALMA data are still not deep enough. There are two major reasons for this. First, if the ALMA observations were only done in Cycle 0 when the array just started to operate and with limited capacity, the images are all much too shallow as compared to the rest of the A$^3$COSMOS data. The second reason could be that the ALMA observations were only done in Band 3 or 4, whose wavelength coverage is around 3 and 2 mm, respectively. These are far away from the wavelength ranges where the FIRGs/SMGs were selected and therefore only sample the long-wavelength tail of the emission that could be too faint to be seen at the existing sensitivity levels. 
                
                However, there are still 227 FIRGs and 13 SMGs (five in common) whose lacking ALMA identifications cannot be explained by the above reasons. On the other hand, we note that this does not necessarily present a new problem. For example, \citet{Stach2019} reported that, out of the 716 AS2UDS ALMA Band 7 maps targeting the S2CLS SMGs in the UDS field, 101 (or 14\%) were “blank” maps that do not contain a counterpart. We will defer the investigation of this problem to a future study.
                
             \subsection{VLA Counterparts of FIRGs/SMGs}
                
                The VLA map covers nearly all of the S2COSMOS area; in fact, only one SMG, which is the northernmost one in the S2COSMOS catalog, is outside of the VLA coverage. However, it only covers $\sim$55\% of the HerMES map and hence only the same fraction of the FIRGs. In summary, the VLA map covers 5884 FIRGs and 573 SMGs, of which 279 sources are in common.
                
                In a similar way, we identified the VLA counterparts for all of the FIRGs/SMGs. We adopted $\theta=$~0\arcsec.75 in Equation (\ref{eqn:equ1}) to determine the $\mathrm{ \sigma_{pos}^{VLA}}$ of the VLA sources and calculated $\mathrm{ \sigma_{pos}= \sqrt{( \sigma_{pos}^{FIRG/SMG})^2+( \sigma_{pos}^{VLA})^2}}$ for the matching. Again, the matches with either $\mathrm{\Delta_{pos}}\leqslant$~3\arcsec \ or $\Delta/ \sigma\leqslant 3$ were deemed to be the corrected counterparts.
                
                In total, we found 3318 FIRGs and 432 SMGs (252 sources in common) matched by 3826 VLA counterparts.  Among these 3498 unique FIRGs/SMGs, 3145 and 353 have single and multiple VLA counterparts, respectively. These include 22 pairs of very close FIRGs that we merged as single FIRGs. For each of these pairs, the two members might have one or multiple VLA counterparts, and they share at least one common VLA counterpart; therefore, we regarded the pair as the result of oversplitting of a single source in HerMES. When we merged the pair, we also grouped the noncommon VLA counterparts with the common ones as the final counterparts of the merged FIRG.
                
                The VLA data, albeit already being very deep, still only identify $\sim$56.4\% of the FIRGs and $\sim$75.2\% of the SMGs. The identification rate for the SMGs is higher than that of \citet[][]{An2019}, who showed that the same VLA data identified $\sim$69\% of the S2COSMOS SMGs. The difference is presumably due to the different methods of identification.

             \subsection{Comparison of Counterparts between A$^3$COSMOS and VLA}
                
                As mentioned in  Section 2.6, we explicitly assume that the ALMA submillimeter/millimeter interferometry provides the most ideal counterpart identification for FIRGs/SMGs. Therefore, we need to assess how reliable the VLA counterpart identification is by comparing to the A$^3$COSMOS identification result. We break this into two questions.
                
                We first answer this question: for the 549 A$^3$COSMOS counterparts of the 509 FIRGs/SMGs identified by A$^3$COSMOS (in  Section 3.4), how many of them are correctly recovered by the VLA identification (in  Section 3.5)? For this purpose, we match the A$^3$COSMOS and the VLA counterparts and calculate $\mathrm{\Delta_{pos}}$ and $\mathrm{ \sigma_{pos}}$ between them. The matches with either $\mathrm{\Delta_{pos}}\leqslant$~1\arcsec \ or $\Delta/ \sigma\leqslant 3$ are deemed to be the corrected matches. 
                
                We find that there are 446 correct VLA matches out of the a total of 549 A$^3$COSMOS counterparts, or a recovery rate of 81.2\%. We can also look at this problem from a different perspective. The 446 correct matches belong to 373 FIRGs and 193 SMGs (127 in common). This means that 89.4\%/85.8\% of the FIRGs/SMGs identified by A$^3$COSMOS can also be identified by the VLA data. There is a subtlety, however, for those that have multiple A$^3$COSMOS counterparts. The VLA data do not always identify all of the multiple counterparts. Among those ``correctly'' identified FIRGs/SMGs, 51 FIRGs and 40 SMGs (31 in common) have multiple A$^3$COSMOS counterparts; 15 FIRGs and seven SMGs (five in common) of them have all of their multiple A$^3$COSMOS counterparts identified by the VLA data. Therefore, we can safely conclude that the VLA identification recovers the A$^3$COSMOS identification at the 80\% level or better.

                As discussed in  Section 3.4, there are FIRGs/SMGs effectively covered by the A$^3$COSMOS data but are not identified by the A$^3$COSMOS. This leads to the second question: how many of those have counterparts claimed by the VLA identification? This question is relevant as we expand the morphological study to the FIRGs/SMGs that are beyond the small coverage of A$^3$COSMOS but can still be identified by the VLA data. 
                
                To make a fair assessment, here we only use those that are not identified by ALMA but have data beyond Cycle 0 and not only in Bands 3 and 4. Among these 227 FIRGs and 13 SMGs (five in common), the VLA data identified 151 FIRGs and six SMGs (three in common).  Recalling that the A$^3$COSMOS identified sources including 417 FIRGs and 225 SMGs (133 in common), the addition  by the VLA identification would be substantial for the FIRGs ($\sim$37.5\% more) but negligible for the SMGs ($\sim$2.8\% more).
        
                As mentioned above, within the A$^3$COSMOS coverage, the VLA identifications recover at least 80\% of the A$^3$COSMOS identifications. Therefore, it is reasonable to believe that the contaminations in this addition are at the $<$20\% level.

             \subsection{Final Sample for Morphological Study}
            
	            \begin{figure*}[htbp!]  
	            	\centering
	            	\includegraphics[width=.85\textwidth]{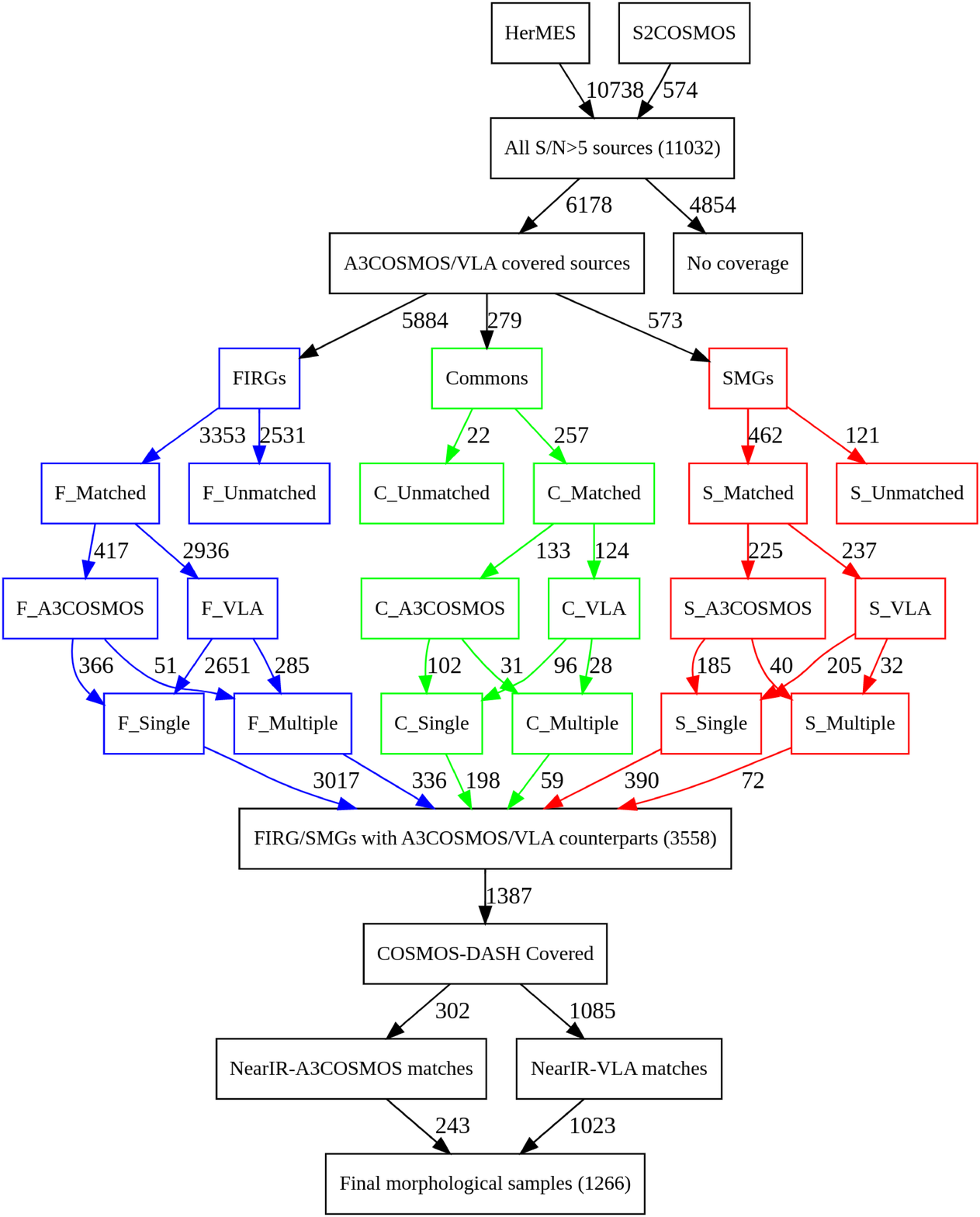} 
	            	\caption{Flowchart demonstrating the counterpart identification process and the construction of the final morphological sample.}
	            	\label{fig:chain}
	            \end{figure*}

                The above procedures resulted in 3353 FIRGs and 462 SMGs (257 in common) that have A$^3$COSMOS and/or VLA identifications, which we searched for the near-IR counterparts in the COSMOS-DASH data. For those that have A$^3$COSMOS identifications, we ignored the VLA ones. As mentioned earlier, these amount to 549 A$^3$COSMOS objects. For those that do not have A$^3$COSMOS identifications, we used the VLA results, which amount to 3329 objects. As discussed above, these VLA objects were all treated as the right counterparts. However, only 302 and 1085 A$^3$COSMOS and VLA objects fall within the COSMOS-DASH coverage, which corresponds to 1157 FIRGs and 232 SMGs (121 in common). 
                
                \begin{figure*}[htbp]  
                	\centering
                	\includegraphics[width= \textwidth]{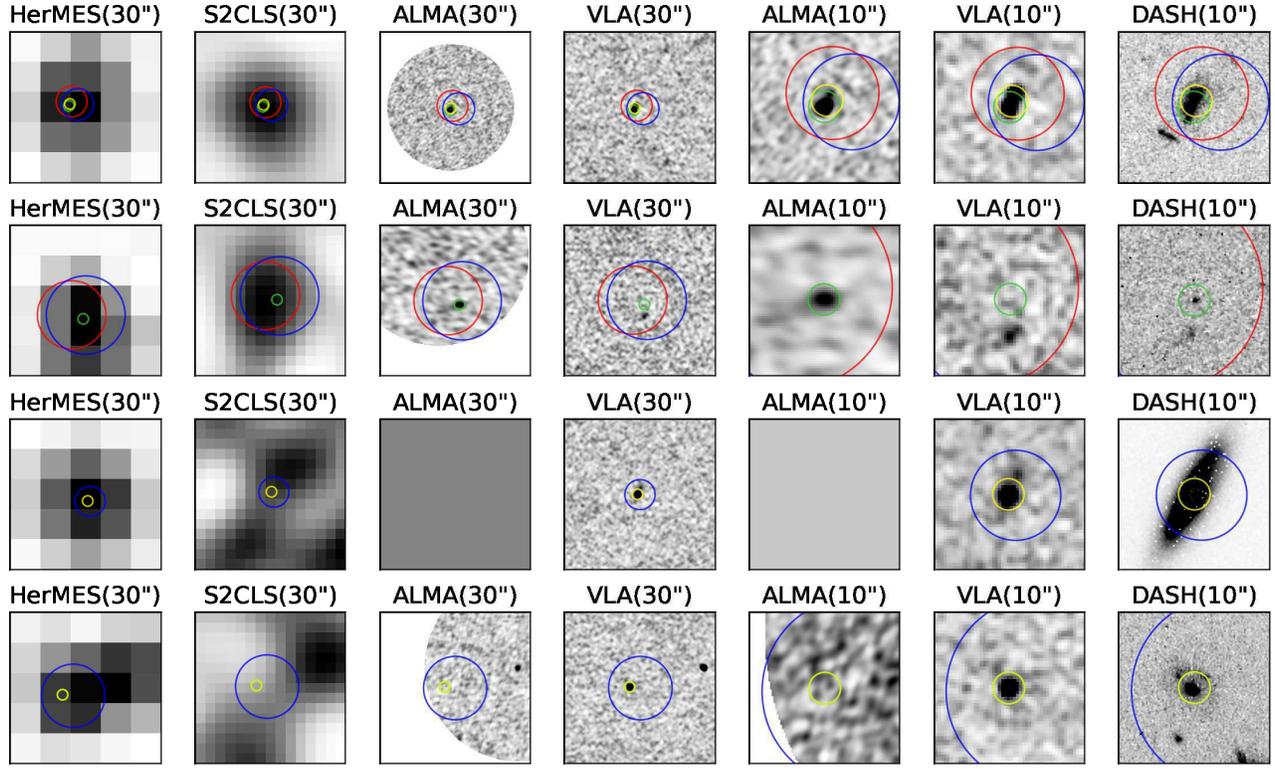} 
                	\caption{Demonstration of FIRG/SMG counterpart identification in various cases. All images are oriented with north to the top and east to the left. Four cases are shown, each in one row. From top to bottom, these are a common FIRG/SMG that has identification in both ALMA and VLA, a common FIRG/SMG that has an ALMA identification that is different from what the VLA would identify (and the former was adopted), an FIRG that is not an SMG and has only a VLA identification due to the lack of ALMA data, and an FIRG that has no ALMA identification within the coverage but has a clear VLA identification. From left to right in each row, the first four images (30\arcsec \ $\times$ 30\arcsec\, in size) are from the maps of the HerMES SPIRE 250~$\mu$m, S2CLS 850~$\mu$m (as the S2COSMOS map is not yet publicly available), A$^3$COSMOS in the adopted band, and VLA 3~GHz, respectively; the next two (10\arcsec \ $\times$ 10\arcsec) are the zoomed-in images of the A$^3$COSMOS and VLA cutouts, respectively; and the last one is the COSMOS-DASH $H_{160}$ cutout (10\arcsec \ $\times$ 10\arcsec). The blue and the red circles are centered on the HerMES and the S2COSMOS positions, respectively, and they show the matching radii when searching for the ALMA/VLA counterpart as described in  Section 3.4 and 3.5, respectively. The green and the yellow circles are centered on the A$^3$COSMOS and the VLA counterparts, respectively, and their size (1\arcsec.05 in radius) indicates the matching radius when searching for the $H_{160}$ counterparts.}
                	\label{fig:source_id}
                \end{figure*}

                The search for near-IR counterparts was done in two passes. First, we matched the A$^3$COSMOS/VLA positions to our COSMOS-DASH catalog using a matching radius of 1\farcs05 (five times the PSF size of the $H_{160}$ mosaic). This generous value was chosen for a reason; as the dust-enshrouded, young star-forming regions do not necessarily coincide with the regions dominated by mature stellar populations, there could be a notable offset between these two regions if the galaxy is large enough in the image. After a few trials, we found that this matching radius was the most appropriate in dealing with such cases and yet only resulting in a low number of incorrect matches. With this criterion, we obtained 1281 matches.
                
                In the second pass, we visually inspected both the matched and unmatched A$^3$COSMOS/VLA objects to ensure that the matches are of sufficient quality and that there were no counterparts missing from the first pass. We used all of the images mentioned in Section 2. We found that we had to remove 25 matches for two reasons: 12 of them are too close to the boundary region of the COSMOS-DASH mosaic, and the other 13 are wrong matches. In the latter case, we knew that these matches were wrong because the correct counterparts were seen in the UltraVISTA $K_s$ image but invisible in the COSMOS-DASH images. Lastly, we need to remove the $z$ = 5.667 galaxy reported by \citet[][]{Pavesi2018}, as it is completely blocked by a foreground galaxy in the optical to near-IR, and the $H_{160}$ image does not reflect the morphology of the real host.
                
                This pass was also where we validated the aforementioned matching radius. If the matching radius were to be reduced to 0\arcsec.63 (3 $\times$ the PSF size), we would miss 13 genuine matches. If it were to be relaxed to 1\arcsec.47 (7 $\times$ the PSF size), we would include 17 additional matches. However, 16 of them are clearly incorrect matches; only one match is valid, a nearby spiral galaxy whose VLA position is on one of its arms and is 1\arcsec.06 away from the $H_{160}$ centroid. Therefore, this object was added to the sample.
                
                Recall that there are 67 radio sources flagged in the VLA catalog as being the merged products of individual components. As discussed in  Section 3.3, some of them should not actually have been merged into single sources. Two of them are in our final morphological sample. Fortunately, the VLA identifications of these two do not matter because one has its individual components separated in the A$^3$COSMOS catalog and the other clearly corresponds to a nearby galaxy. On the other hand, there are 10 additional matches (beyond the aforementioned 67) that each correspond to more than one object in $H_{160}$. These 10 matches are outside of the ALMA coverage. While they are listed as single sources in the VLA catalog, each of them clearly shows two well-separated components in the VLA image that have counterparts in the $H_{160}$ images. Therefore, we should treat these 10 matches as 20 individual $H_{160}$ objects for the morphological study.
                
                We also noticed that 16 pairs of S2COSMOS and HerMES sources have the same A$^3$COSMOS and/or VLA counterparts, while the calculated separations between their HerMES/S2COSMOS positions are larger than the matching radius that is mentioned in  Section 3.1. These S2COSMOS/HerMES pairs are still considered as common sources in the further discussion. Furthermore, nine pairs of HerMES sources have the same VLA counterparts, and these pairs should be due to oversplitting of the same HerMES sources. Therefore, we considered these HerMES pairs as single FIRGs and merged their fluxes in the three SPIRE bands. In the end, our final catalog for morphological study contains 1266 $H_{160}$ counterpart systems from COSMOS-DASH, among which 243 counterparts were identified by the A$^3$COSMOS catalog and 1023 counterparts were identified by the VLA catalog. They belong to 1090 FIRGs and 172 SMGs (117 in common). As our identification process involves many details that could be confusing, we summarize this process in Figure \ref{fig:chain} as a flowchart. Figure \ref{fig:source_id} further demonstrates how the $H_{160}$ counterpart of an FIRG/SMG is located in a few representative (but not exhaustive) cases.
                
                We obtained the redshifts for our sample using a number of spectroscopic ($z_{\rm spec}$) and photometric ($z_{\rm ph}$) redshift  catalogs in the COSMOS field, which are detailed in Appendix \ref{appendix:a}. Among the 1266 $H_{160}$ objects, 1235 have reliable redshift information (623 $z_{\rm spec}$ and 612 $z_{\rm ph}$).

         \section{Morphologies of FIRGs/SMGs}
                
        Our analysis of the morphologies is mainly based on visual classification, which was done on all the 1266 systems described in  Section 3.7. As an independent check on the visual classifications, we fitted analytic profiles to those that have regular light distributions as judged visually, and the agreement is very good. We also applied a number of commonly used nonparametric statistics to derive quantitative morphological indices as an aid to our analysis. These are detailed below.
                
        	 \subsection{Visual Classification}
                
            Our visual classification was mainly based on the COSMOS-DASH $H_{160}$ image. The inspection was done on the $H_{160}$ image cutouts, which are 10\arcsec \ $\times$ 10\arcsec \ in size and are centered on the A$^3$COSMOS/VLA positions. We classified the 1266 systems into five categories as follows.
                
            (1) {\it Elliptical (E)}: These objects are spheroid-like and without additional structures. In total, there are 41 objects in this category.
                
            (2) {\it Disk (D)}: These objects are disk galaxies with or without spiral arms. The decisive feature is a disk of regular, undisturbed shape. Some of them contain a central bulge, which can be of various degrees of prominence. There are 548 objects in this category.
                
            (3) {\it Irregular/Interacting (Irr/Int)}: These objects, some of which clearly consist of multiple members, have features suggestive of interactions or recent mergers. These features include: (a) tidal tail(s), (b) overlapping component(s) from different members, (c) clumpy appearance, and (d) obvious offset of nucleus from the center. The membership of multiples is based on the redshift information as described in Appendix \ref{appendix:a}. If $z_{\rm spec}$ are available, we required that the redshift difference of members be $\Delta z_{\rm spec}\leqslant 0.03$. If only $z_{\rm ph}$ are available, we adopted the criterion of $\Delta z_{\rm ph}\leqslant 0.1$ for membership. At $z = 1.5$, $\Delta z_{\rm ph} = 0.1$ translates to $\Delta z_{\rm ph}/(1+z)=0.04$, which is the typical $\Delta z_{\rm ph}$ accuracy. However, there are systems whose individual members are not resolved in the redshift catalog because they are very close to each other ($<$1\arcsec \ apart). In this case, the system was also deemed as in interaction. In total, there are 472 systems classified as Irr/Int. We emphasize that two members being close to each other would not necessarily make them both be classified as ``Int'', to be classified as such, both members should have at least one of the aforementioned features (see also  Section 5.1).
                
            (4) {\it Compact/Unresolved (C)}: These objects are either star-like quasars or unresolved sources with limited pixels. There are 61 objects in this category.
                                
            (5) {\it Faint (F)}: These objects are very faint sources with low S/N and no reliable classification could be performed. There are 144 objects in this category.

          	\begin{figure}[htbp!]  
          		\centering
          		\includegraphics[width=.45\textwidth]{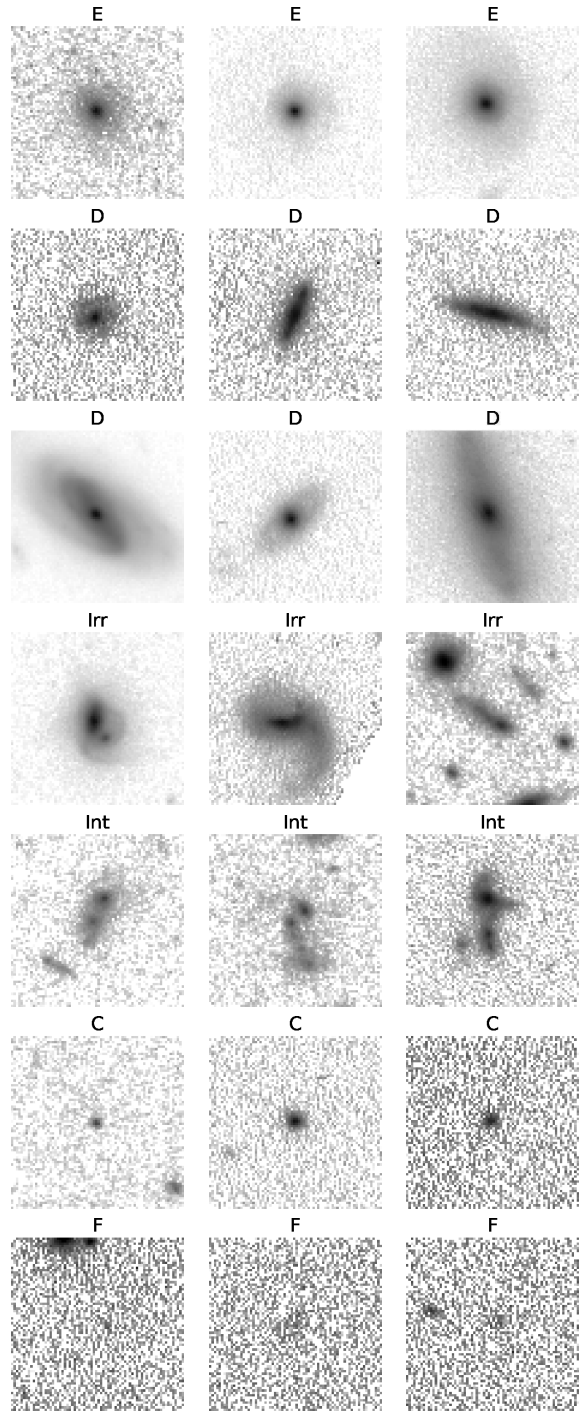} 
          		\caption{Examples of $H_{160}$ images (15\arcsec \ $\times$ 15\arcsec ~ in size) for the five visual classification categories (E, D, Irr/Int, C, and F) as labeled. In particular, two rows are used to show six examples of D galaxies with bulges of various degrees of prominence. The examples for the Irr/Int category are demonstrated in two rows for the Irr and the Int subtypes separately.}
          		\label{fig:stamps}
          	\end{figure}
          
            The classification was done by both authors. We first independently classified a common set of 768 objects and then compared and reconciled the differences. The reconciled results agreed with the two original, individual sets of classifications to 85.4\% and 81.6\%, respectively. One of us (C.L.) then classified the remaining 498 objects alone, using the experience ``calibrated'' based on the reconciliation process. Figure \ref{fig:stamps}  shows the $H_{160}$ image stamps for some random examples in the aforementioned categories.
          
          	Our objects span a wide redshift range of $0.1\lesssim z\lesssim 3$ (only a few are outside of this range). This means that $H_{160}$ samples different rest-frame wavelengths for objects at different redshifts, and therefore the morphological $k$-correction could be a concern. To guard against this possible bias, we further carried out visual classification for the 523 systems at $z<1$ in the sample (30, 342, 145, 5, and 1 systems in the E, D, Irr/Int, C, and F categories, respectively, as described above when classifying based on the $H_{160}$ image) using the ACS $I_{814}$ image, which reflects their morphologies in the rest-frame optical. The consistency between the two sets of results is excellent: the classification agrees to 96.7\%, 92.4\%, 98.6\%, and 40\% in the E, D, Irr/Int, and C categories, respectively. The largest difference is in the F category, which contains only one object at $z<1$ and therefore would not affect any of our follow-up analysis.
          	          	 
        	 \subsection{S\'ersic Profile Fitting}        
                
            To obtain quantitative measurements of galaxy morphologies, it is a common practice to fit the light distributions to the classic S\'ersic profile \citep[][]{Sersic1963},
            
            \begin{equation}
            	I(R) = I_e\exp\left\{ b_n\left[ \left(\frac{R}{R_e}\right)^{1/n}-1\right]\right\},
            \end{equation}
            where $R_e$ is the half-light radius, and $I_e$ is the intensity at $R_e$. When $n=1$, the S\'ersic profile reduces to the exponential profile that describes a pure disk galaxy, while when $n=4$, it becomes the de Vaucouleurs profile that describes an elliptical galaxy.
            
            We used the GALFIT software \citep[version 3;][]{Peng2002, Peng2010} to carry out this analysis. We confined it to only the galaxies in the E and D categories. In particular, we did not fit the Irr/Int galaxies because fitting a regular profile to irregular features would not be meaningful. 
     		
\begin{figure}[htbp!]  
                	\centering
                	\includegraphics[width=.5\textwidth]{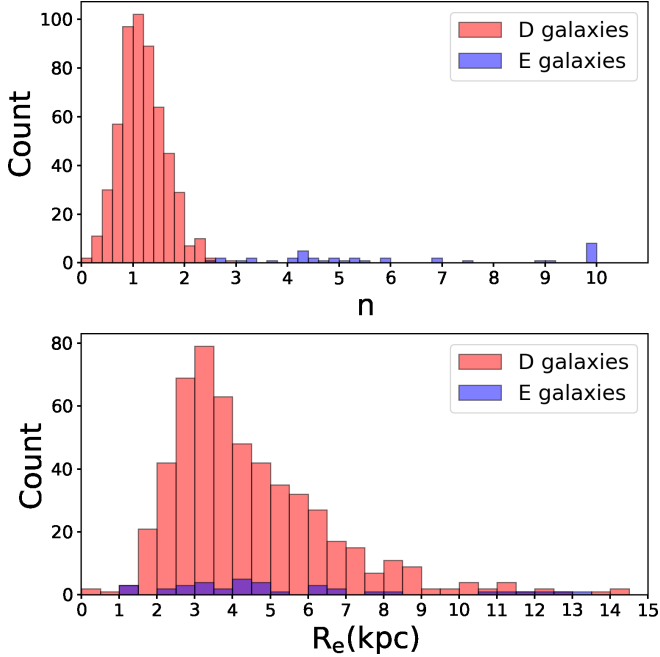} 
                	\caption{S\'ersic index ($n$) and effective radius ($R_e$) distributions of the D and the E galaxies in our visual classifications, shown in red and blue, respectively. The medians are $n=1.1$ and 5.3, respectively. The median $R_e$ of the D and E galaxies are 3.9 and 4.8 kpc, respectively.}
                	\label{fig:sersic}
                \end{figure}

     		The fitting procedures are detailed in Appendix \ref{appendix:b}. The results are shown in Figure \ref{fig:sersic}, which gives histograms of the fitted $n$ value and effective radius $R_e$ for the D  and E galaxies. The latter has been converted from angular size to physical size according to the source redshift. As mentioned already, our focus on the D galaxies is whether they really possess a disk component. Therefore, if a D galaxy is fitted with two components, only the $n$ value for the primary component is shown in this figure. The medians are $n=1.1$ and 5.3 for the D and the E galaxies, respectively, which are fully consistent with our visual classifications. The median $R_e$ of the D and the E galaxies are 3.9 and 4.8 kpc, respectively.
                
             \subsection{Nonparametric Morphologies}
                
            Nonparametric approaches are also widely used in morphological study. As compared to the analytic profile fitting, such approaches are more robust and flexible. Therefore, we also carried out nonparametric analysis utilizing the Python package \texttt{statmorph} \citep[][]{Rodriguez-Gomez2019}. We focused on three popular statistics, namely, Gini-$M_{\rm 20}$, CAS, and MID.
            
            The routine was run on the E, D, and Irr/Int categories. The segmentation map was directly imported from SExtractor, and we manually subtracted the background of our image (see Appendix \ref{appendix:b1}). Most of the galaxies had ``good'' fitting results (with \texttt{Flag = 0} in the output), but some did not. Such failures were due to a variety of reasons (e.g., when there is some artifact, foreground star, or secondary source in the image that was not properly masked, etc.), and they were discarded. In total, we obtained ``good'' results for 864 galaxies (out of the total of 1061).
                  
	             \subsubsection{Gini-$M_{\rm 20}$ Statistics}
	            
	            \begin{figure}[htbp!]  
	            	\centering
	            	\includegraphics[width=.5\textwidth]{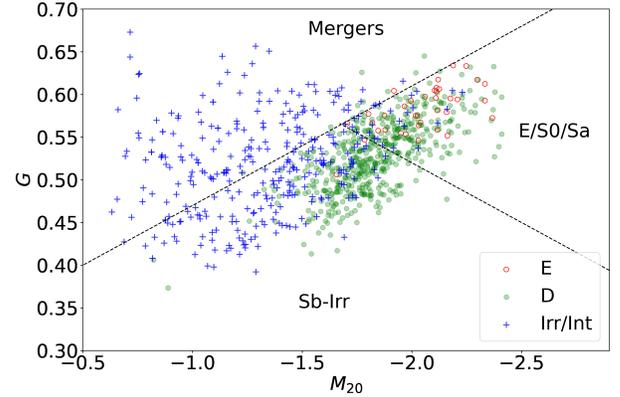} 
	            	\caption{ Gini-$M_{\rm 20}$ statistics of the E (open red circles), D (filled green circles), and Irr/Int (blue plus signs) galaxies in our sample (864 in total). The two dashed lines are taken from \citet[][]{Lotz2008} and separate their galaxies into Mergers, E/S0/Sa and Sb-Irr. Our visual classification is broadly consistent with this analysis.}
	            	\label{fig:Gini-M20}
	            \end{figure}
	                
	            The Gini-$M_{\rm 20}$ classification \citep[][]{Lotz2004} has been used extensively to quantify galaxy morphology. The Gini coefficient measures the relative distribution of the galaxy pixel flux values. For a galaxy occupying $n$ pixels, its Gini coefficient can be calculated as follows:
	            
	            \begin{equation}
	            G = \frac{1}{\overline{f}n(n-1)} \sum_{i=1}^{n}(2i-n-1)f_{i},
	            \end{equation}
	            where $f_i$ is the flux value of the $i$th pixel and is sorted into increasing order, and $\overline{f}$ is the mean flux value. $G$ value ranges from zero to 1, with the low (high) value indicating more uniform (concentrated) distribution.

	            The $M_{\rm 20}$ index measures the normalized second-order moment of the brightest 20\% of the galaxy’s flux. For a galaxy with $n$ pixels and flux $f_i$ in the $i$th pixel, the total second-order moment $M_{\rm tot}$ is calculated as
	            \begin{equation}
	            	M_{\rm tot} \equiv  \sum_{i}^{n}M_{i} =  \sum_{i}^{n}f_i  \left[ (x_i-x_c)^2+(y_i-y_c)^2 \right],
	            \end{equation}          
	            where ($x_c$, $y_c$) is the central position that minimizes $M_{\rm tot}$. To compute $M_{\rm 20}$, the pixels are rank-ordered by flux, and the sum of $M_i$ is done over the brightest pixels until it reaches 20\% of the total galaxy flux $f_{\rm tot}$, and then normalized by $M_{\rm tot}$:
	            \begin{equation}
	            	M_{\rm 20} = \rm log_{10} \it \left(\frac{ \sum_{i}^{n'}M_i}{M_{\rm tot}}\right) \rm while \ \it  \sum_{i}^{n'}f_i \leqslant 0.2 f_{\rm tot},
	            \end{equation}  
	            
	            \figurename{ \ref{fig:Gini-M20}} shows the $G$--$M_{\rm 20}$  distribution of the 864 galaxies that have good results from the \texttt{statmorph} run, where our visual classifications are denoted by different symbols as labeled. The three regions separated by dashed lines are those occupied by mergers, E/S0/Sa, and Sb-Irr, as in \citet[][]{Lotz2008}, where the lines are
	            \begin{equation}
	            	\left\{
	                	\begin{array}{lr}
	                    	G = -0.14M_{\rm 20} + 0.33 &  \\
	                        G = 0.14M_{\rm 20} +0.80, &  \\
	                    \end{array}
	                \right.
	            \end{equation}
		        
        		In general, the results from the Gini-$M_{\rm 20}$ statistics are consistent with those
        		from our visual classification; nearly all of our E galaxies are in the region occupied by the “E/S0/Sa” galaxies of \citet[][]{Lotz2008}, and the vast majority of our D galaxies fall in the place where their disk galaxies would populate.  However, a significant fraction of D galaxies in our visual classification fall in their E/S0/Sa region.

                \subsubsection{CAS Statistics}
               
               \begin{figure*}[htbp!]  
               	\centering
               	\includegraphics[width=.9\textwidth]{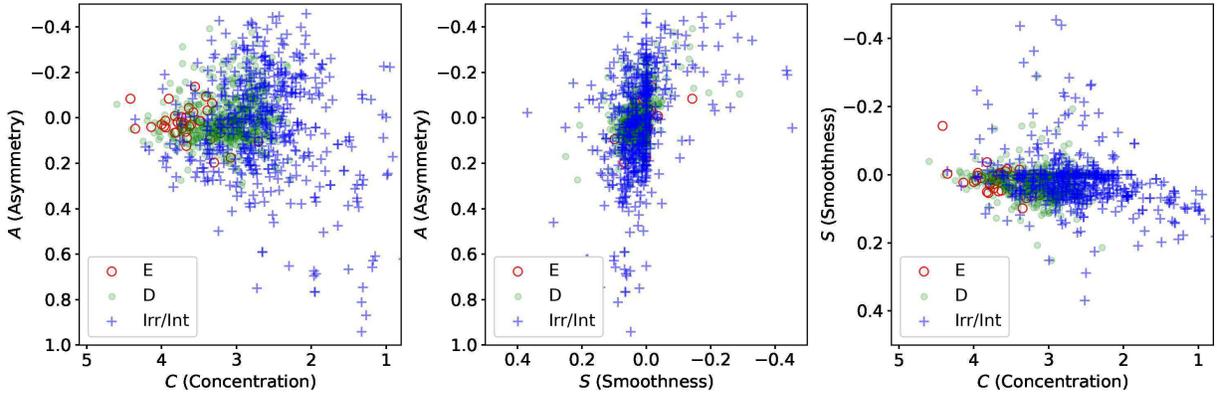} 
               	\caption{The CAS statistics of the galaxies in our sample. Symbols are the same as in \figurename{ \ref{fig:Gini-M20}}.}
               	\label{fig:CAS}
               \end{figure*}
                
               The CAS statistics \citep[][]{Conselice2003} is another widely used nonparametric morphological classification tool. The concentration index, $C$, is defined as
               \begin{equation}
                        C =     5 \rm log_{10} \it \frac{R_{\rm 80}}{R_{\rm 20}},
               \end{equation} where $R_{\rm 20}$ and $R_{\rm 80}$ are the radii that contain 20\% and 80\% of the total light (based on the curve-of-growth analysis), respectively.
           
           	   The asymmetry index, $A$, is obtained by subtracting the galaxy image rotated by 180${^\circ}$ from the original. It can be calculated as
               \begin{equation}
                        A = \frac{ \sum_{i,j} |f_{i,j}-f_{i,j}^{180^\circ}|}{ \sum_{i,j} |f_{i,j}|} -A_{\rm bgr},
               \end{equation}
               where $f_{ij}$ and $f_{i,j}^{180^\circ}$ are the pixel flux values of the original and rotated images, respectively, and $A_{\rm bgr}$ is the average asymmetry of the background.

               The smoothness index, $S$, or ``clumpiness,'' is obtained by subtracting the galaxy image smoothed with a boxcar filter of width $  \sigma = 0.25  r_{\rm Petro}$ from the original image, where $r_{\rm Petro}$ is the Petrosian radius of the galaxy. It can be calculated as
                \begin{equation}
                        S= \frac{ \sum_{i,j} |f_{i,j}-f_{i,j}^{S}|}{ \sum_{i,j} |f_{i,j}|} -S_{\rm bgr},
                \end{equation}          
                where $f_{ij}$ and $f_{i,j}^{S}$ are the pixel flux values of the original and smoothed images, respectively, and $S_{\rm bgr}$ is the average smoothness of the background.
                                
                \figurename{ \ref{fig:CAS}} shows the CAS statistics of our sample. Apparently, this set of indices does not work well in separating the galaxies in our categories.
                
                 \subsubsection{MID Statistics}
                
                 \begin{figure*}[htbp!]  
                	\centering
                	\includegraphics[width=.9\textwidth]{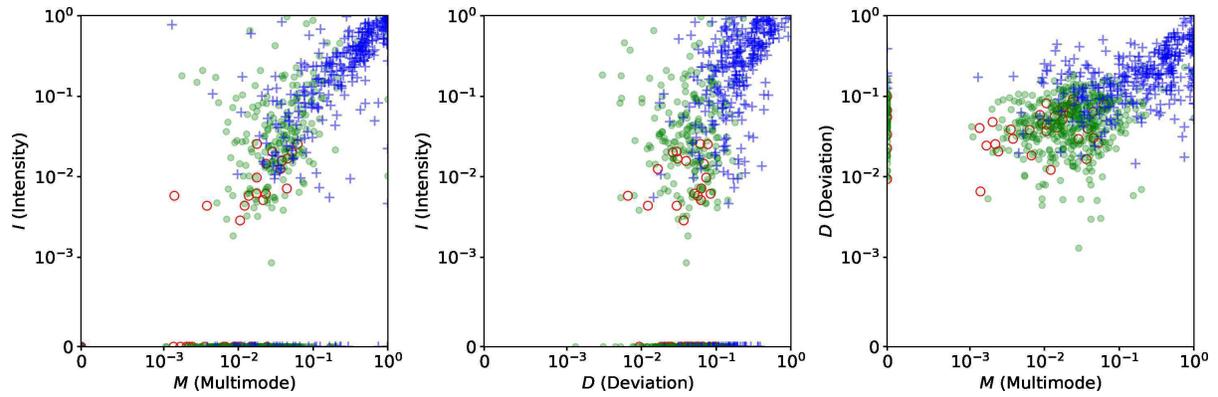} 
                	\caption{The MID statistics of the galaxies in our sample. Symbols are the same as in \figurename{ \ref{fig:Gini-M20}}.}
                	\label{fig:MID}
                \end{figure*}
                
                The MID statistics \citep[][]{Freeman2013,Peth2016} is presumably more sensitive to merger features. The multimode statistics, $M$, measures the ratio of the areas occupied by two brightest regions of a galaxy and works as follows.  For a given galaxy, its pixel fluxes are normalized, and a flux threshold $q$ is chosen to discard any pixel below $q$. The remaining pixels form multiple contiguous areas, which are sorted by size in descending order and denoted as $A_{q,1}$, $A_{q,2}$, … $A_{q,n}$. In practice, $q$ runs from zero to 1 in small steps, and the ratio of the two brightest clumps,  $A_{q,1}/A_{q,2}$, is calculated for each $q$. The $M$ value is set as the maximum of this ratio:
                \begin{equation}
                        M = \max_{q}\frac{A_{q,1}}{A_{q,2}},
                \end{equation}
            
                The intensity statistic, $I$, measures the ratio between the two brightest subregions of a galaxy. To calculate it, the image is first smoothed by a Gaussian kernel with $\sigma =1$ pixel. Then the image is partitioned into distinct pixel groups using the watershed algorithm, so that the pixel intensity gradient in any direction within each group would lead to a local maximum. The summed intensities of all groups are sorted in descending order, $I_1$, $I_2$,...$I_n$, and the $I$ index can be calculated as
                \begin{equation}
                        I = \frac{I_2}{I_1},
                \end{equation}
            
                The deviation statistic, $D$, measures the distance between the image centroid, ($x_c$,$y_c$), and the brightest peak found with the $I$ statistics, ($x_{I_1}$,$y_{I_1}$), which is
                \begin{equation}
                        D =  \sqrt{\frac{\pi}{n_{\rm seg}}} \sqrt{(x_c-x_{I_1})^2+(y_c-y_{I_1})^2},
                \end{equation}
                where $n_{\rm seg}$ is the number of pixels in the segmentation map.
                
		        \figurename{ \ref{fig:MID}} shows the MID statistics of our sample. This set of indices (particularly the multimode) works comparably to the Gini-$M_{\rm 20}$ statistics in separating the galaxies in terms of our categories. We could see it works better to separate the D and Irr/Int galaxies, especially for the $M$--$D$ statistics.
		        
		         \begin{figure*}[htbp]  
		        	\centering
		        	\includegraphics[width=\textwidth]{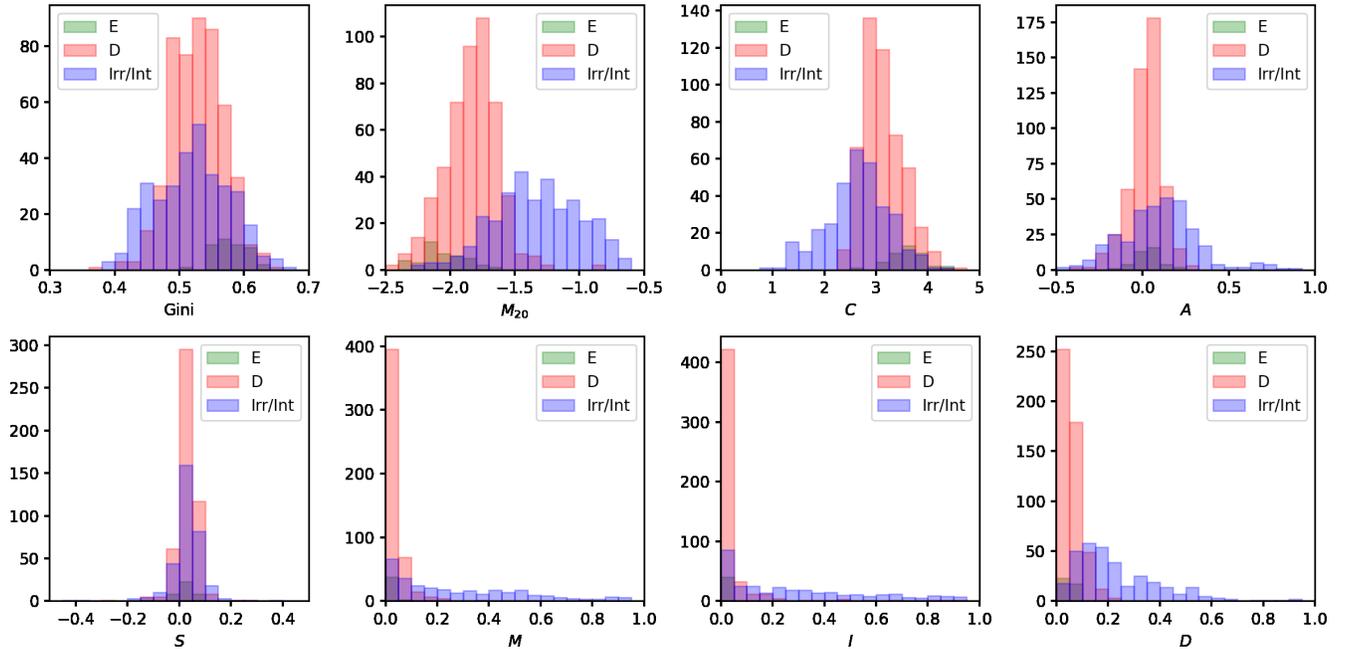} 
		        	\caption{Distributions of our visual E, D, and Irr/Int galaxies in terms of the eight nonparametric indices discussed in  Section 4.3.}
		        	\label{fig:non_hist}
		        \end{figure*}

		         \subsubsection{Comparison of Visual  and Nonparametric Morphologies}
		        
		        To investigate whether we could make further use of the nonparametric morphologies in our study, we compare the visual E, D, and Irr/Int galaxies against the six indices derived above. This is shown in Figure \ref{fig:non_hist} as histograms. While Gini and CAS obviously cannot separate the D and Int/Irr galaxies, $M_{20}$ and $MID$ seem promising for this purpose. For $M_{20}$, a good separation can be achieved at $-1.6$: 90.1\% of the D galaxies have $M_{20} \leqslant -1.6$, while 79.3\% of the Irr/Int galaxies have $M_{20}>-1.6$. Similar results can be achieved for the $M$, $I$ and $D$ indices: (1) 93.7\% of the D galaxies have $M\leqslant 0.1$, while 69.0\% of the Irr/Int galaxies have $M>0.1$; (2) 91.9\% of the D galaxies have $I\leqslant 0.1$, while 66.3\% of the Irr/Int galaxies have $I>0.1$; and (3) 87.1\% of the D galaxies have $D\leqslant 0.1$, while 79.3\% of the Irr/Int galaxies have $D>0.1$. The best result is obtained when we use $M$ and $D$ jointly: 82.2\% of the D galaxies satisfy $M\leqslant 0.1$ and $D\leqslant 0.1$, while 88.1\% of the Irr/Int galaxies have either $M>0.1$ or $D>0.1$. 
		        	
		        For this reason, we will also use the results based on $M_{20}$ as well as $M$--$D$ (as described above), in some of our analysis below. We note that there are two caveats with this approach: (1) only 864 out of 1061 galaxies (i.e., 81.4\%) have  these indices derived, and (2) E and D galaxies cannot be separated. On the other hand, neither would create a severe bias in this particular study; the lesser amount of classified galaxies only decreases the significance of the statistics, and the contamination of E galaxies in D galaxies is small because    	there are only very few E galaxies in our sample.

         \section{Discussion}
        
        In this section, we discuss the implications of our results. 
        
	         \subsection{Counterpart Multiplicity}

\begin{figure}[htbp]  
	        	\centering
	        	\includegraphics[width=.45\textwidth]{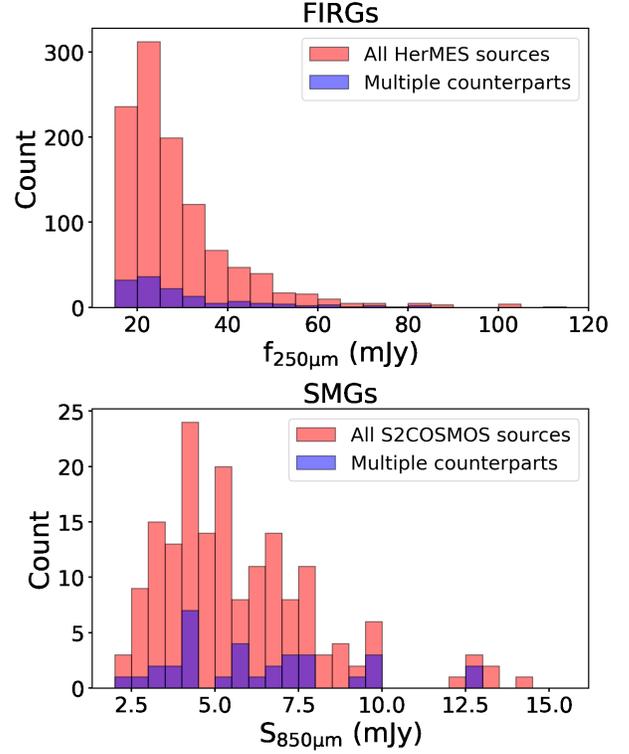} 
	        	\caption{Flux density distribution of FIRGs in 250~$\mu$m (top) and SMGs in 850~$\mu$m (bottom). The distribution of the FIRGs/SMGs with multiple counterparts are coded in blue, while those of their parent samples are coded in red. The 850~$\mu$m flux densities are the deboosted values.}
	        	\label{fig:multiple-hist}
	        \end{figure}

	        As summarized in Section 3.7, our final morphological sample contains 1090 FIRGs and 172 SMGs (117 objects in common). Among them, 12.3\% of the FIRGs (134 out of 1090) and 19.2\% of the SMGs (33 out of 172) have two or more counterparts. For the common FIRGs/SMGs, this fraction is 23.1\% (27 out of 117). The SMG multiplicity problem has been investigated in recent years using ALMA data \citep[e.g.,][]{Hodge2013, Karim2013, Chen2015, Simpson2015a, Stach2018, Simpson2020}. Depending on the source brightness and the sensitivity of the ALMA data in use, the quoted SMG multiplicity fraction ranges from $\sim$10\% to 80\%. Our result (19.2\%) falls in the range of 11\% $\pm$ 1\% to 26\% $\pm$ 2\% as derived by \citet[][]{Stach2018}. At face value, our FIRG multiplicity fraction (12.3\%) seems to be slightly lower. Figure \ref{fig:multiple-hist} shows the flux density distributions of the FIRGs/SMGS that have multiple ALMA/VLA counterparts and compares them to those of their respective parent samples.	        
	        
	        A key question is whether such multiple counterparts are physically associated or due to alignment by chance. The redshifts of these counterparts can shed light on this question. We follow the same method as detailed in  Section 4.1 when discussing the Irr/Int systems: the criterion for physical association is $\Delta z_{\rm spec}<0.03$ (if $z_{\rm spec}$ are available for all counterparts) or $\Delta z_{\rm ph}<0.1$. From now on, we will use the term ``multiple-component'' and the like to refer to the cases of  multiplicity where the multiple counterparts are physically associated.
	        
	        Among the 134 cases of multiplicity in the FIRGs, three have three counterparts and 131 have two counterparts. The first one of the three-counterpart cases has two counterparts at $z_{\rm spec}=0.36$ (both are E galaxies) but the third at $z_{\rm spec}=0.69$ (Irr galaxy). The second one has two counterparts at $z_{\rm ph}=1.44$ and 1.48, respectively (one D galaxy and one Irr galaxy), but with the third at $z_{\rm ph}=0.66$ (D galaxy). The third one has two C galaxies at redshift $z_{\rm spec}=1.24$ and $z_{\rm phot}=1.68$ and one faint galaxy without an $H_{\rm 160}$ counterpart. For the sake of simplicity, these three systems are hereafter excluded from our analysis. Among the 131 two-counterpart cases, 39 have counterparts at the same redshifts (18 by $z_{\rm spec}$ and 21 by $z_{\rm ph}$, respectively), and 75 have counterparts at different redshifts (11 by $z_{\rm spec}$ and 64 by $z_{\rm ph}$, respectively). The remaining 17 cases are undecided because the redshifts are not available for both counterparts. Therefore, we conclude that at least 29.1\% of the FIRG multiplicity cases (39 out of 134) are made of physically associated multiple components, and that at least 56.0\% (75 out of 134) are due to chance alignment. Among the 39 two-component FIRGs, 20 show distinct interacting features between components (Int), seven consist of one Irr galaxy and one D galaxy (Irr+D), two consist of one Irr galaxy and one F galaxy (Irr+F), four consist of two D galaxies (D+D), three consist of one D galaxy and one F galaxy (D+F), one consists of one D galaxy and one E galaxy (D+E), one consists of two E galaxies (E+E), and one consists of two F galaxies (F+F).
	        
 			The 33 cases of multiplicity among the SMGs all have two counterparts. Among them, three cases contain counterparts of nearly the same $z_{\rm spec}$, while six contain counterparts with $\Delta z_{\rm ph}<0.1$. There are 16 cases where the counterparts are not associated, judged either by their $z_{\rm spec}$ or $z_{\rm ph}$. The remaining nine cases only have redshift information for one of the counterparts and therefore are undecided. In conclusion, among the SMG multiplicity cases, at least 27.3\% (nine out of 33, three by $z_{\rm spec}$ and six by $z_{\rm ph}$) have physically associated multiple components, and at least 48.5\% (16 out of 33) are due to chance alignment. Among the nine two-component SMGs, three are Int systems, three are Irr+D, two are D+F, and one is Irr+F. In other words, 55.6\% (10 out of 18) of the individual galaxies are in the Irr/Int category, while 27.8\% (five out of 18) are in the D category.

   			It is also worthwhile to single out the 27 multiplicity cases among the common FIRGs/SMGs from the above and comment on them collectively. Of these cases, 33.3\% (nine out of 27) have two physically associated components (three by $z_{\rm spec}$ and six by $z_{\rm ph}$), 44.4\% (12 out of 27) are due to  chance alignment (by $z_{\rm ph}$), and 22.2\% (six out of 27) are undecided due to the lack of redshifts. As it turns out, the nine two-component systems are exactly the same nine Int SMG systems already commented on above.	        
	        	                
	         \subsection{Redshift Distribution by Morphological Types}	
	        
	        \begin{figure}[htbp]  
	        	\centering
	        	\includegraphics[width=.45\textwidth]{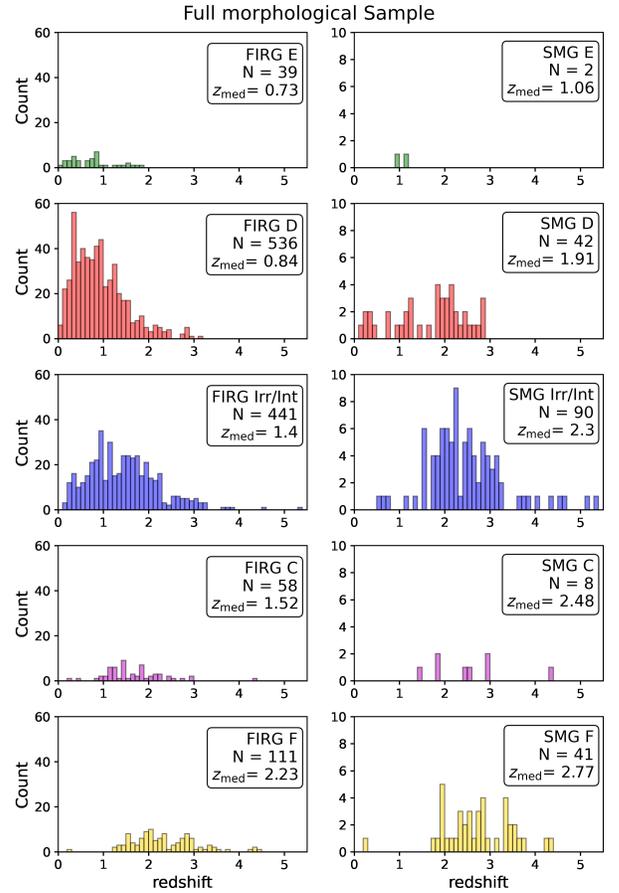} 
	        	\caption{Redshift distribution of FIRG and SMG counterparts in different morphological types: ellipticals (E), disky galaxies (D), irregular and/or interacting systems (Irr/Int), compact galaxies (C), and objects too faint to tell the details (F). See  Section 4.1 for the designation details. The total number of objects ($N$) and the median redshifts ($z_{\rm med}$) in each category are also labeled.}
	        	\label{fig:z-distribution}
	        \end{figure}
	        	        
	        After solving the multiplicity problem, we are in a position to discuss the redshift distributions of our samples. 
	        
	        \begin{figure*}[htbp]  
	        	\centering
	        	\includegraphics[width=1\textwidth,]{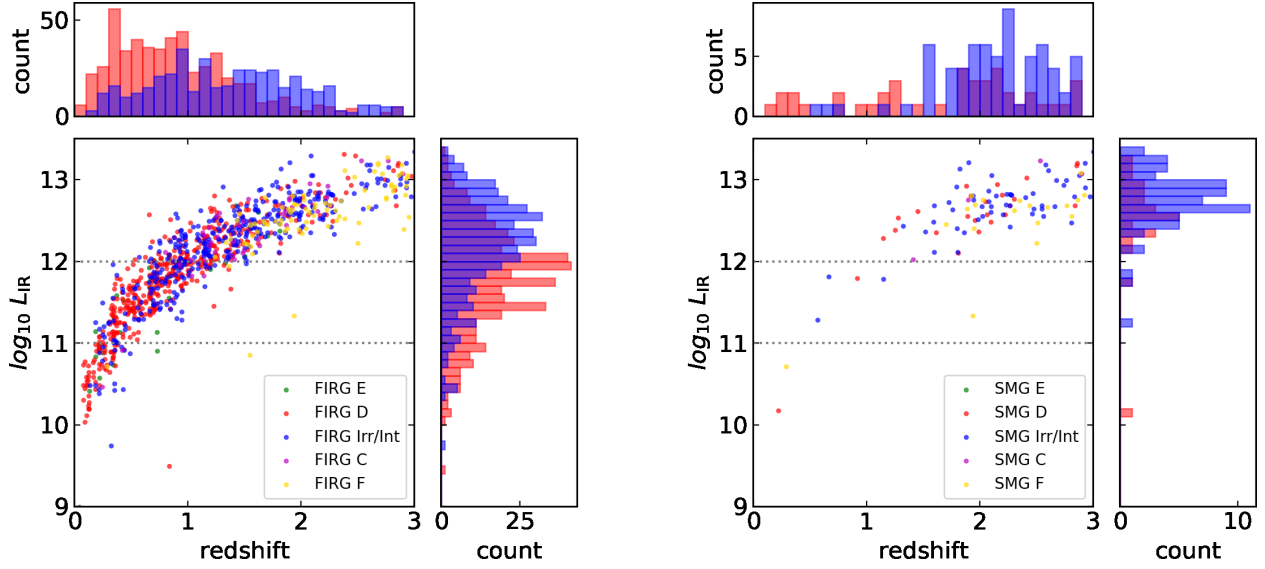} 
	        	\caption{Distribution of $L_{\rm IR}$ with respect to redshift for the FIRGs (left) and the SMGs (right) in our sample, plotted for different morphological categories using different symbols. The lack of low $L_{\rm IR}$ objects at high redshifts is due to the Malmquist bias. The distributions of the D and Irr/Int galaxies with respect to redshift and $L_{\rm IR}$ are also shown in histograms.}
	        	\label{fig:SED}
	        \end{figure*}
	        
	        Statistically, the FIRGs are at systematically lower redshifts as compared to the SMGs; the median redshifts ($z_{\rm med}$) of the two samples are 1.16 and 2.27, respectively. In each morphological category, the FIRGs are also at systematically lower redshifts than the SMGs. However, it is more appropriate to say that the SMGs are at the high-redshift tail of the FIRG distribution (see also  Section 3.1). This is shown in Figure \ref{fig:z-distribution}, where $z_{\rm med}$ and the total number of objects ($N$) in each category are also labeled. The vast majority of our galaxies are in the Irr/Int and D categories. Interestingly, the D galaxies seem to ``lag behind'' the Irr/Int galaxies among both the FIRGs and SMGs; the median redshifts of the FIRGs are 0.84 and 1.40 in the D and Irr/Int categories, respectively, while those of the SMGs are 1.91 and 2.30 in these two categories, respectively.
	        
	         \subsection{Infrared Luminosity}

	       \begin{figure*}[htbp]  
	       	\begin{minipage}{.5\textwidth}
	       		\centering
	       		\includegraphics[width=0.9\textwidth]{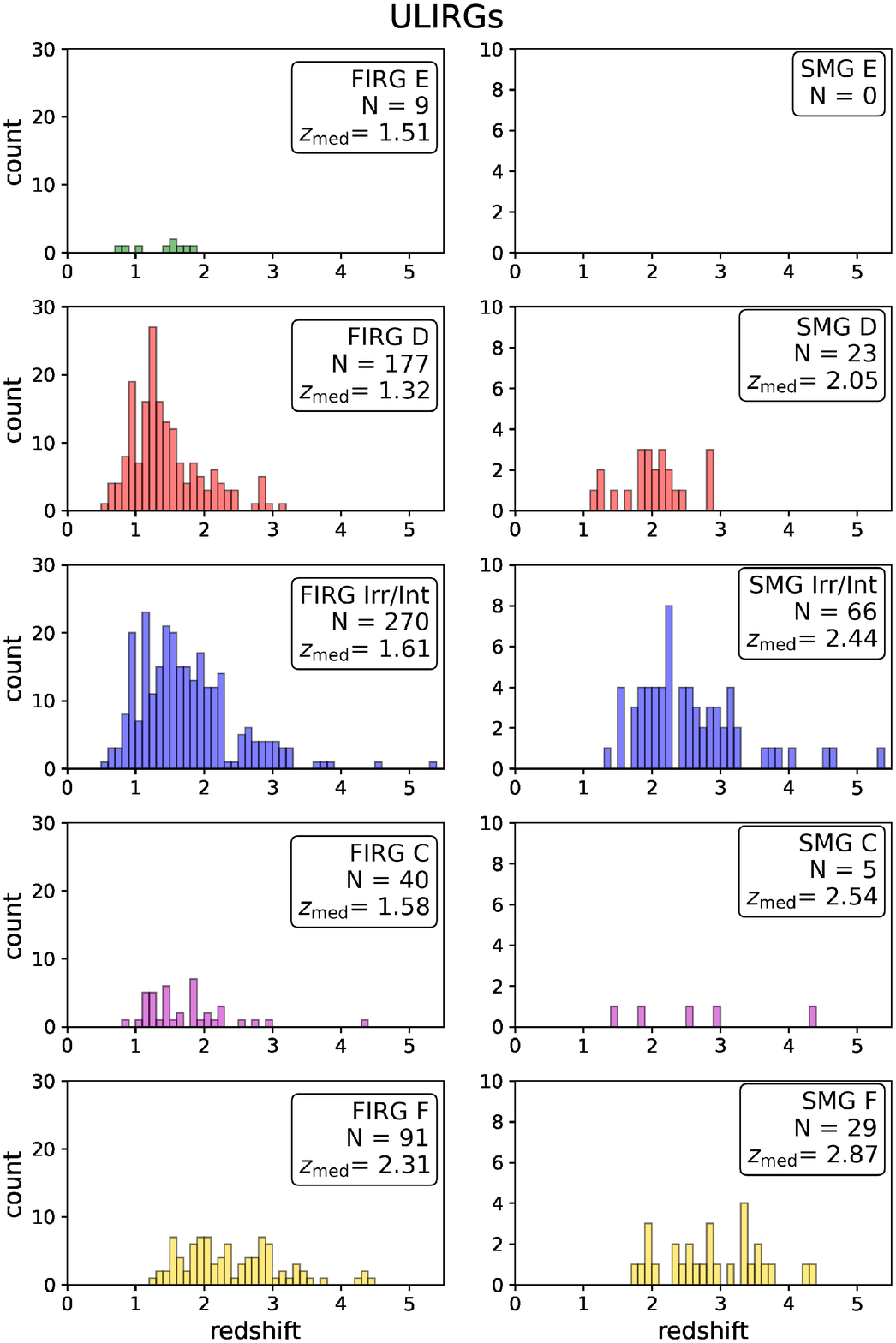} 
	       		\caption{Redshift distribution of FIRG and SMG ULIRGs in different morphological types, Legends are the same as in Figure \ref{fig:z-distribution}.}
	       		\label{fig:z-dist_ulirg}
	       	\end{minipage}
	       	\begin{minipage}{.5\textwidth}
	       		\centering
	       		\includegraphics[width=0.9\textwidth]{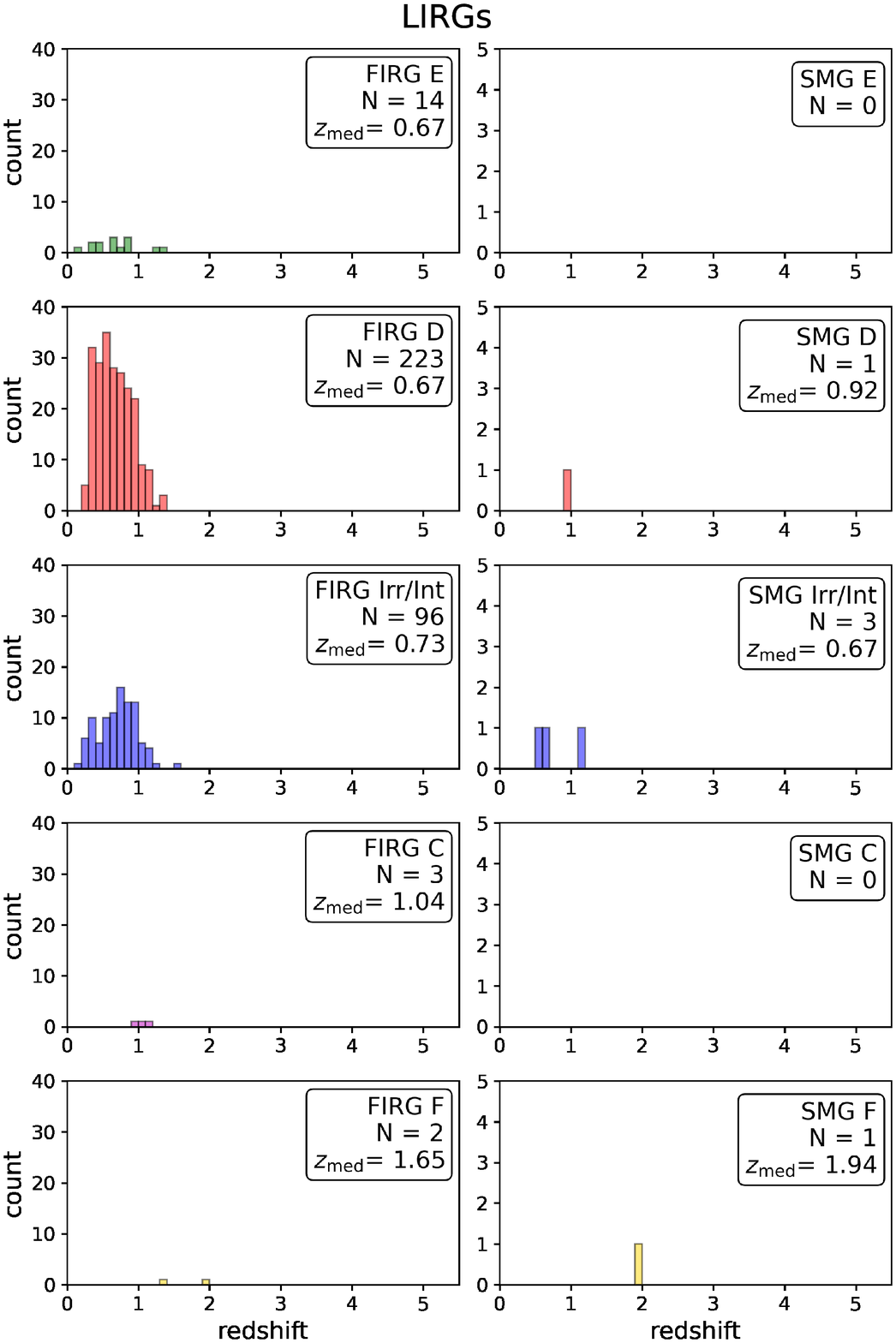} 
	       		\caption{Redshift distribution of FIRG and SMG LIRGs in different morphological types, Legends are the same as in Figure \ref{fig:z-distribution}.}
	       		\label{fig:z-dist_lirg}
	       	\end{minipage}
	       \end{figure*}

			As our goal is to study the host morphologies of ULIRGs and LIRGs, we must obtain the total infrared luminosities ($L_{\rm IR}$, integrated over rest-frame 8--1000~$\mu$m) for the objects in our sample. For this purpose, we fit the spectral energy distributions (SEDs) in the FIR-to-millimeter regime, following the method of \citet[][]{Ma2015}.
	        
	        In this analysis, we discarded those multiple-counterpart cases that are due to chance alignment or are undecided. In other words, we only considered the single- and two-component FIRGs/SMGs. Due to the well-known degeneracy between dust temperature and redshift, we must confine ourselves to the objects that have reliable redshifts. In total, there are 986 single-component and 38 two-component FIRGs/SMGs that have $z_{\rm spec}$ or $z_{\rm ph}$.  In order to best constrain $L_{\rm IR}$, we constructed the SEDs incorporating the 100 and 160~$\mu$m data from the PACS Evolutionary Probe \citep[][]{Lutz2011} by including the photometry with S/N $\geqslant 3$ in either band, as in the catalog of \citet[][]{Magnelli2009}. The matching was done in the same way as described in  Section 3.1. For the FIRGs, the matching radii were 3\farcs13 and 3\farcs97 to 100 and 160~$\mu$m, respectively. For the SMGs, these were 3\farcs71 and 4\farcs43 to these two bands, respectively. In addition, we  incorporated the SCUBA2 850 $\mu$m  photometry by relaxing the S/N requirement to $\geqslant 3$ in the S2COSMOS catalog. Furthermore, the ALMA data (in Bands 3, 4, 6, 7, and 9) were also included whenever available. As mentioned in  Section 3.2, the A$^3$COSMOS catalog gives photometry from all available programs separately, and it often happens that the same source has multiple photometry in the same band. In this case, we adopted the one with the highest S/N. 
	        
	        We fit the SEDs to the  theoretical models of \citet[][]{Siebenmorgen2007}. For the objects whose SEDs have photometry in at least three bands, the fitting was done with the scaling factor as a free parameter \citep[see][]{Ma2015}. For those that have photometry in two bands (38 FIRGs/SMGs in total), we followed the common practice by adopting the $L_{\rm IR}$ of the template that would give the closest values to the flux densities in the given bands \citep[see, e.g.,][]{Runge2018}. There are 21 objects with photometry in only one band, and they are not considered in this analysis.
	        
	        We obtained reliable fits for the vast majority of the FIRGs/SMGs. For the 38 two-component FIRGs/SMGs, we attributed the derived $L_{\rm IR}$ to the individual components according to their flux density ratio either in the ALMA band or in the VLA 3~GHz.\footnote{We had to exclude one member of a two-component SMG from further analysis, becaue this member falls in a blank region in the $H_{160}$ mosaic and hence is not in the morphological sample.} There is a complication with the 10 two-counterpart cases as identified based on the VLA map. From their redshifts, these are all two-component systems by our criterion. As described in  Section 3.7, the catalog of \citet[][]{Smolcic2017a} only quotes the merged fluxes of the whole systems. To properly split the derived $L_{\rm IR}$ among the two components in these cases, we carried out our own photometry on the VLA map using SExtractor and obtained the flux density ratio of the two components for all these cases.
	        
	        In the end, we obtained $L_{\rm IR}$ for 1040 individual galaxies (1013 are FIRG counterparts and 130 are SMG counterparts, among which 103 are in common) in our sample, which correspond to 977 FIRGs and 122 SMGs (95 in common). 
	        
	        \begin{figure*}[htbp]  
	        	\centering
	        	\includegraphics[width=0.9\textwidth]{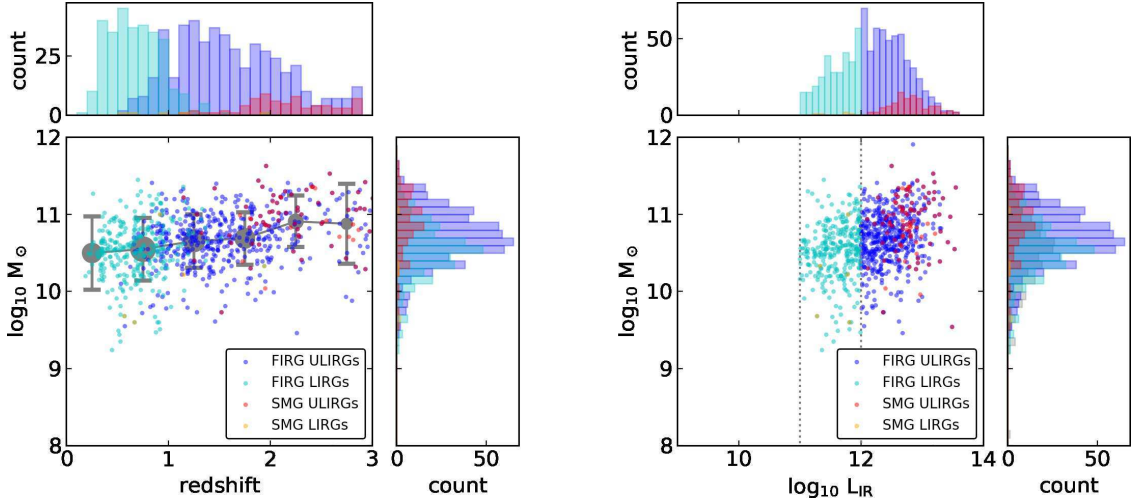} 
	        	\caption{ Distribution of $M_{\odot}$ with respect to redshift (left) and $L_{\rm IR}$ (right) in our sample. The FIRG (U)LIRGs and SMG (U)LIRGs are plotted in blue, cyan, red and orange, respectively. The distribution of FIRG (U)LIRGs and SMG (U)LIRGs with respect to redshift, $M_{\odot}$ and $L_{IR}$ are also shown in histograms. The grey symbols in the left panel are the median values at successive redshifts (step-size $\Delta z=0.5$), and the associated error bars represent the 1$\sigma$ dispersion.}
	        	\label{fig:mass}
	        \end{figure*}	
	        
	        Figure \ref{fig:SED} shows the $L_{\rm IR}$ distribution with respect to redshift for FIRG and SMG hosts, respectively. Among the 1013 FIRG counterparts, 587 belong to ULIRGs ($L_{\rm IR} \geqslant 10^{12} L_\odot$), and 338 belong to LIRGs ($10^{11} L_\odot \leqslant L_{\rm IR} \le 10^{12} L_\odot$). Among the 130 SMG counterparts, 123 and five are ULIRGs and LIRGs, respectively. 
	        
	        Figures \ref{fig:z-dist_ulirg} and \ref{fig:z-dist_lirg} show the redshift distributions of the morphological types of these ULIRGs and LIRGs.  The D galaxies ``lagging behind'' the Irr/Int galaxies are still obvious among the FIRG ULIRGs, with $z_{\rm med}$ of 1.32 and 1.61, respectively (or $\sim$740~Myr of difference in time in our adopted cosmology). This becomes insignificant among the FIRG LIRGs, with $z_{\rm med}$ of 0.67 and 0.73, respectively ($\sim$340~Myr of time difference). It is not obvious whether the ``lagging behind'' exists among the SMGs (ULIRGs or LIRGs) due to the less sufficient statistics.

        	 \subsection{Lack of Correlation with Stellar Mass}

        	\begin{figure*}[htbp]  
        		\centering
        		\includegraphics[width=0.9\textwidth]{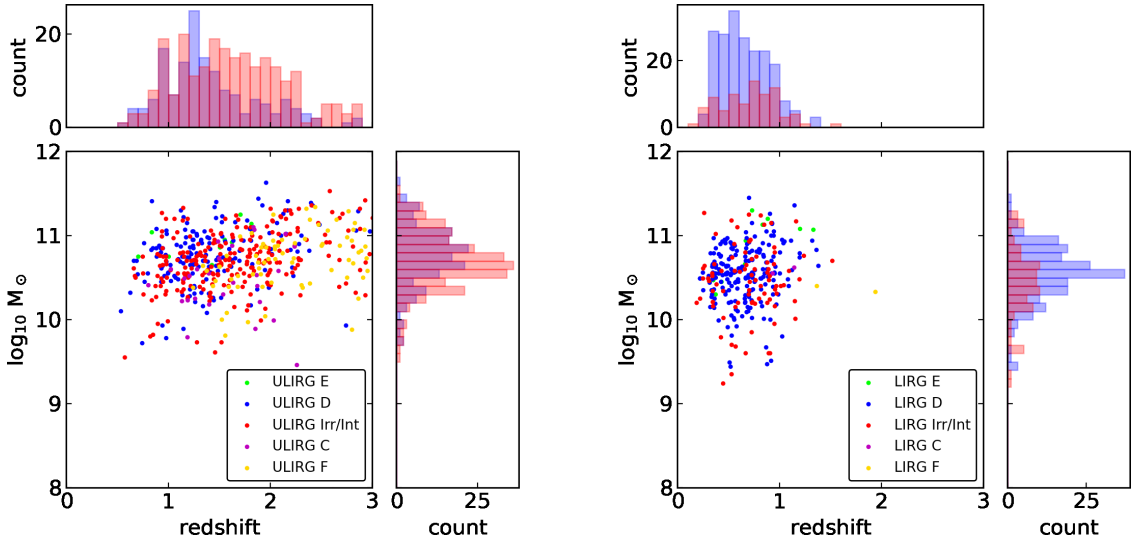} 
        		\caption{ Distribution of $M_{\odot}$ with redshift for different morphological subtypes of ULIRGs (left) and LIRGs (right) in our sample. The color-coding for different morphological categories is the same as in Figure \ref{fig:z-distribution}. The distriubiton of the D and Irr/Int galaxies with respect to redshift and $M_{\odot}$ are also shown in histograms.}
        		\label{fig:mass-z}
        	\end{figure*}

        	It is interesting to understand whether the starburst activity in (U)LIRGs is related to the past star formation history (SFH) of their hosts. The most basic diagnostic is to study the stellar masses ($M_*$) of the existing stellar populations in the hosts. For this purpose, we derived the stellar masses of the galaxies in our sample by fitting their SEDs to the stellar population synthesis model of \citet[][]{BC03} with the initial mass function of \citet[][]{Chabrier2003}. The SEDs cover the range from 0.3 to 2.2 ~$\mu$m and include 28 optical--to--near--IR bands. These were constructed by matching the $H_{\rm 160}$ positions of our sources to the catalog of \citet[][matching radius of 1\arcsec]{Laigle2016}, from which we extracted the photometry in 21 optical bands (CFHT/Megacam $u^*$; Subaru/Suprime-Cam $B$, $V$, $r$, $i^{+}$, $z^{+}$, and $z^{++}$; 12 medium bands from IA427 to IA827; and two narrow bands, NB711 and NB816) and seven near-IR bands (Subaru/HSC $Y$; VISTA/VIRCAM $Y$, $J$, $H$, and $K_s$; and CFHT/WIRCam $H$ and $K_s$). In total, we obtained the SEDs for 1191 sources, and these SEDs were fitted using the Le Phare program \citep[][]{Arnouts2002, Ilbert2006} with the galaxy library ``\texttt{BC03\_COMM}''. This set of templates is constructed assuming solar metallicity and the exponentially declining star formation histories in the form of SFR $\propto e^{- t/\tau}$, where $\tau$ ranges from zero to 13~Gyr. The source redshifts were fixed to their $z_{\rm spec}$ (when available) or $z_{\rm ph}$ as described in Appendix \ref{appendix:a}. We adopted the extinction law of \citet[][]{Calzetti2000} with $E(B-V)$ ranging from zero to 0.7 mag. The Le Phare outputs include several flavors of stellar mass estimates, and we adopted MASS\_{MED}. 
        	
			\begin{figure*}[htbp]  
				\centering
				\includegraphics[width=0.9\textwidth]{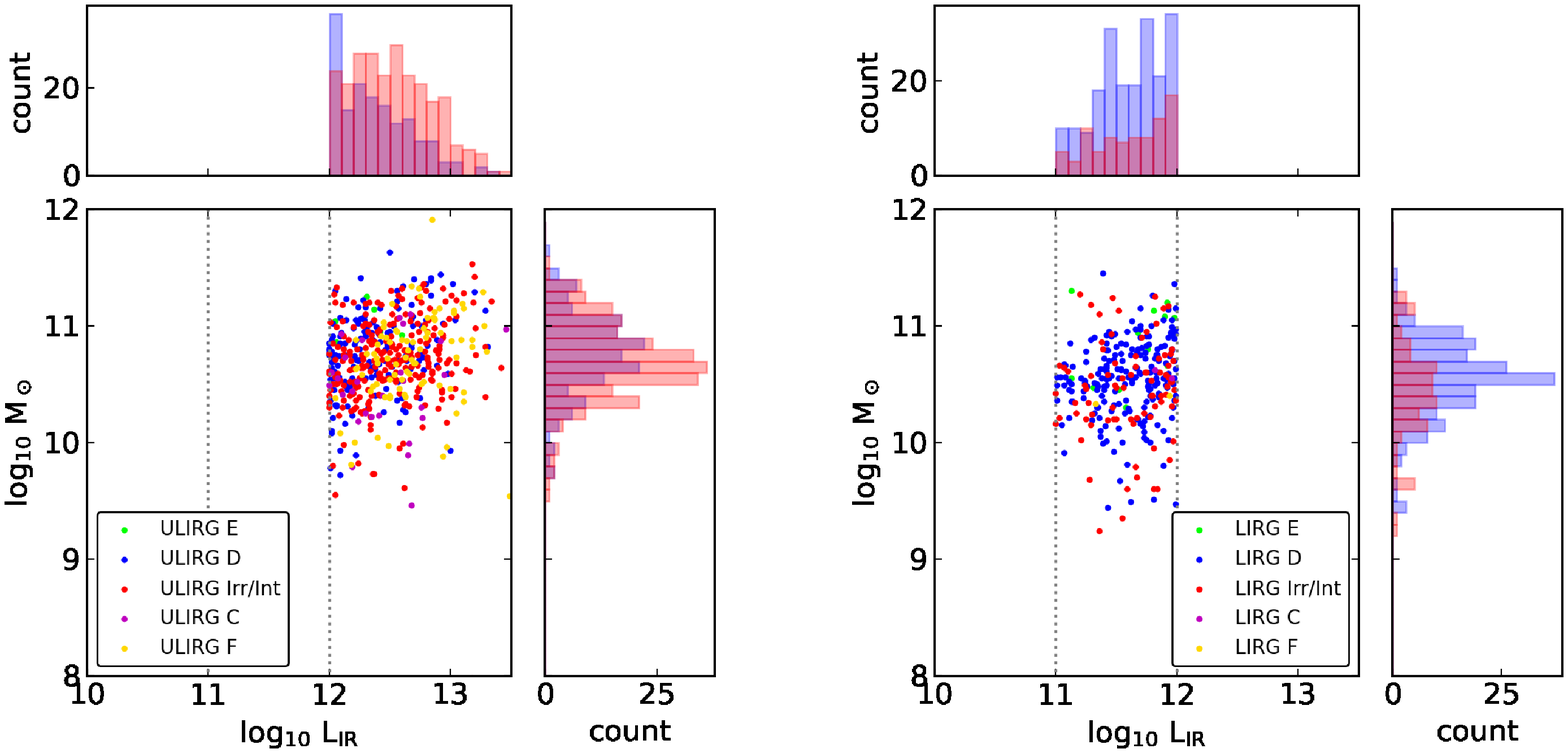} 
				\caption{Distribution of $M_{\odot}$ with $L_{\rm IR}$ for different morphological  ULIRGs (left) and LIRGs (right) in our sample. The color-coding for different morphological categories is the same as in Figure \ref{fig:z-distribution}. The distriubiton of the D and Irr/Int galaxies with respect to $L_{\rm IR}$ and $M_{\odot}$ are also shown in histograms.}
				\label{fig:mass-lir}
			\end{figure*}         	
        	In the end, we obtained satisfactory fits for 1085 sources. Figure \ref{fig:mass} shows their stellar mass distributions with respect to redshift and $L_{\rm IR}$. Among them, 514 and 96 sources are FIRG ULIRGs and SMG ULIRGs, respectively (76 in common), and 302 and five are FIRG LIRGs and SMG LIRGs, respectively (four in common). The vast majority of these sources have masses in the range of $10^{9.5} M_\odot \leqslant M_* \leqslant 10^{11.5} M_\odot$. The weak increasing trend with redshift is due to the Malmquist bias that higher-mass ones are easier to see at higher redshifts. The median stellar masses of SMG ULIRGs, FIRG ULIRGs, and FIRG LIRGs are $10^{10.9}$, $10^{10.7}$, and $10^{10.6} M_\odot$, respectively. While the SMG ULIRGs seem to be of systematically higher masses (by $\sim$1.5 times) than the FIRG ULIRGs, it is due to the selection bias of the two samples that the SMGs are at systematically higher redshifts than the FIRGs. The difference between the FIRG ULIRGs and LIRGs is small and can be explained similarly (very few LIRGs are seen at high redshifts due to selection bias).

        	We further investigate the stellar mass distributions of different morphological types among ULIRGs and LIRGs (regardless of being FIRGs or SMGs), which are shown in Figures \ref{fig:mass-z} and \ref{fig:mass-lir}, respectively. Similar to Figure \ref{fig:mass}, the histograms highlight the distributions of the D and Irr/Int galaxies, which are the dominant categories. Again, there is no obvious trend between stellar  mass and $L_{\rm IR}$. On the other hand, the D galaxies seem to be of higher stellar masses than the Irr/Int galaxies; the median $M_*$ values are  $10^{10.8}$ and $10^{10.7}$ $M_\odot$ for the ULIRG D and Irr/Int galaxies, respectively, and $10^{10.6}$ and $10^{10.5}$ $M_\odot$ for the LIRG D and Irr/Int galaxies, respectively. As the D galaxies are at systematically lower redshifts than the Irr/Int galaxies, such small differences probably can be accounted for by the overall growth of the galaxy population across the cosmic time.

        	In summary, from the stellar mass analysis, we do not find an indication that would suggest a correlation between the (U)LIRG activities and the past SFH of the host galaxies.

			 \subsection{Morphological Evolution of (U)LIRG Hosts}
			
			Section 5.2 and Figure \ref{fig:z-dist_ulirg} already hint that different types of (U)LIRG host galaxies change their relative importance over cosmic time. Here we further address this question using both visual morphologies and those determined by $M_{20}$, as well as $M$--$D$ analysis. While the latter suffer from an $\sim$18\% reduction in the sample size,  they provide a valuable cross-check.

			 \subsubsection{Evolutionary Trend Based on Visual Classification}

\begin{figure*}[htbp]
				\centering
				\includegraphics[width= \textwidth]{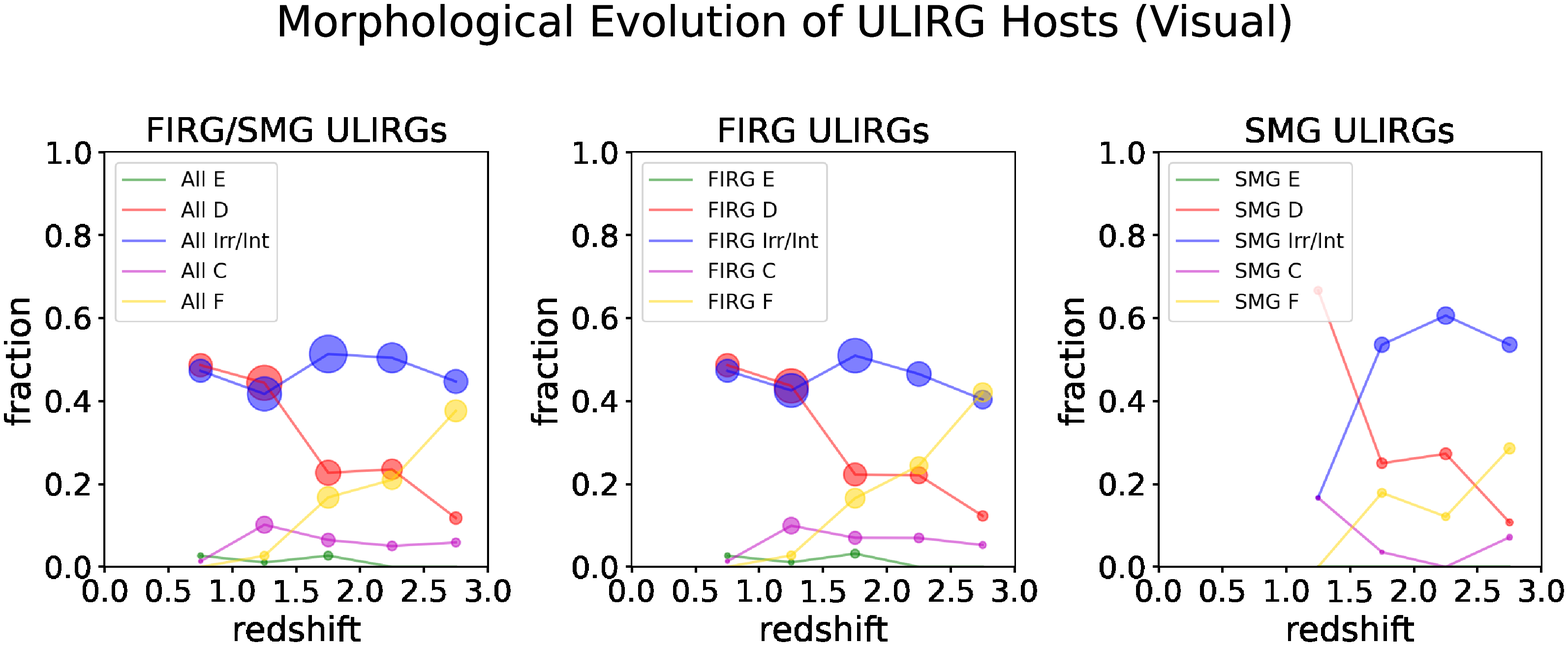} 
				\caption{We show the morphological evolution of ULIRG hosts in terms of the fractional contribution of each morphological type among to the total. Different colors represent different types, as explained in the legends. The sizes of the symbols are proportional to the number of objects, which are given in Table \ref{table:table1}. From left to right, the plots are for the combined FIRG/SMG ULIRG, the FIRG ULIRG, and the SMG ULIRG samples.}
				\label{fig:z_distri}
			\end{figure*}

			\figurename{ \ref{fig:z_distri}} shows the morphological evolution of these ULIRG hosts in terms of the fraction of each category among the total at different redshifts. The $z=$0.5--1 bin is not included, as there are no ULIRG hosts in this bin. The detailed breakdown is given in Table \ref{table:table1}. The E galaxies make the smallest contribution, but they do exist. While this might seem surprising, we point out that a small fraction of the brightest cluster galaxies, which are all ellipticals, are ULIRGs \citep[][]{Runge2018}. Due to the insufficient statistics, we refrain from discussing them further here. The F galaxies steadily increase in fraction toward high redshift, and we can only hope that the James Webb Space Telescope will be able to discern their details in the near future. The C category, while only a small fraction, persists at all redshifts. 
		
			\begin{figure*}[htbp]
				\centering
				
				\includegraphics[width= 0.7\textwidth]{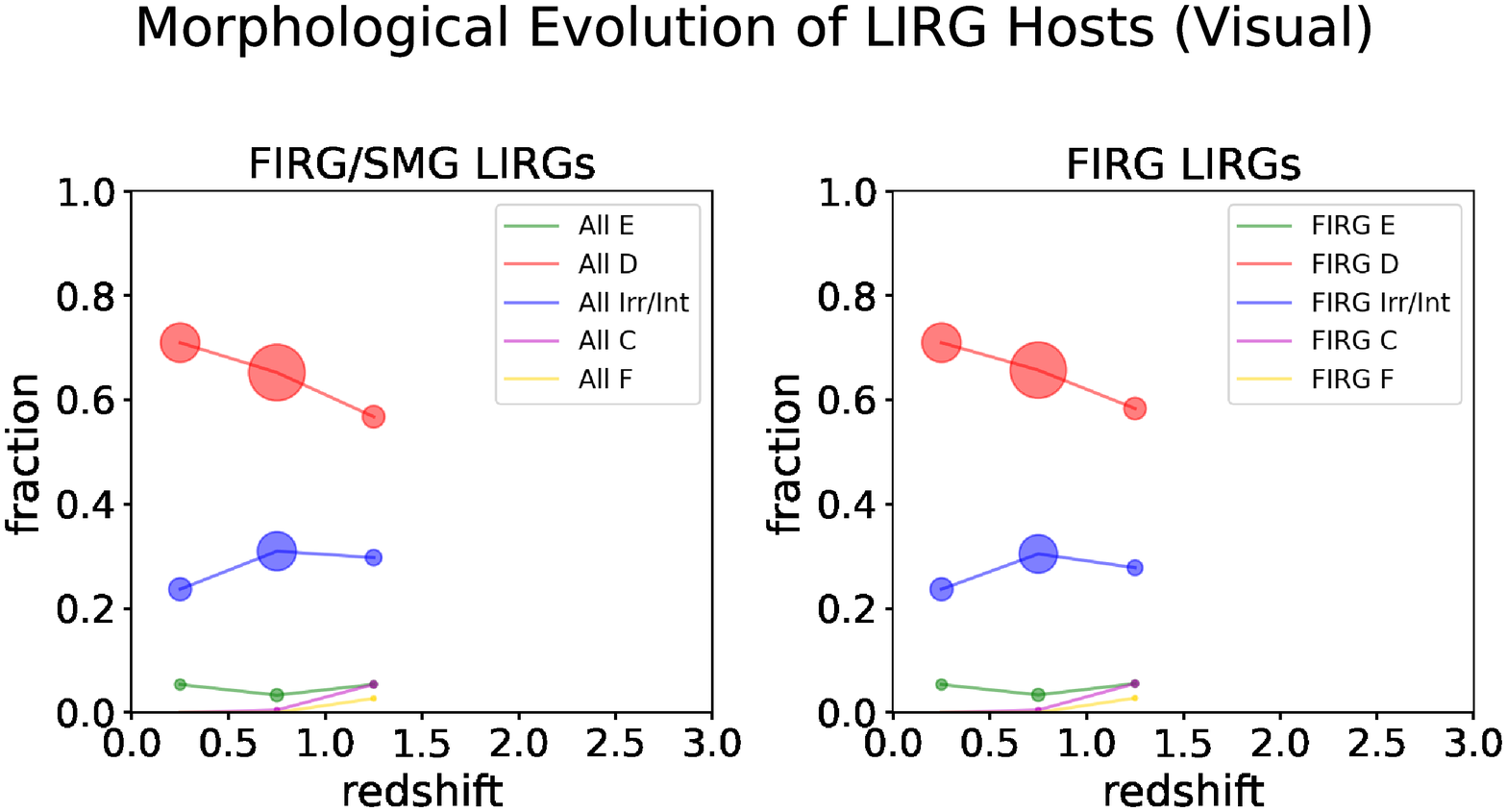} 
				\caption{Morphological evolution of LIRG hosts in terms of fractional contribution of each morphological type among the total. Legends are the same as in Figure \ref{fig:z_distri}.}
				\label{fig:z_distri_LIRG}
			\end{figure*} 	     		
						
			The most important point is regarding the D and Irr/Int galaxies, which contribute predominantly over the full redshift range. Their relative importance seems to have a transition in between $z$ = 1.25 and 1.75. For the FIRG ULIRGs, the turning point happens in the $z$ = 1.0--1.5 bin. Above this bin, the Irr/Int galaxies outnumber the D galaxies by a factor of 2.2--4$\times$, while from this bin to lower redshifts, the two categories reach nearly the same fraction. For the SMG ULIRGs, the situation is similar; the Irr/Int galaxies also outnumber the D galaxies by a factor of 2--5$\times$ above this bin, although the number of our SMG ULIRGs in this bin and below is too small to robustly determine the behavior at lower redshifts. We note that \citet[][]{Zavala2018} pointed out a transition of SMG morphological type from ``irregular disk'' galaxies to ``disk + spheroid'' galaxies at $z\approx 1.4$. Their study is based on $\sim$71 SCUBA2 sources in the Extended Groth Strip field detected in either 450 or 850~$\mu$m and having optical/near-IR counterparts identified through the bridging of the deep VLA 1.4~GHz data. As SMGs are predominantly ULIRGs (see our Figure \ref{fig:SED}), comparing our result to theirs is appropriate. It can be seen from their Figure 17 that their irregular disks (similar to our Irr/Int galaxies) outnumber the sum of the other two regular disks (pure disks and disks with spheroids; similar to our D galaxies) by a factor of $\sim$2--3 at $z\gtrsim 1.7$, and that the trend is reversed at $z\lesssim 1.2$. This is very similar to our result, despite that theirs also suffers from small number statistics at the lowest redshifts. We also note that \citet[][]{Kartaltepe2012} carried out a similar host morphological study of 52 $z \sim 2$ ULIRGs selected using the Herschel PACS data in the GOODS-South field. They found that $\sim$72\% of their objects showed irregular/interacting features, a fraction that is even higher than what we see in our sample. However, when they compared to ULIRGs at $z \sim 1$, they did not see any significant difference between the two redshifts. This is different from our result; we believe that their smaller sample size might be a major reason for the discrepancy.
			
			\begin{deluxetable*}{@{\extracolsep{10pt}}lcccccccccccc}
				\centering
				\tablecaption{Redshift distribution of (U)LIRG hosts by visual morphological types
				\label{table:table1}}
				\tablewidth{0pc}
				\tabletypesize{ \scriptsize}
				\tablehead
				{\colhead{}&\multicolumn{2}{c}{$0 \leqslant z \le 0.5$} &  \multicolumn{2}{c}{$0.5 \leqslant z \le 1$} & \multicolumn{2}{c}{$1.0 \leqslant z \le 1.5$}& \multicolumn{2}{c}{$1.5 \leqslant z \le 2$}& \multicolumn{2}{c}{$2 \leqslant z \le 2.5$}& \multicolumn{2}{c}{$2.5 \leqslant z \le 3$}  \\
					\cline{2-3} \cline{4-5} \cline{6-7} \cline{8-9} \cline{10-11} \cline{12-13}
					\colhead{Subtypes}& \colhead{$N$}& \colhead{$P$}& 
					\colhead{$N$}& \colhead{$P$}&  \colhead{$N$}& \colhead{$P$}& 
					\colhead{$N$}& \colhead{$P$}&  \colhead{$N$}& \colhead{$P$}&
					\colhead{$N$}& \colhead{$P$} 
				}
				 \startdata	
				FIRG ULIRG E           &0   & ... & 2   &2.7  \%& 2   &1.1  \%& 5   &3.2  \%& 0   &0.0  \%& 0   &0.0  \% \\
				FIRG ULIRG D           &0   & ... & 36  &48.6 \%& 79  &43.6 \%& 35  &22.3 \%& 19  &22.1 \%& 7   &12.3 \% \\
				FIRG ULIRG Irr/Int     &0   & ... & 35  &47.3 \%& 77  &42.5 \%& 80  &51.0 \%& 40  &46.5 \%& 23  &40.4 \% \\
				FIRG ULIRG C           &0   & ... & 1   &1.4  \%& 18  &9.9  \%& 11  &7.0  \%& 6   &7.0  \%& 3   &5.3  \% \\
				FIRG ULIRG F           &0   & ... & 0   &0.0  \%& 5   &2.8  \%& 26  &16.6 \%& 21  &24.4 \%& 24  &42.1 \% \\
				$\rm  \sum_{ FIRG \ ULIRG}$  &0   & ... & 74  &100  \%& 181 &100  \%& 157 &100  \%& 86  &100  \%& 57  &100  \% \\
				\midrule
				SMG ULIRG  E           &0   & ... & 0   & ... & 0   &0.0  \%& 0   &0.0  \%& 0   &0.0  \%& 0   &0.0  \% \\
				SMG ULIRG  D           &0   & ... & 0   & ... & 4   &66.7 \%& 7   &25.0 \%& 9   &27.3 \%& 3   &10.7 \% \\
				SMG ULIRG  Irr/Int     &0   & ... & 0   & ... & 1   &16.7 \%& 15  &53.6 \%& 20  &60.6 \%& 15  &53.6 \% \\
				SMG ULIRG  C           &0   & ... & 0   & ... & 1   &16.7 \%& 1   &3.6  \%& 0   &0.0  \%& 2   &7.1  \% \\
				SMG ULIRG  F           &0   & ... & 0   & ... & 0   &0.0  \%& 5   &17.9 \%& 4   &12.1 \%& 8   &28.6 \% \\
				$\rm  \sum_{SMG \ ULIRG}$   &0   & ... & 0   & ... & 6   &100  \%& 28  &100  \%& 33  &100  \%& 28  &100  \% \\
				\midrule
				FIRG LIRG E            &5     & 5.4  \% & 7     & 3.4  \% & 2     & 5.6  \% & 0     & 0.0  \% & 0     & ... & 0     & ...  \\
				FIRG LIRG D            &66    & 71.0 \% & 136   & 65.7 \% & 21    & 58.3 \% & 0     & 0.0  \% & 0     & ...  & 0     & ... \\
				FIRG LIRG Irr/Int      &22    & 23.7 \% & 63    & 30.4 \% & 10    & 27.8 \% & 1     & 50.0 \% & 0     & ...  & 0     & ... \\
				FIRG LIRG C            &0     & 0.0  \% & 1     & 0.5  \% & 2     & 5.6  \% & 0     & 0.0  \% & 0     & ...  & 0     & ... \\
				FIRG LIRG F            &0     & 0.0  \% & 0     & 0.0  \% & 1     & 2.8  \% & 1     & 50.0 \% & 0     & ...  & 0     & ... \\
				$\rm  \sum_{FIRG \ LIRG}$   &93    & ... & 207   & ... & 36    & ... & 2     & ... & 0     & ... & 0     & ... \\
				\midrule
				SMG LIRG E             &0     & ... & 0     & 0.0  \% & 0     & 0.0  \% & 0     & 0.0  \% & 0     & ... & 0     & ...  \\
				SMG LIRG D             &0     & ... & 1     & 33.3 \% & 0     & 0.0  \% & 0     & 0.0  \% & 0     & ... & 0     & ...  \\
				SMG LIRG Irr/Int       &0     & ... & 2     & 66.7 \% & 1     & 100.0\% & 0     & 0.0  \% & 0     & ... & 0     & ...  \\
				SMG LIRG C             &0     & ... & 0     & 0.0  \% & 0     & 0.0  \% & 0     & 0.0  \% & 0     & ... & 0     & ...  \\
				SMG LIRG F             &0     & ... & 0     & 0.0  \% & 0     & 0.0  \% & 1     & 100.0\% & 0     & ... & 0     & ...  \\
				$\rm  \sum_{SMG \ LIRG}$    &0     & ... & 3     & ... & 1     & ... & 1     & ... & 0     & ... & 0     & ... \\
				\enddata  
				\tabletypesize{ \small}
				\tablecomments{The top two blocks of the table list the total number of objects  (``$N$'') and the corresponding percentage (``$P$'') for the ULIRG hosts of each morphological subtype in different redshift bins, with FIRGs and SMGs listed separately. The bottom two blocks list the same for the LIRG hosts.}
			\end{deluxetable*}

			We further explored the morphological evolution of LIRGs in our sample. As shown in Figure \ref{fig:z_distri_LIRG}, the fraction of FIRG D galaxies is generally 2--3$\times$ higher than that of FIRG Irr/Int galaxies over the range $0.5\leqslant z \leqslant 1.5$ where there is sufficient statistics. On the other hand, no conclusion could be drawn for the SMG LIRGs because there are only five objects.

			 \subsubsection{Evolutionary Trend Based on  $\rm M_{20}$ and $\rm M$--$\rm D$ Diagnostics}
			
\begin{figure*}[htbp]
       			\centering
       			\includegraphics[width= \textwidth]{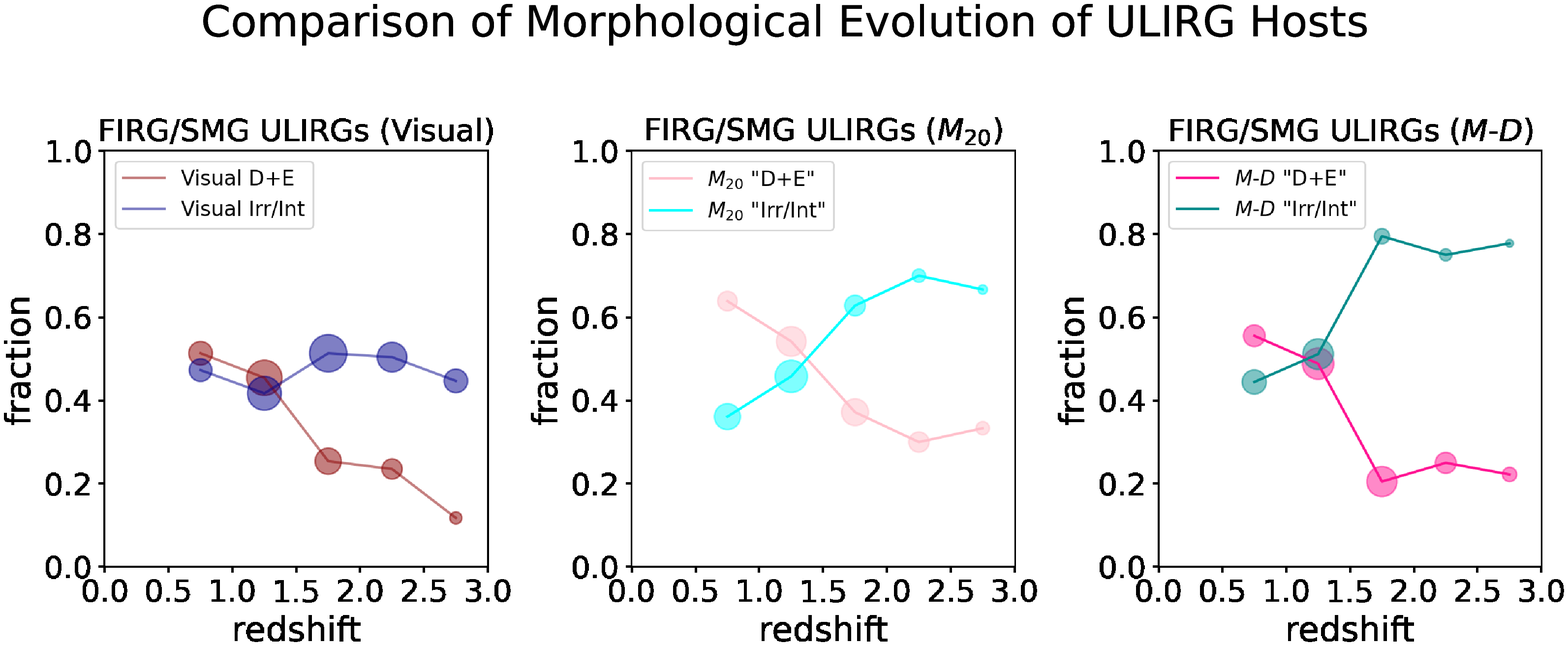} 
       			\caption{Morphological evolutionary trends of ULIRG hosts, derived using $M_{20}$ (middle) and $M$-$D$ indices (right) to mimic the separation of the visual D+E galaxies and Irr/Int galaxies. For comparison, the left panel shows the result in Figure \ref{fig:z_distri} by combining the visual D and E galaxies. The legends are explained in the insets.}
       			\label{fig:z_distri2}
       		\end{figure*}

			Figure \ref{fig:z_distri2} shows the morphological evolution of ULIRG hosts based on $M_{20}$ (middle) and $M$--$D$ diagnostics (right), using the criteria as mentioned in  Section 4.3.4. The detailed breakdown is given in Table \ref{table:table2}. As already noted, these indices only separate galaxies into ``regular'' (corresponding to our visual D and E types) and ``irregular'' (corresponding to our visual Irr/Int type) shapes. To make a fair comparison, we show the evolution based on the visual classification again (left) but plot the E and D galaxies together.  The contamination caused by the E galaxies in the D galaxies is negligible only because there is a small number of E galaxies in the whole sample.
			
			While the trends based on $M_{20}$ and $M$--$D$ differ from the trend based on the visual classification in some details, the transition from Irr/Int to D+E remains the same. 
			
\begin{figure*}[htbp]
       			\centering
       			\includegraphics[width= \textwidth]{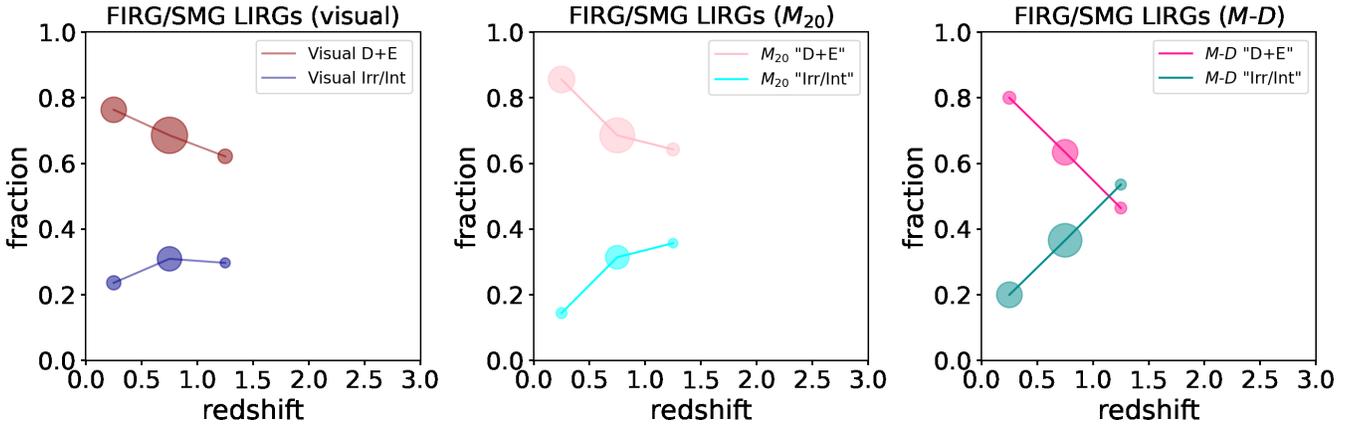} 
       			\caption{Similar to Figure \ref{fig:z_distri2}, but for LIRG hosts.}
       			\label{fig:z_distri_LIRG2}
       		\end{figure*}

			For the sake of completeness, Figure \ref{fig:z_distri_LIRG2} shows the evolutionary trends for the LIRG hosts based on these two sets of indices. These are similar to Figure \ref{fig:z_distri_LIRG}, with the slight difference that there also seems to be a transition at $z\approx 1.25$ in the $M$--$D$ analysis. However, this is highly uncertain due to the small number statistics.
			
			\begin{deluxetable*}{@{\extracolsep{10pt}}lcccccccccccc}
				\centering
				\tablecaption{Redshift distribution of (U)LIRG hosts by morphological types using $M_{20}$ and M-D statistics \label{table:table2}}
				\tablewidth{0.9\textwidth}
				\tabletypesize{ \scriptsize}
				\tablehead
				{\colhead{}&\multicolumn{2}{c}{$0 \leqslant z \le 0.5$} &  \multicolumn{2}{c}{$0.5 \leqslant z \le 1$} & \multicolumn{2}{c}{$1.0 \leqslant z \le 1.5$}& \multicolumn{2}{c}{$1.5 \leqslant z \le 2$}& \multicolumn{2}{c}{$2 \leqslant z \le 2.5$}& \multicolumn{2}{c}{$2.5 \leqslant z \le 3$}  \\
					\cline{2-3} \cline{4-5} \cline{6-7} \cline{8-9} \cline{10-11} \cline{12-13}
					\colhead{Subtypes}& \colhead{$N$}& \colhead{$P$}& 
					\colhead{$N$}& \colhead{$P$}&  \colhead{$N$}& \colhead{$P$}& 
					\colhead{$N$}& \colhead{$P$}&  \colhead{$N$}& \colhead{$P$}&
					\colhead{$N$}& \colhead{$P$} 
				}
				 \startdata	
				ULIRG D+E ($M_{20}$) &0   & ... & 46  &63.9 \%& 71  &54.2 \%& 29  &37.2 \%& 12  &30.0 \%& 6   &33.3 \% \\
				ULIRG Irr/Int ($M_{20}$)&0   & ... & 26  &36.1 \%& 60  &45.8 \%& 49  &62.8 \%& 28  &70.0 \%& 12  &66.7 \% \\
				\midrule
				ULIRG D+E ($M-D$)    &0   & ... & 32  &44.4 \%& 67  &51.1 \%& 62  &79.5 \%& 30  &75.0 \%& 14  &77.8 \% \\
				ULIRG Irr/Int ($M-D$)&0   & ... & 40  &55.6 \%& 64  &48.9 \%& 16  &20.5 \%& 10  &25.0 \%& 4   &22.2 \%\\
				\midrule
				LIRG D+E ($M_{20}$)  &77  &85.6 \%& 133 &68.6 \%& 18  &64.3 \%& 0   &0.0  \%& 0   & ... & 0   & ...  \\
				LIRG Irr/Int ($M_{20}$)&13  &14.4 \%& 61  &31.4 \%& 10  &35.7 \%& 1   &100.0\%& 0   & ... & 0   & ...  \\
				\midrule
				LIRG D+E ($M-D$)     &18  &20.0 \%& 71  &36.6 \%& 15  &53.6 \%& 1   &100.0\%& 0   & ... & 0   & ...  \\
				LIRG Irr/Int ($M-D$) &72  &80.0 \%& 123 &63.4 \%& 13  &46.4 \%& 0   &0.0  \%& 0   & ... & 0   & ...  \\
				\enddata  
				\tabletypesize{ \small}
				\tablecomments{Similar to Table \ref{table:table1}, but for the host types derived using $M_{20}$ and $M$--$D$ criteria, as described in  Section 4.3.4. These indices only separate galaxies into the regular shape that mimics the visual D and E types (but cannot separate between the two) and the irregular shape that mimics the visual Irr/Int type.}
			\end{deluxetable*}

			 \subsection{Locations of (U)LIRG Regions in the Hosts}

            The accurate ALMA and VLA pinpointing of the (U)LIRG regions, together with the high-resolution COSMOS-DASH imagery, offers a unique opportunity to investigate the locations of the (U)LIRG regions with respect to their galaxy hosts as a whole. To minimize the possible ambiguity in determining the centroid, we limit this investigation to only the disk galaxies at $z\leqslant 3$, which have defined shapes and are of a large enough number for statistics.
            
\begin{figure*}[htbp]  
       			\centering
       			\includegraphics[width=\textwidth]{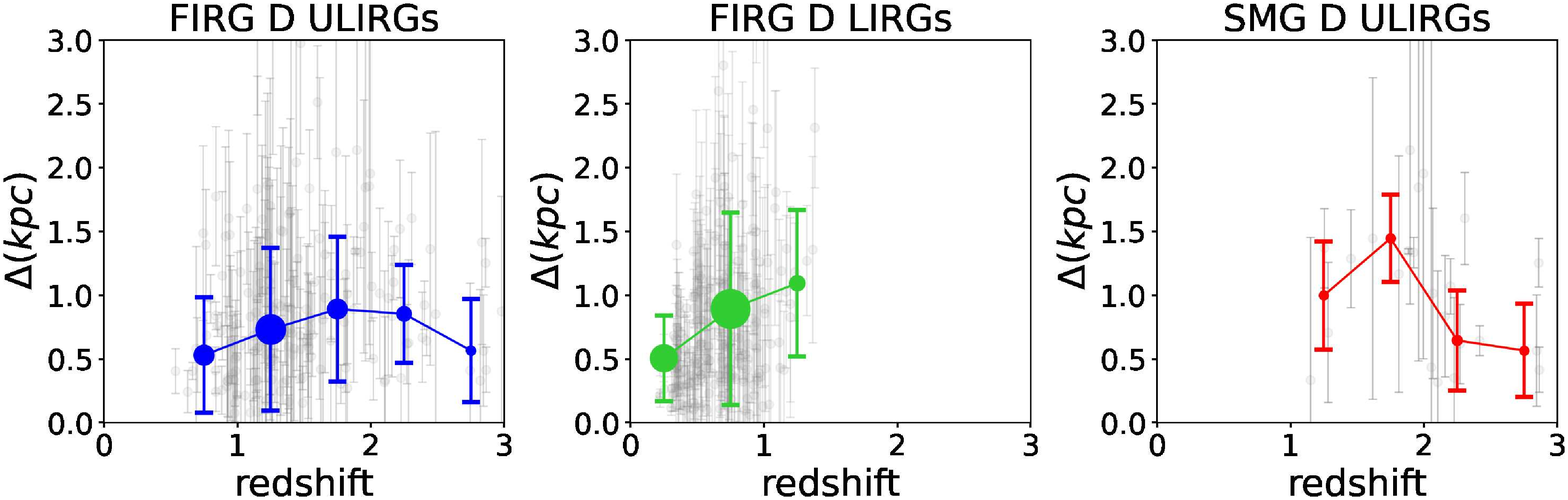} 
       			\caption{Physical separations (in kiloparsecs) between the (U)LIRG regions as pinpointed by the ALMA/VLA positions and the centroids of their galaxy hosts as determined in the COSMOS-DASH $H_{160}$ images, shown for the disk galaxies at $z\leqslant 3$ in our sample. From left to right, the results are shown for the FIRG ULIRGs, FIRG LIRGs, and SMG ULIRGs. The gray symbols are the individual objects, and the error bars represent the ALMA/VLA positional uncertainties. The colored symbols are the median values in the redshift bins as in Figure \ref{fig:z_distri} and \ref{fig:z_distri_LIRG} and Table \ref{table:table1}, and the attached error bars represent the 1 $\sigma$ dispersion of the data points.}
       			\label{fig:dist_center}
       		\end{figure*}

            There are 176, 23, and 223 D galaxies among the FIRG ULIRG, SMG ULIRG, and FIRG LIRG hosts, respectively. We calculated the separations between their ALMA/VLA positions and their centroids in the $H_{160}$ images, the latter of which were based on our SExtractor run as described in  Section 2.4.   {The small systematic offset between the ALMA/VLA and the $H_{160}$ positions has been corrected as explained in Appendix \ref{appendix:d}.} These angular separations (in arcseconds) were then converted to physical separations (in kiloparsecs) according to the source redshifts. The results are shown in Figure \ref{fig:dist_center} for the FIRG ULIRGs, FIRG LIRGs, and SMG ULIRGs. The error bars represent the ALMA/VLA positional errors as described in  Sections 3.4 and 3.5, while the errors of the $H_{160}$ centroids are ignored because they are negligible. The mean values at the various redshift bins (the same as those in Figure \ref{fig:z_distri}) are also given as colored circles, and their error bars are the 1$\sigma$ dispersions of the data points. The median physical separations are   {0.75, 0.62, and 1.07} kpc for the FIRG ULIRGs, FIRG LIRGs, and SMG ULIRGs, respectively. In other words, the (U)LIRG regions are not concentric with other stellar populations in their hosts.
            
       		The offset for the SMG ULIRGs is consistent with the result of \citet[][]{Lang2019} based on 20 SMGs (median offset of 1.1 kpc) and is also in general agreement with the conclusion of \citet[][]{Chen2015} based on 48 SMGs in which that they saw a large scatter of positional differences between the ALMA centroids and the $H_{160}$ centroids. The morphologies of the objects in these two latter studies are mostly irregular. In contrast, our measurements here are limited to disk galaxies for both the FIRGs and the SMGs; therefore, the centroid determinations are more secure.
       		
       		The offset can also be viewed as a piece of evidence against the suggestion that the intense star formation in high-$z$ (U)LIRGs is a galaxy-wide activity extending over a few kiloparsecs but not confined to a number of discrete clumps, as observed in local ULIRGs. Such a suggestion was put forward by a number of SMG size measurements using ALMA and/or VLA images \citep[see, e.g.,][and references therein]{Simpson2015a}. If this is true, the dust emission in high-$z$ (U)LIRGs should tend to be concentric with their starlight emission. However, this is not what we observe. As pointed out by \citet[][]{Yan2016}, the Stefan-Boltzmann law limits the sizes of dusty starburst regions to $\lesssim$2~kpc, and the seemingly extended dust emission could simply be due to blending of discrete clumps under insufficient spatial resolution. In addition to the examples in \citet[][]{Yan2016}, this is also supported by the very high resolution ALMA observations of \citet[][]{Iono2016} on two of the brightest SMGs in COSMOS. The offset as shown in Figure \ref{fig:dist_center} can be easily understood if the dust-embedded starburst regions in high-$z$ (U)LIRGs are clumpy but not diffuse and widespread over the entire hosts.
       		
        \section{Summary} 	   
   
	   	    In this work, we have studied the rest-frame optical morphologies of the FIRG and SMG host galaxies in the COSMOS field. The FIRGs are from the HerMES SPIRE 250~$\mu$m detections, while the SMGs are the SCUBA2 850~$\mu$m sources from the S2COSMOS and S2CLS programs. The SMGs do not constitute a subset of FIRGs, but this is mainly due to the S/N requirements in selecting the samples. Only a minority of the SMGs are genuine nondetections in the HerMES SPIRE maps, while the latter have detected nearly an order of magnitude more FIRGs than SMGs.
	   	    
	   	    The exact locations of the FIRGs/SMGs are pinpointed by the deep VLA 3~GHz data in this field, as well as the ALMA data from the compilation of the A$^3$COSMOS program. The agreement between the ALMA and VLA identifications is $>$80\%, which gives us the confidence to trust the VLA identifications in the much wider area where there are not yet public ALMA data. While the VLA data still only identify 56.4\% and 75.2\% of the FIRGs and  SMGs, respectively, their use has enabled the construction of the largest FIRG/SMG sample to date for morphological studies. The morphologies are mostly based on the HST WFC3 $H_{160}$ images from the COSMOS-DASH program and are also aided by using the ACS $I_{814}$ image when checking for the ``morphological $k$-correction.'' Here we summarize the main results of our work.
	   	   \begin{enumerate}
	   	     \item Our final morphological sample contains 1266 $H_{160}$ counterparts that correspond to 1090 FIRGs and 172 SMGs (117 in common). Among them, 1235 have reliable redshifts (623 $z_{\rm spec}$ and 612 $z_{\rm ph}$). The vast majority of these objects are at $z<3$, therefore, $H_{160}$ samples their rest-frame optical--to--near--IR emissions that are dominated by the mature stellar populations of the hosts.
	   	    
	   	     \item About 12.3\% of the FIRGs and 19.2\% of the SMGs in our sample have multiple counterparts identified. Among these FIRG and SMG multiplicity cases, at least 29.2\% and 27.2\% are made of physically associated multiple components, and at least 56.0\% and 48.5\% are due to chance alignment, respectively.
	   	    
	   	     \item Based on our visual classification using their $H_{160}$ images, these 1266 counterparts are put into five morphological categories, namely, Elliptical (E), Disk (D), Irregular/Interacting (Irr/Int), Compact/Unresolved (C), and Faint (F). The majority are in the D and Irr/Int categories. The objects at $z < 1$ (523 in total) also have visual classifications based on their $I_{814}$ images, and the agreement between the two sets is very good. This means that the morphological $k$-correction is minimal in our case. Interestingly, the D galaxies are systematically at lower redshifts than (``lagging behind'') the Irr/Int galaxies among both the FIRGs and SMGs.
	   	    
	   	     \item The D and E galaxies also have quantitative measurements from the fits of their light distributions to the S\'ersic profile. The results are very consistent with our visual classifications. A single profile does not fit well some D galaxies that have prominent bulges, in which case an additional profile must be used to fit the bulge component. Nevertheless, all galaxies visually classified as D galaxies have S\'ersic indices consistent with possessing a disk, i.e., having a profile with $n \approx 1$. The E galaxies have $n \approx 5$, also as expected. No attempt was made to fit galaxies in other types, especially the Irr/Int category.
	   	    
	   	     \item Three different types of nonparametric morphology classifications, namely Gini-$\rm M_{20}$, CAS, and MID, were carried out for the E, D and Irr/Int galaxies, and reliable results were obtained for 864 of them, among the total of 1061. Using $M_{20}$ or $M$-$D$ analysis, we can achieve better than 80\% consistency with the visual classification in separating regular shape (the visual D + E type) and irregular shape (the visual Irr/Int type) galaxies.
	   	    
	   	     \item About 82\% of the objects in our sample (1040 out of 1266) have reliable $L_{\rm IR}$ estimates from their FIR-to-mm SEDs and redshifts. We focus on the ULIRGs (587 FIRGs and 123 and SMGs, 97 in common) and LIRGs (338 FIRGs and five SMGs, four in common). The D galaxies are lagging behind the Irr/Int galaxies in both ULIRGs and LIRGs.
	   	    
	   	     \item We were able to derive stellar masses for 1085 objects using their 28-band optical-to-near-IR SEDs. Among them, 514 and 96 are FIRG and SMG ULIRG hosts, respectively (76 in common), and 302 and five are FIRG andSMG LIRG hosts, respectively (four in common). The median $M_*$ for the hosts of the SMG ULIRGs, FIRG ULIRGs and FIRG LIRGs are $7.9\times 10^{10}$, $5.0\times 10^{10}$ and $4.0\times 10^{10} M_\odot$, respectively. Such differences can be largely explained by the selection bias, which is also seen in their redshift distributions. Among the hosts of the FIRG ULIRGs, FIRG LIRGs, and SMG ULIRGs, the D galaxies also seem to lag behind the Irr/Int galaxies in redshift and are of slightly higher stellar masses.
	   	    
	   	     \item For the ULIRG hosts, there is a strong evolution of the fractional contributions of different morphological types as a function of redshift. For the FIRG ULIRG hosts, the transitional redshift is $z\approx 1.25$, above which the Irr/Int galaxies dominate and below which the D and Irr/Int galaxies have nearly the same contributions. The SMG ULIRG hosts also seem to exhibit a similar transition, with the difference that the D galaxies become dominant over (instead of having the same contribution as) the Irr/Int galaxies at $z\approx 1.25$. Among the FIRG LIRG hosts, which are mostly at $z\leqslant 1.25$ in our sample, the D galaxies always dominate over the Irr/Int galaxies ($\sim$70\% versus $\sim$20\%).
	   	    
	   	     \item The ALMA/VLA and $H_{160}$ positions of our objects do not coincide exactly. Even for the D galaxies whose regular light profiles make their positions the least prone to error, the offset between the two sets is obvious (median value of $\sim$0.6--1.1 kpc). Such differences can be understood if the starburst activities in high-$z$ (U)LIRGs are confined in a few discrete dusty clumps instead of happening over the entire hosts in diffused form.
	   	\end{enumerate}

   			We thank the referee for the constructive comments. We acknowledge the support of University of Missouri Research Council grant URC-21-005.

        \appendix

         \section{Redshift Information}
         \label{appendix:a}
        
        We searched through all of the public redshift catalogs in the COSMOS field for our 1266 systems. These include 32 spectroscopic ($z_{\rm spec}$) and four photometric ($z_{\rm ph}$) catalogs. The matching radius was set to 1\farcs05. Among the 32 $z_{\rm spec}$ catalogs, 15 have a quality flag, and we only consider the redshifts with secure flags. The other 17 do not have such a flag, and we regard all of their quoted redshifts as being secure. A number of our objects have different, secure $z_{\rm spec}$ from different catalogs, and we adopted the ones with the lowest errors. In total, we obtained 613 secure $z_{\rm spec}$.
                            
        The 32 $z_{\rm spec}$ catalogs are taken from the following papers: \citet[][SDSS DR14]{Abolfathi2018}; \citet[][SDSS DR16]{Ahumada2020}; \citet[][GEMINI-S]{Balogh2014}; \citet[][Magellan-FIRE]{Calabro2018}; \citet[][LRIS/DEIMOS]{Casey2012b}; \citet[][MOSFIRE]{Casey2015}; \citet[][LRIS/DEIMOS]{Casey2017}; \citet[][PRIMUS]{Coil2011}; \citet[][FORS2]{Comparat2015}; \citet[][hCOSMOS]{Damjanov2018}; \citet[][]{Gomez-Guijarro2018}; \citet[][DEIMOS]{Hasinger2018}; \citet[][ALMA]{Kaasinen2019}; \citet[][FMOS]{Kartaltepe2015}; \citet[][FMOS]{Kashino2019}; \citet[][MOSFIRE]{Kriek2015}; \citet[][WFC3-Grism]{Krogager2014}; \citet[][]{Lee2016}; \citet[][zCOSMOS-DR3]{Lilly2007,Lilly2009}; \citet[][]{Liu2019a}; \citet[][NIRSPEC]{Marsan2017}; \citet[][]{Marchesi2016}; \citet[][DEIMOS-C3R2]{Masters2017}; \citet[][ZFIRE]{Nanayakkara2016}; \citet[][]{Roseboom2012}; \citet[][MOSFIRE]{Schreiber2018}; \citet[][FMOS]{Silverman2015a}; \citet[][IMACS]{Trump2007,Trump2009}; \citet[][KMOS/VLA/NOEMA]{Wang2016}; \citet[][LMT]{Yun2015}; \citet[][LEGA-C]{vanderWel2016}.

        For those that do not have $z_{\rm spec}$, we had to rely on $z_{\rm ph}$. The main $z_{\rm ph}$ catalog that we used was the COSMOS2015 catalog of \citet[][]{Laigle2016}, from which we found 526 matches. However, we had to reject 21 of them that were identified as X-ray AGNs, and we replaced them with $z_{\rm ph}$ in \citet[][]{Salvato2011}. We also replaced six other matches in COSMOS2015 with those from \citet[][]{Davidzon2017} because they were preferred in the Robust Redshift Catalog of \citet[][]{Liu2019a}. For those that were not matched by COSMOS2015, we searched the UltraVISTA catalog \citep[][]{Muzzin2013} and obtained 86 matches.
        
        Therefore, we obtained redshifts for 1235 of our objects (623 $z_{\rm spec}$ and 612 $z_{\rm ph}$). Among them, five were clearly in interacting systems, and their redshifts were obtained through their companions. There are 31 objects in our morphological samples that do not have reliable redshift information; we still study their morphology, but we do not include them in the final discussion.
                   
        \begin{figure*}[htbp]  
        	\centering
        	\includegraphics[width=\textwidth]{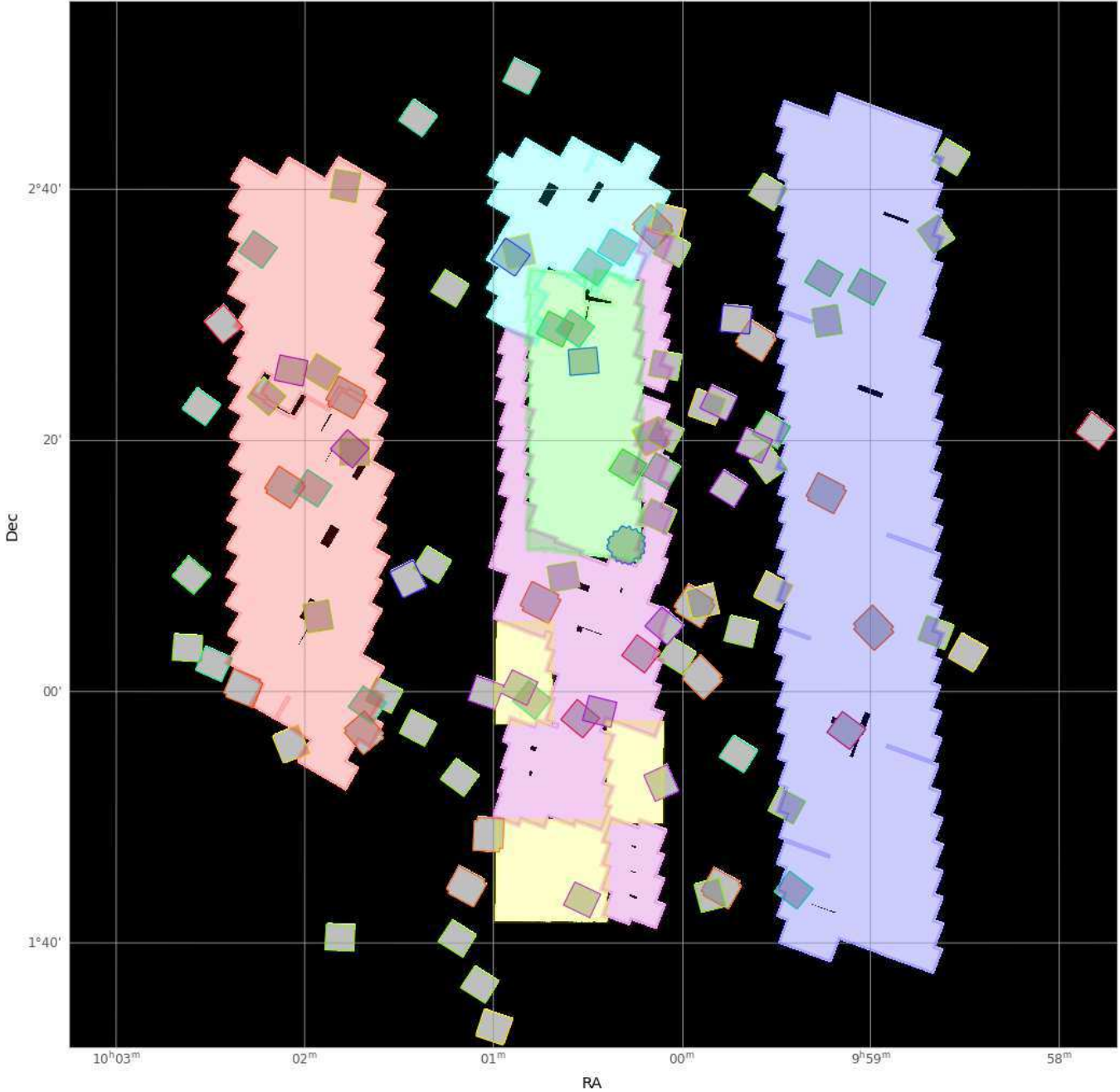} 
        	\caption{Subregion splitting of the COSMOS-DASH mosaic for the PSF constructions. The entire mosaic is split into 16 subregions, each with its own PSF. The area coded in green represents the CANDELS-COSMOS subregion. Those coded in red, cyan, magenta, yellow, and blue represent the five shallow (but wide) subregions covered by the DASH's own observations. They are separated as such because of their sky positions and the orientations of the camera. The small, discrete areas (coded in gray) spreading over the whole field are those observed by other GO programs, and they are grouped into individual subregions based on the corresponding GO programs that they belong to. There are 10 such subregions, each coded with a different boundary color.}
        	\label{fig:subregions}
        \end{figure*}

         \section{Details of S\'ersic Profile Fitting}
        \label{appendix:b}
        
        	Here we provide the details of the S\'ersic profile fitting, which was done for the galaxies that are visually classified as D and E galaxies.
        	
        	 \subsection{ Preparation of Image Files}
        	\label{appendix:b1}
        	
	        The $10^{\prime\prime}\times10^{\prime\prime}$ $H_{\rm 160}$ cutouts were large enough to run GALFIT on most objects. A few low-redshift galaxies, however, needed larger sizes. For these objects, their cutouts were made with a size that was at least 6 $\times$~KRON\_RADIUS~$\times$~A\_IMAGE on a side, where KRON\_RADIUS is the Kron radius and A\_IMAGE is the semimajor axis, both obtained through our SExtractor run. For all sources, we created cutouts from both the science image and the rms map, as both would be needed by GALFIT.
	        
	        While the COSMOS-DASH mosaic was made to be free of background, we still estimated the background value independently for each object because GALFIT is known to be sensitive to this value. This was done around each object in an area twice the size of its cutout on each side. We masked all of the detected sources, ran 3$\sigma$ clipping on the remaining pixels, and took the average as the background value. These values are all very close to zero (ranging from $-0.01$ to $0.01$), which is expected for the background-free COSMOS-DASH image. The pixel distributions all follow a Gaussian function, which further validates our approach.
	        
	         \subsection{Building PSFs}
	        
	        The COSMOS-DASH mosaic is a combination of observations from multiple programs and therefore has significant variation of depth and instrument orientation in different regions of the mosaic. As the GALFIT performance critically depends on the quality of the PSF, we need to consider the possible impact of the PSF differences in different regions. To this end, we split the whole mosaic into 16 subregions: the CANDELS-COSMOS subregion, 10 other deep subregions observed by various GO programs, and five shallow (but wide) regions covered by DASH's own observations (see Figure \ref{fig:subregions}). We derived the PSF in each subregion individually. We attribute the overlapping parts of different subregions to the deepest subregions. In this process, we made extensive use of the \texttt{AstroPy} package \citep{Astropy2013, Astropy2018}.

	        We selected point sources for PSF construction using the method of \citet[][]{Momcheva2017}. The basic idea of the method is that the flux ratio of point sources in different circular apertures is more or less a fixed value, while that of extended sources should have a large dispersion. \citet[][]{Momcheva2017} showed that point sources in the COSMOS-DASH mosaic follow a tight sequence on the flux ratio versus \texttt{MAG\_AUTO} plane, where the flux ratio is measured using two circular apertures of 2\farcs0 and 0\farcs5 in diameter. We made the same plot and selected the sources with a flux ratio between 1.28 and 1.38 and \texttt{MAG\_AUTO} $<22.0$ as in \citet[][]{Momcheva2017}. We visually inspected all of the selected sources to make sure that they are good stars for stacking. The number of selected stars in different subregions ranged from 347 stars in the CANDELS field to 11 stars in the field of the GO-12461 project.
	        
	        The PSFs were constructed as follows. We first cut out a square region of $201\times 201$ pixels around each star and masked all other sources in this region according to our source catalog. We then subsampled the cutout to a finer grid by 10 times in both dimensions using linear interpolation. The subsampled image was then recentered on the star. All of the subsampled images in a given subregion were normalized to the unity flux and stacked by average. This stacked image was rebinned by a factor of 10 in both dimensions to restore the original resolution. Finally, the central region of $101\times 101$ pixels was cut out to create the PSF image of this subregion. The same procedure was carried out to create the rms map of the PSF image as well.

	         \subsection{Running GALFIT}
	        	              
	        We wrote a wrapper program to create the GALFIT input files to run GALFIT. For each object, it first reran SExtractor to get the initial values of the necessary parameters, including the total magnitude ($m$), the half-light radius ($R_e$) measured along the semimajor axis, the axis ratio ($b/a$), the position angle (PA), and the central position ($x_0$, $y_0$). The initial S\'ersic index ($n$) was set to 1. The constraint file was constructed to keep $n$ between 0.1 and 10,  $R_e$ between 0\farcs01 and 20\farcs0 \ (0.1 and 200 pixels), $b/a$ between 0.1 and 1, $m$ between $-$3 and +3 mag from the input value, and the position uncertainty between $-$0\farcs1 and 0\farcs1 \ ($-$1 and 1 pixel). The wrapper then decided on what to do with the neighbors of the target object following a general rule: if the neighbors are far away from the target, they should be masked, and if the neighbors are so close that they could potentially contaminate the target, they should be fitted simultaneously with the target. GALFIT was then run accordingly. The procedure often needed to be run iteratively because of two complications. One was due to very close neighbor(s), which severely contaminate(s) the target. In this case, we masked the target and first fitted the neighbor(s). The residual map, which contained only the target, was then fitted again. The other complication was that some disk galaxies could not be fitted by a single S\'{e}rsic profile due to their prominent bulges. Such galaxies were fitted again using two profiles. The primary profile (for the disk component) took the results from the single profile as the initial input. If the bulge was very compact, the second profile would simply be the PSF. If it was resolved, an  $n=4$ profile would be used instead. The second profile would share the same initial position as the primary one, and its initial magnitude would be set to 2.0 mag fainter than the primary. If it was an $n=4$ profile added, its initial $R_e$ value would be set to 2$\times$ smaller than the primary.

	        Finally, we tested the robustness of the GALFIT results against the PSF variations. We randomly chose 10 galaxies over the entire COSMOS-DASH mosaic and fit each of them using our 16 PSFs. As it turned out, the resulting S\'{e}rsic index varied by about 5\%--10\%. We further used 50 PSF images provided by \citet[][]{Mowla2019} for the COSMOS-DASH mosaic and repeated the test for these 10 galaxies. We found that the results were 3\%--5\% smaller than those based on our own PSFs. Therefore, we conclude that our GALFIT results are robust even if our PSFs in any given subregion might not be perfect.

	        \section{Discussion of the Varying Depth of the CANDELS-DASH Mosaic}
	       \label{appendix:c}
	       
	       \begin{figure*}[htbp]  
	       	\centering
	       	\includegraphics[width=\textwidth]{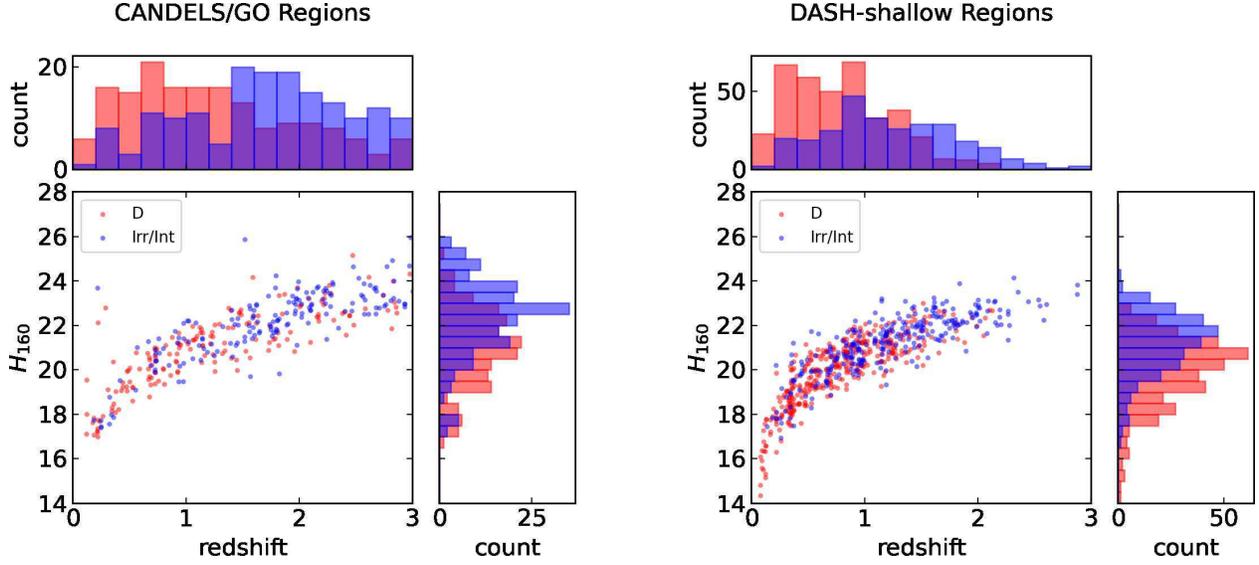} 
	       	\caption{  {Redshift and $H_{160}$ magnitude distributions of the D and Irr/Int galaxies in the CANDELS/GO regions (left) and the DASH-shallow (right) regions. The red and blue points represent the D galaxies and the Irr/Int galaxies in our sample, respectively.}}
	       	\label{fig:mag-redshift}
	       \end{figure*} 	
	       
	       As discussed in  Section 2.3, the COSMOS-DASH mosaic was created by including both the data from its own shallow observations (DASH-shallow) and the deeper archival data from CANDELS and 10 other GO programs (CANDELS/GO). The 5$\sigma$ limits are $H_{160}=25.1$ and approximately $H_{160}\approx 26.9$ mag for the DASH-shallow and CANDELS/GO data, respectively. Such a large difference in sensitivity raises a concern as to galaxy morphology is assessed on fair grounds in the shallow and deep regions and whether this could affect our conclusion on the evolutionary trend.
	       
	       We argue that our approach largely avoids any possible systematic biases that could be caused by the varying sensitivity. Most importantly, we quantify the evolution in terms of the fraction of a certain morphological type among the total. As galaxies of a certain type (e.g., D or Irr/Int) would not fall preferentially in the shallow or deep regions, using fractions would not have any intrinsic bias. In addition, it is unlikely that either D or Irr/Int galaxies would be preferentially classified as one versus the other due to the lack of sensitivity. We can look at this from a different angle. It is true that the F (faint) or even some of the C (compact) galaxies in the shallow regions would have been identified with distinctive features should they be found in the deep regions; however, it is unlikely that they would become preferentially either D or Irr/Int galaxies.
	       
	       Using fractions also avoids the complexity of addressing the   completeness issue.   {Figure \ref{fig:mag-redshift}  shows the redshift and the $H_{160}$ magnitude distributions} of the D and Irr/Int galaxies (based on visual classification) in the CANDELS/GO regions and  DASH-shallow regions. Clearly, the DASH-shallow data suffer from significant in  completeness at $z>2$ as compared to the CANDELS/GO data. However, as we are using fractions, the shallower data still add significantly to the statistics.

	        \section{Correction of Astrometric Offset}
	        \label{appendix:d}
	        
	          {\citet[][]{Smolcic2017b} and \citet[][]{Liu2019a} discussed the systematic differences in astrometry between the radio/submillimeter/millimeter data and the optical/near-IR data. \citet[][]{Liu2019a} reported that the A$^3$COSMOS and VLA positions agreed very well. However, they both have offsets with respect to the UltraVISTA data, which amount to 0\arcsec.10 and 0\arcsec.088 in the R.A. direction (negligible in the decl. direction) for the VLA and A$^3$COSMOS positions, respectively.} 

\begin{figure}[htbp]  
	        	\centering
	        	\includegraphics[width=\textwidth]{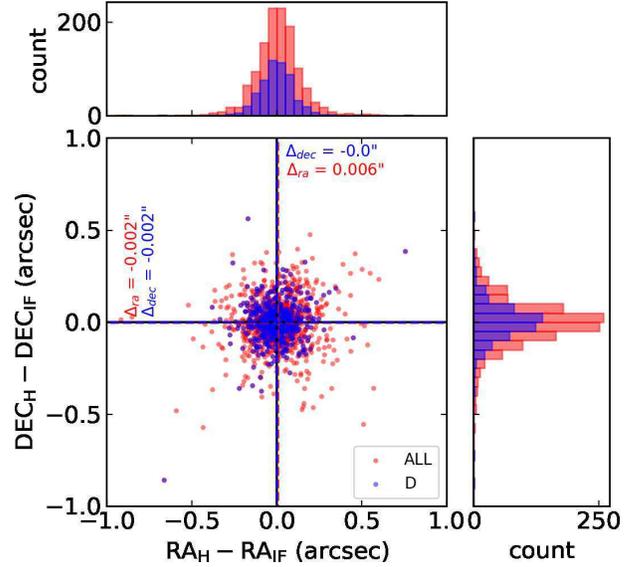} 
	        	\caption{  {Elimination of systematic difference in astrometry between the $H_{160}$ (labeled as ``H'') and the VLA/A$^3$COSMOS interferometric (labeled as ``IF'') positions after the correction of 0\arcsec.065 in the R.A. direction. The blue and the red points represent the D-type galaxies and all galaxies in our sample, respectively. The dashed lines indicate the mean offset values in the R.A. and the decl. directions, which are also labeled in the figure.}}
	        	\label{fig:sep-cor}
	        \end{figure}

		      {For the objects in our sample, we compared the $H_{160}$ positions to the A$^3$COSMOS and VLA positions and found similar offsets. The difference in the decl. direction is negligible, and the mean offset in the R.A. direction is 0\arcsec.065 and 0\arcsec.071 for the D category and for all objects, respectively. In  Section 5.5.1, we present the discussion of the locations of the (U)LIRG sites in their hosts, which is based on the $H_{160}$ positions of the D galaxies after correcting for this systematic offset of 0\arcsec.065  in R.A. Figure \ref{fig:sep-cor} shows the positional differences between the $H_{160}$ and VLA/A$^3$COSMOS positions after this correction.}
	        
            \section{summary of online data table}

        	The header information for the online data table is given in Table \ref{table:table3}. This table includes all of the 1266 sources as summarized in  Section 3.7. 
        	
        	In most cases, the IDs of our objects are those taken from their respective parent catalogs, i.e., \texttt{XIDs} in the HerMES catalog and \texttt{Nickname} in the S2COSMOS catalog. If a source is a common source, its ID is constructed by connecting its HerMES ID and S2COSMOS ID using an underscore. In a few cases, the source is oversplit in the HerMES catalog and therefore has two \texttt{XIDs}; the ID of such a source is constructed by connecting the two \texttt{XIDs} using a plus sign.
        	
        	As discussed in  Section 3.2, the same ALMA source can have multiple entries in the A$^3$COSMOS catalog if it has multiple observations. For such objects, we chose to show only two entries: one is what we adopt for the position (denoted as ``A3Pos'' in our catalog) because it has the smallest positional error, and the other is what we adopt for the ALMA flux measurement because it has the largest S/N (denoted as ``A3MaxSN'' in our catalog).

           \startlongtable
        	\begin{deluxetable}{lll}
        		\tabletypesize{ \scriptsize}
        		\tablewidth{0.45\textwidth} 
        		\tablecaption{Summary of the Online Data Table \label{table:table3}}
        		\tablecolumns{3}
        		\tablehead
        		{
        			\colhead{Column} & \colhead{Name} &  \multicolumn{1}{l}{Description}
        		}
        		 \startdata	
        		1	&	ID	&	Source ID	\\
        		2	&	RAdeg-250	&	HerMES R.A. in deg (J2000)	\\
        		3	&	DEdeg-250	&	HerMES decl. in deg (J2000)	\\
        		4	&	SNR-250	&	HerMES S/N	\\
        		5	&	Sigma-Pos-250	& 1 $\sigma$ error of HerMES position		\\
        		6	&	RAdeg-850	&	S2COSMOS R.A. in deg (J2000)	\\
        		7	&	DEdeg-850	&	S2COSMOS decl. in deg (J2000)	\\
        		8	&	SNR-850	&	S2COSMOS S/N	\\
        		9	&	Sigma-Pos-850	&	1 $\sigma$ error of S2COSMOS position	\\
        		10	&	ID-3GHz	&	VLA 3 GHz ID	\\
        		11	&	RAdeg-3GHz	&	VLA 3 GHz R.A. in deg (J2000)	\\
        		12	&	DEdeg-3GHz	&	VLA 3 GHz decl. in deg (J2000)	\\
        		13	&	SNR-3GHz	&	VLA 3 GHz S/N	\\
        		14	&	Sigma-Pos-3GHz	&	1 $\sigma$ error of VLA 3 GHz Position (arcsec)	\\
        		15  &	flag-multi-3GHz &	VLA 3 GHz multiflag \tablenotemark{ \scriptsize a}	\\
        		16	&	RAdeg-A3	&	A$^3$COSMOS R.A. in deg (J2000)	\\
        		17	&	DEdeg-A3	&	A$^3$COSMOS decl. in deg (J2000)	\\
        		18	&	A3PosEntry	& Entry no. of A3Pos in the blind cat.  \tablenotemark{ \scriptsize b}\\
        		19	&	A3PosSN	&	A3Pos S/N	\\
        		20	&	A3PosWave	&	A3Pos central wavelength (mm)	\\
        		21	&	A3PosBmaj	&	A3Pos major beam size (arcsec)	\\
        		22	&	A3PosBmin	&	A3Pos minor beam size (arcsec)	\\
        		23	&	A3MaxSNEntry	&	Entry no. of A3MaxSN in the blind cat. \tablenotemark{ \scriptsize c} \	\\
        		24	&	A3MaxSN	&	S/N of A3MaxSN	\\
        		25	&	A3MaxSN-Flux	&	Total\_flux\_pbcor of A3MaxSN (mJy)	\\
        		26	&	e\_A3MaxSN-Flux	&	E\_Total\_flux\_pbcor of A3MaxSN (mJy)\\
        		27	&	RAdeg-H160	&	$H_{\rm 160}$ counterpart R.A. in deg (J200)		\\
        		28	&	DEdeg-H160	&	$H_{\rm 160}$ counterpart decl. in deg (J200)		\\
        		29	&	f250	&	250 $\mu $m flux density (mJy)	\\
        		30	&	e\_f250	&	250 $\mu $m flux density error (mJy)	\\
        		31	&	f350	&	350 $\mu $m flux density (mJy)	\\
        		32	&	e\_f350	&	350 $\mu $m flux density error (mJy)	\\
        		33	&	f500	&	500 $\mu $m flux density (mJy)	\\
        		34	&	e\_f500	&	500 $\mu $m flux density  error (mJy)	\\
        		35	&	S850-deb	&	850 $\mu $m deboosted flux density (mJy)	\\
        		36	&	e\_S850-deb	&	850 $\mu $m deboosted flux density error (mJy)	\\
        		37	&	S3GHz	&	VLA 3 GHz flux density (mJy)	\\
        		38	&	e\_S3GHz	&	VLA 3 GHz flux density error (mJy)	\\
        		39	&	H160	&	MAG\_AUTO of  $H_{\rm 160}$ counterpart	\\
        		40	&	e\_H160	&	MAG\_AUTOERR of  $H_{\rm 160}$ counterpart	\\
        		41	&	Log10(LIR)	&	$\rm log_{10}$(LIR) in solar units	\\
        		42	&	e\_Log10(LIR)	&	Error of Log10(LIR)	\\
        		43	&	AGE	&	Age in yr	\\
        		44	&	e\_AGE	&	Error of AGE	\\
        		45	&	Log10(Mass)	&	$\rm log_{10}$(stellar mass) in solar units	\\
        		46	&	e\_Log10(Mass)	&	Error of Log10(Mass)	\\
        		47	&	EBV	&	$E(B$--$V)$	\\
        		48	&	z	&	Adopted redshift	\\
        		49	&	z-type	&	Type of redshift (spec/phot) 	\\
        		50	&	r\_z		&	Reference of redshift catalog	\\
        		51	&	VClass-I814	&	$I_{\rm 814}$ visual classification type\\
        		52	&	VClass-H160	&	$H_{\rm 160}$ visual classification type	\\
        		53 	&	flag-Galfit-method & GALFIT method flag \tablenotemark{ \scriptsize d}\\
        		54	&	n	&	S\'ersic index	\\
        		55	&	e\_n	&	S\'ersic index error	\\
        		56	&	Re	&	Effective radius (arcsec)	\\
        		57	&	e\_Re	&	Effective radius error (arcsec)	\\
        		58	&	Re-kpc	&	Effective radius (kpc)	\\
        		59	&	Gini	&	Gini index 	\\
        		60	&	M20	&	$M_{\rm 20}$ index 	\\
        		61	&	C	&	Concentration index	\\
        		62	&	A	&	Asymmetry index	\\
        		63	&	S	&	Smoothness index	\\
        		64	&	M	&	Multimode index	\\
        		65	&	I	&	Intensity index	\\
        		66	&	D	&	Deviation index	\\
        		67	&	flag-statmorph	&	\texttt{starmorph} flag	 \tablenotemark{ \scriptsize e}\\
        		\enddata
        		\tablenotetext{a}{flag\_Multi\_3 GHz: 0 for no multiple components,  1 for multiple components identified by \citet[][]{Smolcic2017a}, 2 for multiple components identified by our visual check.}
        		\tablenotetext{b}{A3Pos stands for the adopted A$^3$COSMOS source for position and the relevant observation. The entry number is the row number in the A$^3$COSMOS blind catalog.}
        		\tablenotetext{c}{A3MaxSN stands for the adopted A$^3$COSMOS source for flux density and the relevant observation. The entry number is the row number in the A$^3$COSMOS blind catalog.}
        		\tablenotetext{d}{flag\_Galfit\_method: 1 for fitting with a single S\'ersic profile, 2 for fitting with one S\'ersic profile and one PSF profile, 3 for fitting with two S\'ersic profiles.}
        		\tablenotetext{e}{flag\_statmorph: 0 for good fitting, 1 for bad fitting.}
        \end{deluxetable}
	\newpage  
	
	\bibliography{manuscript.bib}    

\begin{thebibliography}{}
\expandafter\ifx\csname natexlab\endcsname\relax\def\natexlab#1{#1}\fi
\providecommand{\url}[1]{\href{#1}{#1}}
\providecommand{\dodoi}[1]{doi:~\href{http://doi.org/#1}{\nolinkurl{#1}}}
\providecommand{\doeprint}[1]{\href{http://ascl.net/#1}{\nolinkurl{http://ascl.net/#1}}}
\providecommand{\doarXiv}[1]{\href{https://arxiv.org/abs/#1}{\nolinkurl{https://arxiv.org/abs/#1}}}

\bibitem[{{Aaronson} \& {Olszewski}(1984)}]{Aaronson1984}
{Aaronson}, M., \& {Olszewski}, E.~W. 1984, \nat, 309, 414,
  \dodoi{10.1038/309414a0}

\bibitem[{{Abolfathi} {et~al.}(2018){Abolfathi}, {Aguado}, {Aguilar}, {Allende
  Prieto}, {Almeida}, {Ananna}, {Anders}, {Anderson}, {Andrews}, {Anguiano}, \&
  et~al.}]{Abolfathi2018}
{Abolfathi}, B., {Aguado}, D.~S., {Aguilar}, G., {et~al.} 2018, \apjs, 235, 42,
  \dodoi{10.3847/1538-4365/aa9e8a}

\bibitem[{{Ahumada} {et~al.}(2020){Ahumada}, {Prieto}, {Almeida}, {Anders},
  {Anderson}, {Andrews}, {Anguiano}, {Arcodia}, {Armengaud}, {Aubert}, \&
  et~al.}]{Ahumada2020}
{Ahumada}, R., {Prieto}, C.~A., {Almeida}, A., {et~al.} 2020, \apjs, 249, 3,
  \dodoi{10.3847/1538-4365/ab929e}

\bibitem[{{An} {et~al.}(2019){An}, {Simpson}, {Smail}, {Swinbank}, {Ma}, {Liu},
  {Lang}, {Schinnerer}, {Karim}, {Magnelli}, {Leslie}, {Bertoldi}, {Chen},
  {Geach}, {Matsuda}, {Stach}, {Wardlow}, {Gullberg}, {Ivison}, {Ao}, {Coogan},
  {Thomson}, {Chapman}, {Wang}, {Wang}, {Yang}, {Asquith}, {Bourne}, {Coppin},
  {Hine}, {Ho}, {Hwang}, {Kato}, {Lacaille}, {Lewis}, {Oteo}, {Scholtz},
  {Sawicki}, \& {Smith}}]{An2019}
{An}, F.~X., {Simpson}, J.~M., {Smail}, I., {et~al.} 2019, \apj, 886, 48,
  \dodoi{10.3847/1538-4357/ab4d53}

\bibitem[{{Arnouts} {et~al.}(2002){Arnouts}, {Moscardini}, {Vanzella},
  {Colombi}, {Cristiani}, {Fontana}, {Giallongo}, {Matarrese}, \&
  {Saracco}}]{Arnouts2002}
{Arnouts}, S., {Moscardini}, L., {Vanzella}, E., {et~al.} 2002, \mnras, 329,
  355, \dodoi{10.1046/j.1365-8711.2002.04988.x}

\bibitem[{{Astropy Collaboration} {et~al.}(2013){Astropy Collaboration},
  {Robitaille}, {Tollerud}, {Greenfield}, {Droettboom}, {Bray}, {Aldcroft},
  {Davis}, {Ginsburg}, {Price-Whelan}, {Kerzendorf}, {Conley}, {Crighton},
  {Barbary}, {Muna}, {Ferguson}, {Grollier}, {Parikh}, {Nair}, {Unther},
  {Deil}, {Woillez}, {Conseil}, {Kramer}, {Turner}, {Singer}, {Fox}, {Weaver},
  {Zabalza}, {Edwards}, {Azalee Bostroem}, {Burke}, {Casey}, {Crawford},
  {Dencheva}, {Ely}, {Jenness}, {Labrie}, {Lim}, {Pierfederici}, {Pontzen},
  {Ptak}, {Refsdal}, {Servillat}, \& {Streicher}}]{Astropy2013}
{Astropy Collaboration}, {Robitaille}, T.~P., {Tollerud}, E.~J., {et~al.} 2013,
  \aap, 558, A33, \dodoi{10.1051/0004-6361/201322068}

\bibitem[{{Astropy Collaboration} {et~al.}(2018){Astropy Collaboration},
  {Price-Whelan}, {Sip{\H{o}}cz}, {G{\"u}nther}, {Lim}, {Crawford}, {Conseil},
  {Shupe}, {Craig}, {Dencheva}, {Ginsburg}, {VanderPlas}, {Bradley},
  {P{\'e}rez-Su{\'a}rez}, {de Val-Borro}, {Aldcroft}, {Cruz}, {Robitaille},
  {Tollerud}, {Ardelean}, {Babej}, {Bach}, {Bachetti}, {Bakanov}, {Bamford},
  {Barentsen}, {Barmby}, {Baumbach}, {Berry}, {Biscani}, {Boquien}, {Bostroem},
  {Bouma}, {Brammer}, {Bray}, {Breytenbach}, {Buddelmeijer}, {Burke},
  {Calderone}, {Cano Rodr{\'\i}guez}, {Cara}, {Cardoso}, {Cheedella}, {Copin},
  {Corrales}, {Crichton}, {D'Avella}, {Deil}, {Depagne}, {Dietrich}, {Donath},
  {Droettboom}, {Earl}, {Erben}, {Fabbro}, {Ferreira}, {Finethy}, {Fox},
  {Garrison}, {Gibbons}, {Goldstein}, {Gommers}, {Greco}, {Greenfield},
  {Groener}, {Grollier}, {Hagen}, {Hirst}, {Homeier}, {Horton}, {Hosseinzadeh},
  {Hu}, {Hunkeler}, {Ivezi{\'c}}, {Jain}, {Jenness}, {Kanarek}, {Kendrew},
  {Kern}, {Kerzendorf}, {Khvalko}, {King}, {Kirkby}, {Kulkarni}, {Kumar},
  {Lee}, {Lenz}, {Littlefair}, {Ma}, {Macleod}, {Mastropietro}, {McCully},
  {Montagnac}, {Morris}, {Mueller}, {Mumford}, {Muna}, {Murphy}, {Nelson},
  {Nguyen}, {Ninan}, {N{\"o}the}, {Ogaz}, {Oh}, {Parejko}, {Parley}, {Pascual},
  {Patil}, {Patil}, {Plunkett}, {Prochaska}, {Rastogi}, {Reddy Janga},
  {Sabater}, {Sakurikar}, {Seifert}, {Sherbert}, {Sherwood-Taylor}, {Shih},
  {Sick}, {Silbiger}, {Singanamalla}, {Singer}, {Sladen}, {Sooley},
  {Sornarajah}, {Streicher}, {Teuben}, {Thomas}, {Tremblay}, {Turner},
  {Terr{\'o}n}, {van Kerkwijk}, {de la Vega}, {Watkins}, {Weaver}, {Whitmore},
  {Woillez}, {Zabalza}, \& {Astropy Contributors}}]{Astropy2018}
{Astropy Collaboration}, {Price-Whelan}, A.~M., {Sip{\H{o}}cz}, B.~M., {et~al.}
  2018, \aj, 156, 123, \dodoi{10.3847/1538-3881/aabc4f}

\bibitem[{{Balogh} {et~al.}(2014){Balogh}, {McGee}, {Mok}, {Wilman},
  {Finoguenov}, {Bower}, {Mulchaey}, {Parker}, \& {Tanaka}}]{Balogh2014}
{Balogh}, M.~L., {McGee}, S.~L., {Mok}, A., {et~al.} 2014, \mnras, 443, 2679,
  \dodoi{10.1093/mnras/stu1332}

\bibitem[{{Bertin} \& {Arnouts}(1996)}]{Bertin1996}
{Bertin}, E., \& {Arnouts}, S. 1996, \aaps, 117, 393,
  \dodoi{10.1051/aas:1996164}

\bibitem[{{Bruzual} \& {Charlot}(2003)}]{BC03}
{Bruzual}, G., \& {Charlot}, S. 2003, \mnras, 344, 1000,
  \dodoi{10.1046/j.1365-8711.2003.06897.x}

\bibitem[{{Calabr{\`o}} {et~al.}(2018){Calabr{\`o}}, {Daddi}, {Cassata},
  {Onodera}, {Gobat}, {Puglisi}, {Jin}, {Liu}, {Amor{\'\i}n}, {Arimoto},
  {Boquien}, {Carraro}, {Elbaz}, {Ibar}, {Juneau}, {Mannucci}, {M{\'e}ndez
  Hern{\'a}nez}, {Oliva}, {Rodighiero}, {Valentino}, \&
  {Zanella}}]{Calabro2018}
{Calabr{\`o}}, A., {Daddi}, E., {Cassata}, P., {et~al.} 2018, \apjl, 862, L22,
  \dodoi{10.3847/2041-8213/aad33e}

\bibitem[{{Calzetti} {et~al.}(2000){Calzetti}, {Armus}, {Bohlin}, {Kinney},
  {Koornneef}, \& {Storchi-Bergmann}}]{Calzetti2000}
{Calzetti}, D., {Armus}, L., {Bohlin}, R.~C., {et~al.} 2000, \apj, 533, 682,
  \dodoi{10.1086/308692}

\bibitem[{{Capak} {et~al.}(2007){Capak}, {Aussel}, {Ajiki}, {McCracken},
  {Mobasher}, {Scoville}, {Shopbell}, {Taniguchi}, {Thompson}, {Tribiano},
  {Sasaki}, {Blain}, {Brusa}, {Carilli}, {Comastri}, {Carollo}, {Cassata},
  {Colbert}, {Ellis}, {Elvis}, {Giavalisco}, {Green}, {Guzzo}, {Hasinger},
  {Ilbert}, {Impey}, {Jahnke}, {Kartaltepe}, {Kneib}, {Koda}, {Koekemoer},
  {Komiyama}, {Leauthaud}, {Le Fevre}, {Lilly}, {Liu}, {Massey}, {Miyazaki},
  {Murayama}, {Nagao}, {Peacock}, {Pickles}, {Porciani}, {Renzini}, {Rhodes},
  {Rich}, {Salvato}, {Sanders}, {Scarlata}, {Schiminovich}, {Schinnerer},
  {Scodeggio}, {Sheth}, {Shioya}, {Tasca}, {Taylor}, {Yan}, \&
  {Zamorani}}]{Capak2007}
{Capak}, P., {Aussel}, H., {Ajiki}, M., {et~al.} 2007, \apjs, 172, 99,
  \dodoi{10.1086/519081}

\bibitem[{{Casey} {et~al.}(2014){Casey}, {Narayanan}, \& {Cooray}}]{Casey2014}
{Casey}, C.~M., {Narayanan}, D., \& {Cooray}, A. 2014, \physrep, 541, 45,
  \dodoi{10.1016/j.physrep.2014.02.009}

\bibitem[{{Casey} {et~al.}(2012{\natexlab{a}}){Casey}, {Berta},
  {B{\'e}thermin}, {Bock}, {Bridge}, {Burgarella}, {Chapin}, {Chapman},
  {Clements}, {Conley}, {Conselice}, {Cooray}, {Farrah}, {Hatziminaoglou},
  {Ivison}, {le Floc'h}, {Lutz}, {Magdis}, {Magnelli}, {Oliver}, {Page},
  {Pozzi}, {Rigopoulou}, {Riguccini}, {Roseboom}, {Sanders}, {Scott},
  {Seymour}, {Valtchanov}, {Vieira}, {Viero}, \& {Wardlow}}]{Casey2012a}
{Casey}, C.~M., {Berta}, S., {B{\'e}thermin}, M., {et~al.} 2012{\natexlab{a}},
  \apj, 761, 139, \dodoi{10.1088/0004-637X/761/2/139}

\bibitem[{{Casey} {et~al.}(2012{\natexlab{b}}){Casey}, {Berta},
  {B{\'e}thermin}, {Bock}, {Bridge}, {Budynkiewicz}, {Burgarella}, {Chapin},
  {Chapman}, {Clements}, {Conley}, {Conselice}, {Cooray}, {Farrah},
  {Hatziminaoglou}, {Ivison}, {le Floc'h}, {Lutz}, {Magdis}, {Magnelli},
  {Oliver}, {Page}, {Pozzi}, {Rigopoulou}, {Riguccini}, {Roseboom}, {Sanders},
  {Scott}, {Seymour}, {Valtchanov}, {Vieira}, {Viero}, \&
  {Wardlow}}]{Casey2012b}
---. 2012{\natexlab{b}}, \apj, 761, 140, \dodoi{10.1088/0004-637X/761/2/140}

\bibitem[{{Casey} {et~al.}(2015){Casey}, {Cooray}, {Capak}, {Fu}, {Kovac},
  {Lilly}, {Sanders}, {Scoville}, \& {Treister}}]{Casey2015}
{Casey}, C.~M., {Cooray}, A., {Capak}, P., {et~al.} 2015, \apjl, 808, L33,
  \dodoi{10.1088/2041-8205/808/2/L33}

\bibitem[{{Casey} {et~al.}(2017){Casey}, {Cooray}, {Killi}, {Capak}, {Chen},
  {Hung}, {Kartaltepe}, {Sanders}, \& {Scoville}}]{Casey2017}
{Casey}, C.~M., {Cooray}, A., {Killi}, M., {et~al.} 2017, \apj, 840, 101,
  \dodoi{10.3847/1538-4357/aa6cb1}

\bibitem[{{Chabrier}(2003)}]{Chabrier2003}
{Chabrier}, G. 2003, \pasp, 115, 763, \dodoi{10.1086/376392}

\bibitem[{{Chapman} {et~al.}(2005){Chapman}, {Blain}, {Smail}, \&
  {Ivison}}]{Chapman2005}
{Chapman}, S.~C., {Blain}, A.~W., {Smail}, I., \& {Ivison}, R.~J. 2005, \apj,
  622, 772, \dodoi{10.1086/428082}

\bibitem[{{Chapman} {et~al.}(2003){Chapman}, {Windhorst}, {Odewahn}, {Yan}, \&
  {Conselice}}]{Chapman2003}
{Chapman}, S.~C., {Windhorst}, R., {Odewahn}, S., {Yan}, H., \& {Conselice}, C.
  2003, \apj, 599, 92, \dodoi{10.1086/379120}

\bibitem[{{Chen} {et~al.}(2015){Chen}, {Smail}, {Swinbank}, {Simpson}, {Ma},
  {Alexander}, {Biggs}, {Brandt}, {Chapman}, {Coppin}, {Danielson},
  {Dannerbauer}, {Edge}, {Greve}, {Ivison}, {Karim}, {Menten}, {Schinnerer},
  {Walter}, {Wardlow}, {Wei{\ss}}, \& {van der Werf}}]{Chen2015}
{Chen}, C.-C., {Smail}, I., {Swinbank}, A.~M., {et~al.} 2015, \apj, 799, 194,
  \dodoi{10.1088/0004-637X/799/2/194}

\bibitem[{{Coil} {et~al.}(2011){Coil}, {Blanton}, {Burles}, {Cool},
  {Eisenstein}, {Moustakas}, {Wong}, {Zhu}, {Aird}, {Bernstein}, {Bolton}, \&
  {Hogg}}]{Coil2011}
{Coil}, A.~L., {Blanton}, M.~R., {Burles}, S.~M., {et~al.} 2011, \apj, 741, 8,
  \dodoi{10.1088/0004-637X/741/1/8}

\bibitem[{{Comparat} {et~al.}(2015){Comparat}, {Richard}, {Kneib}, {Ilbert},
  {Gonzalez-Perez}, {Tresse}, {Zoubian}, {Arnouts}, {Brownstein}, {Baugh},
  {Delubac}, {Ealet}, {Escoffier}, {Ge}, {Jullo}, {Lacey}, {Ross}, {Schlegel},
  {Schneider}, {Steele}, {Tasca}, {Yeche}, {Lesser}, {Jiang}, {Jing}, {Fan},
  {Fan}, {Ma}, {Nie}, {Wang}, {Wu}, {Zhang}, {Zhou}, {Zhou}, \&
  {Zou}}]{Comparat2015}
{Comparat}, J., {Richard}, J., {Kneib}, J.-P., {et~al.} 2015, \aap, 575, A40,
  \dodoi{10.1051/0004-6361/201424767}

\bibitem[{{Condon}(1992)}]{Condon1992}
{Condon}, J.~J. 1992, \araa, 30, 575,
  \dodoi{10.1146/annurev.aa.30.090192.003043}

\bibitem[{{Conselice} {et~al.}(2003){Conselice}, {Chapman}, \&
  {Windhorst}}]{Conselice2003}
{Conselice}, C.~J., {Chapman}, S.~C., \& {Windhorst}, R.~A. 2003, \apjl, 596,
  L5, \dodoi{10.1086/379109}

\bibitem[{{Damjanov} {et~al.}(2018){Damjanov}, {Zahid}, {Geller}, {Fabricant},
  \& {Hwang}}]{Damjanov2018}
{Damjanov}, I., {Zahid}, H.~J., {Geller}, M.~J., {Fabricant}, D.~G., \&
  {Hwang}, H.~S. 2018, \apjs, 234, 21, \dodoi{10.3847/1538-4365/aaa01c}

\bibitem[{{Dav{\'e}} {et~al.}(2010){Dav{\'e}}, {Finlator}, {Oppenheimer},
  {Fardal}, {Katz}, {Kere{\v{s}}}, \& {Weinberg}}]{Dave2010}
{Dav{\'e}}, R., {Finlator}, K., {Oppenheimer}, B.~D., {et~al.} 2010, \mnras,
  404, 1355, \dodoi{10.1111/j.1365-2966.2010.16395.x}

\bibitem[{{Davidzon} {et~al.}(2017){Davidzon}, {Ilbert}, {Laigle}, {Coupon},
  {McCracken}, {Delvecchio}, {Masters}, {Capak}, {Hsieh}, {Le F{\`e}vre},
  {Tresse}, {Bethermin}, {Chang}, {Faisst}, {Le Floc'h}, {Steinhardt}, {Toft},
  {Aussel}, {Dubois}, {Hasinger}, {Salvato}, {Sanders}, {Scoville}, \&
  {Silverman}}]{Davidzon2017}
{Davidzon}, I., {Ilbert}, O., {Laigle}, C., {et~al.} 2017, \aap, 605, A70,
  \dodoi{10.1051/0004-6361/201730419}

\bibitem[{{Dekel} {et~al.}(2009){Dekel}, {Sari}, \& {Ceverino}}]{Dekel2009}
{Dekel}, A., {Sari}, R., \& {Ceverino}, D. 2009, \apj, 703, 785,
  \dodoi{10.1088/0004-637X/703/1/785}

\bibitem[{{Elbaz} {et~al.}(2011){Elbaz}, {Dickinson}, {Hwang},
  {D{\'\i}az-Santos}, {Magdis}, {Magnelli}, {Le Borgne}, {Galliano},
  {Pannella}, {Chanial}, {Armus}, {Charmandaris}, {Daddi}, {Aussel}, {Popesso},
  {Kartaltepe}, {Altieri}, {Valtchanov}, {Coia}, {Dannerbauer}, {Dasyra},
  {Leiton}, {Mazzarella}, {Alexander}, {Buat}, {Burgarella}, {Chary}, {Gilli},
  {Ivison}, {Juneau}, {Le Floc'h}, {Lutz}, {Morrison}, {Mullaney}, {Murphy},
  {Pope}, {Scott}, {Brodwin}, {Calzetti}, {Cesarsky}, {Charlot}, {Dole},
  {Eisenhardt}, {Ferguson}, {F{\"o}rster Schreiber}, {Frayer}, {Giavalisco},
  {Huynh}, {Koekemoer}, {Papovich}, {Reddy}, {Surace}, {Teplitz}, {Yun}, \&
  {Wilson}}]{Elbaz2011}
{Elbaz}, D., {Dickinson}, M., {Hwang}, H.~S., {et~al.} 2011, \aap, 533, A119,
  \dodoi{10.1051/0004-6361/201117239}

\bibitem[{{Freeman} {et~al.}(2013){Freeman}, {Izbicki}, {Lee}, {Newman},
  {Conselice}, {Koekemoer}, {Lotz}, \& {Mozena}}]{Freeman2013}
{Freeman}, P.~E., {Izbicki}, R., {Lee}, A.~B., {et~al.} 2013, \mnras, 434, 282,
  \dodoi{10.1093/mnras/stt1016}

\bibitem[{{Geach} {et~al.}(2017){Geach}, {Dunlop}, {Halpern}, {Smail}, {van der
  Werf}, {Alexander}, {Almaini}, {Aretxaga}, {Arumugam}, {Asboth}, {Banerji},
  {Beanlands}, {Best}, {Blain}, {Birkinshaw}, {Chapin}, {Chapman}, {Chen},
  {Chrysostomou}, {Clarke}, {Clements}, {Conselice}, {Coppin}, {Cowley},
  {Danielson}, {Eales}, {Edge}, {Farrah}, {Gibb}, {Harrison}, {Hine}, {Hughes},
  {Ivison}, {Jarvis}, {Jenness}, {Jones}, {Karim}, {Koprowski}, {Knudsen},
  {Lacey}, {Mackenzie}, {Marsden}, {McAlpine}, {McMahon}, {Meijerink},
  {Micha{\l}owski}, {Oliver}, {Page}, {Peacock}, {Rigopoulou}, {Robson},
  {Roseboom}, {Rotermund}, {Scott}, {Serjeant}, {Simpson}, {Simpson}, {Smith},
  {Spaans}, {Stanley}, {Stevens}, {Swinbank}, {Targett}, {Thomson}, {Valiante},
  {Wake}, {Webb}, {Willott}, {Zavala}, \& {Zemcov}}]{Geach2017}
{Geach}, J.~E., {Dunlop}, J.~S., {Halpern}, M., {et~al.} 2017, \mnras, 465,
  1789, \dodoi{10.1093/mnras/stw2721}

\bibitem[{{Giavalisco} {et~al.}(2004){Giavalisco}, {Ferguson}, {Koekemoer},
  {Dickinson}, {Alexander}, {Bauer}, {Bergeron}, {Biagetti}, {Brandt},
  {Casertano}, {Cesarsky}, {Chatzichristou}, {Conselice}, {Cristiani}, {Da
  Costa}, {Dahlen}, {de Mello}, {Eisenhardt}, {Erben}, {Fall}, {Fassnacht},
  {Fosbury}, {Fruchter}, {Gardner}, {Grogin}, {Hook}, {Hornschemeier}, {Idzi},
  {Jogee}, {Kretchmer}, {Laidler}, {Lee}, {Livio}, {Lucas}, {Madau},
  {Mobasher}, {Moustakas}, {Nonino}, {Padovani}, {Papovich}, {Park},
  {Ravindranath}, {Renzini}, {Richardson}, {Riess}, {Rosati}, {Schirmer},
  {Schreier}, {Somerville}, {Spinrad}, {Stern}, {Stiavelli}, {Strolger},
  {Urry}, {Vandame}, {Williams}, \& {Wolf}}]{Giavalisco2004}
{Giavalisco}, M., {Ferguson}, H.~C., {Koekemoer}, A.~M., {et~al.} 2004, \apjl,
  600, L93, \dodoi{10.1086/379232}

\bibitem[{{G{\'o}mez-Guijarro} {et~al.}(2018){G{\'o}mez-Guijarro}, {Toft},
  {Karim}, {Magnelli}, {Magdis}, {Jim{\'e}nez-Andrade}, {Capak}, {Fraternali},
  {Fujimoto}, {Riechers}, {Schinnerer}, {Smol{\v{c}}i{\'c}}, {Aravena},
  {Bertoldi}, {Cortzen}, {Hasinger}, {Hu}, {Jones}, {Koekemoer}, {Lee},
  {McCracken}, {Micha{\l}owski}, {Navarrete}, {Povi{\'c}}, {Puglisi},
  {Romano-D{\'\i}az}, {Sheth}, {Silverman}, {Staguhn}, {Steinhardt},
  {Stockmann}, {Tanaka}, {Valentino}, {van Kampen}, \&
  {Zirm}}]{Gomez-Guijarro2018}
{G{\'o}mez-Guijarro}, C., {Toft}, S., {Karim}, A., {et~al.} 2018, \apj, 856,
  121, \dodoi{10.3847/1538-4357/aab206}

\bibitem[{{Griffin} {et~al.}(2010){Griffin}, {Abergel}, {Abreu}, {Ade},
  {Andr{\'e}}, {Augueres}, {Babbedge}, {Bae}, {Baillie}, {Baluteau}, {Barlow},
  {Bendo}, {Benielli}, {Bock}, {Bonhomme}, {Brisbin}, {Brockley-Blatt},
  {Caldwell}, {Cara}, {Castro-Rodriguez}, {Cerulli}, {Chanial}, {Chen},
  {Clark}, {Clements}, {Clerc}, {Coker}, {Communal}, {Conversi}, {Cox},
  {Crumb}, {Cunningham}, {Daly}, {Davis}, {de Antoni}, {Delderfield}, {Devin},
  {di Giorgio}, {Didschuns}, {Dohlen}, {Donati}, {Dowell}, {Dowell}, {Duband},
  {Dumaye}, {Emery}, {Ferlet}, {Ferrand}, {Fontignie}, {Fox}, {Franceschini},
  {Frerking}, {Fulton}, {Garcia}, {Gastaud}, {Gear}, {Glenn}, {Goizel},
  {Griffin}, {Grundy}, {Guest}, {Guillemet}, {Hargrave}, {Harwit}, {Hastings},
  {Hatziminaoglou}, {Herman}, {Hinde}, {Hristov}, {Huang}, {Imhof}, {Isaak},
  {Israelsson}, {Ivison}, {Jennings}, {Kiernan}, {King}, {Lange}, {Latter},
  {Laurent}, {Laurent}, {Leeks}, {Lellouch}, {Levenson}, {Li}, {Li},
  {Lilienthal}, {Lim}, {Liu}, {Lu}, {Madden}, {Mainetti}, {Marliani}, {McKay},
  {Mercier}, {Molinari}, {Morris}, {Moseley}, {Mulder}, {Mur}, {Naylor},
  {Nguyen}, {O'Halloran}, {Oliver}, {Olofsson}, {Olofsson}, {Orfei}, {Page},
  {Pain}, {Panuzzo}, {Papageorgiou}, {Parks}, {Parr-Burman}, {Pearce},
  {Pearson}, {P{\'e}rez-Fournon}, {Pinsard}, {Pisano}, {Podosek}, {Pohlen},
  {Polehampton}, {Pouliquen}, {Rigopoulou}, {Rizzo}, {Roseboom}, {Roussel},
  {Rowan-Robinson}, {Rownd}, {Saraceno}, {Sauvage}, {Savage}, {Savini},
  {Sawyer}, {Scharmberg}, {Schmitt}, {Schneider}, {Schulz}, {Schwartz},
  {Shafer}, {Shupe}, {Sibthorpe}, {Sidher}, {Smith}, {Smith}, {Smith},
  {Spencer}, {Stobie}, {Sudiwala}, {Sukhatme}, {Surace}, {Stevens}, {Swinyard},
  {Trichas}, {Tourette}, {Triou}, {Tseng}, {Tucker}, {Turner}, {Vaccari},
  {Valtchanov}, {Vigroux}, {Virique}, {Voellmer}, {Walker}, {Ward}, {Waskett},
  {Weilert}, {Wesson}, {White}, {Whitehouse}, {Wilson}, {Winter}, {Woodcraft},
  {Wright}, {Xu}, {Zavagno}, {Zemcov}, {Zhang}, \& {Zonca}}]{Griffin2010}
{Griffin}, M.~J., {Abergel}, A., {Abreu}, A., {et~al.} 2010, \aap, 518, L3,
  \dodoi{10.1051/0004-6361/201014519}

\bibitem[{{Grogin} {et~al.}(2011){Grogin}, {Kocevski}, {Faber}, {Ferguson},
  {Koekemoer}, {Riess}, {Acquaviva}, {Alexander}, {Almaini}, {Ashby}, {Barden},
  {Bell}, {Bournaud}, {Brown}, {Caputi}, {Casertano}, {Cassata}, {Castellano},
  {Challis}, {Chary}, {Cheung}, {Cirasuolo}, {Conselice}, {Roshan Cooray},
  {Croton}, {Daddi}, {Dahlen}, {Dav{\'e}}, {de Mello}, {Dekel}, {Dickinson},
  {Dolch}, {Donley}, {Dunlop}, {Dutton}, {Elbaz}, {Fazio}, {Filippenko},
  {Finkelstein}, {Fontana}, {Gardner}, {Garnavich}, {Gawiser}, {Giavalisco},
  {Grazian}, {Guo}, {Hathi}, {H{\"a}ussler}, {Hopkins}, {Huang}, {Huang},
  {Jha}, {Kartaltepe}, {Kirshner}, {Koo}, {Lai}, {Lee}, {Li}, {Lotz}, {Lucas},
  {Madau}, {McCarthy}, {McGrath}, {McIntosh}, {McLure}, {Mobasher},
  {Moustakas}, {Mozena}, {Nandra}, {Newman}, {Niemi}, {Noeske}, {Papovich},
  {Pentericci}, {Pope}, {Primack}, {Rajan}, {Ravindranath}, {Reddy}, {Renzini},
  {Rix}, {Robaina}, {Rodney}, {Rosario}, {Rosati}, {Salimbeni}, {Scarlata},
  {Siana}, {Simard}, {Smidt}, {Somerville}, {Spinrad}, {Straughn}, {Strolger},
  {Telford}, {Teplitz}, {Trump}, {van der Wel}, {Villforth}, {Wechsler},
  {Weiner}, {Wiklind}, {Wild}, {Wilson}, {Wuyts}, {Yan}, \& {Yun}}]{Grogin2011}
{Grogin}, N.~A., {Kocevski}, D.~D., {Faber}, S.~M., {et~al.} 2011, \apjs, 197,
  35, \dodoi{10.1088/0067-0049/197/2/35}

\bibitem[{{Hasinger} {et~al.}(2018){Hasinger}, {Capak}, {Salvato}, {Barger},
  {Cowie}, {Faisst}, {Hemmati}, {Kakazu}, {Kartaltepe}, {Masters}, {Mobasher},
  {Nayyeri}, {Sanders}, {Scoville}, {Suh}, {Steinhardt}, \&
  {Yang}}]{Hasinger2018}
{Hasinger}, G., {Capak}, P., {Salvato}, M., {et~al.} 2018, \apj, 858, 77,
  \dodoi{10.3847/1538-4357/aabacf}

\bibitem[{{Helou} {et~al.}(1985){Helou}, {Soifer}, \&
  {Rowan-Robinson}}]{Helou1985}
{Helou}, G., {Soifer}, B.~T., \& {Rowan-Robinson}, M. 1985, \apjl, 298, L7,
  \dodoi{10.1086/184556}

\bibitem[{{Hodge} {et~al.}(2013){Hodge}, {Karim}, {Smail}, {Swinbank},
  {Walter}, {Biggs}, {Ivison}, {Weiss}, {Alexander}, {Bertoldi}, {Brandt},
  {Chapman}, {Coppin}, {Cox}, {Danielson}, {Dannerbauer}, {De Breuck},
  {Decarli}, {Edge}, {Greve}, {Knudsen}, {Menten}, {Rix}, {Schinnerer},
  {Simpson}, {Wardlow}, \& {van der Werf}}]{Hodge2013}
{Hodge}, J.~A., {Karim}, A., {Smail}, I., {et~al.} 2013, \apj, 768, 91,
  \dodoi{10.1088/0004-637X/768/1/91}

\bibitem[{{Holland} {et~al.}(2013){Holland}, {Bintley}, {Chapin},
  {Chrysostomou}, {Davis}, {Dempsey}, {Duncan}, {Fich}, {Friberg}, {Halpern},
  {Irwin}, {Jenness}, {Kelly}, {MacIntosh}, {Robson}, {Scott}, {Ade},
  {Atad-Ettedgui}, {Berry}, {Craig}, {Gao}, {Gibb}, {Hilton}, {Hollister},
  {Kycia}, {Lunney}, {McGregor}, {Montgomery}, {Parkes}, {Tilanus}, {Ullom},
  {Walther}, {Walton}, {Woodcraft}, {Amiri}, {Atkinson}, {Burger}, {Chuter},
  {Coulson}, {Doriese}, {Dunare}, {Economou}, {Niemack}, {Parsons},
  {Reintsema}, {Sibthorpe}, {Smail}, {Sudiwala}, \& {Thomas}}]{Holland2013}
{Holland}, W.~S., {Bintley}, D., {Chapin}, E.~L., {et~al.} 2013, \mnras, 430,
  2513, \dodoi{10.1093/mnras/sts612}

\bibitem[{{Houck} {et~al.}(1985){Houck}, {Schneider}, {Danielson}, {Beichman},
  {Lonsdale}, {Neugebauer}, \& {Soifer}}]{Houck1985}
{Houck}, J.~R., {Schneider}, D.~P., {Danielson}, G.~E., {et~al.} 1985, \apjl,
  290, L5, \dodoi{10.1086/184431}

\bibitem[{{Houck} {et~al.}(1984){Houck}, {Soifer}, {Neugebauer}, {Beichman},
  {Aumann}, {Clegg}, {Gillett}, {Habing}, {Hauser}, {Low}, {Miley},
  {Rowan-Robinson}, \& {Walker}}]{Houck1984}
{Houck}, J.~R., {Soifer}, B.~T., {Neugebauer}, G., {et~al.} 1984, \apjl, 278,
  L63, \dodoi{10.1086/184224}

\bibitem[{{Hung} {et~al.}(2014){Hung}, {Sanders}, {Casey}, {Koss}, {Larson},
  {Lee}, {Li}, {Lockhart}, {Shih}, {Barnes}, {Kartaltepe}, \&
  {Smith}}]{Hung2014}
{Hung}, C.-L., {Sanders}, D.~B., {Casey}, C.~M., {et~al.} 2014, \apj, 791, 63,
  \dodoi{10.1088/0004-637X/791/1/63}

\bibitem[{{Ilbert} {et~al.}(2006){Ilbert}, {Arnouts}, {McCracken},
  {Bolzonella}, {Bertin}, {Le F{\`e}vre}, {Mellier}, {Zamorani}, {Pell{\`o}},
  {Iovino}, {Tresse}, {Le Brun}, {Bottini}, {Garilli}, {Maccagni}, {Picat},
  {Scaramella}, {Scodeggio}, {Vettolani}, {Zanichelli}, {Adami}, {Bardelli},
  {Cappi}, {Charlot}, {Ciliegi}, {Contini}, {Cucciati}, {Foucaud}, {Franzetti},
  {Gavignaud}, {Guzzo}, {Marano}, {Marinoni}, {Mazure}, {Meneux}, {Merighi},
  {Paltani}, {Pollo}, {Pozzetti}, {Radovich}, {Zucca}, {Bondi}, {Bongiorno},
  {Busarello}, {de La Torre}, {Gregorini}, {Lamareille}, {Mathez}, {Merluzzi},
  {Ripepi}, {Rizzo}, \& {Vergani}}]{Ilbert2006}
{Ilbert}, O., {Arnouts}, S., {McCracken}, H.~J., {et~al.} 2006, \aap, 457, 841,
  \dodoi{10.1051/0004-6361:20065138}

\bibitem[{{Ilbert} {et~al.}(2009){Ilbert}, {Capak}, {Salvato}, {Aussel},
  {McCracken}, {Sanders}, {Scoville}, {Kartaltepe}, {Arnouts}, {Le Floc'h},
  {Mobasher}, {Taniguchi}, {Lamareille}, {Leauthaud}, {Sasaki}, {Thompson},
  {Zamojski}, {Zamorani}, {Bardelli}, {Bolzonella}, {Bongiorno}, {Brusa},
  {Caputi}, {Carollo}, {Contini}, {Cook}, {Coppa}, {Cucciati}, {de la Torre},
  {de Ravel}, {Franzetti}, {Garilli}, {Hasinger}, {Iovino}, {Kampczyk},
  {Kneib}, {Knobel}, {Kovac}, {Le Borgne}, {Le Brun}, {Le F{\`e}vre}, {Lilly},
  {Looper}, {Maier}, {Mainieri}, {Mellier}, {Mignoli}, {Murayama}, {Pell{\`o}},
  {Peng}, {P{\'e}rez-Montero}, {Renzini}, {Ricciardelli}, {Schiminovich},
  {Scodeggio}, {Shioya}, {Silverman}, {Surace}, {Tanaka}, {Tasca}, {Tresse},
  {Vergani}, \& {Zucca}}]{Ilbert2009}
{Ilbert}, O., {Capak}, P., {Salvato}, M., {et~al.} 2009, \apj, 690, 1236,
  \dodoi{10.1088/0004-637X/690/2/1236}

\bibitem[{{Iono} {et~al.}(2016){Iono}, {Yun}, {Aretxaga}, {Hatsukade},
  {Hughes}, {Ikarashi}, {Izumi}, {Kawabe}, {Kohno}, {Lee}, {Matsuda},
  {Nakanishi}, {Saito}, {Tamura}, {Ueda}, {Umehata}, {Wilson}, {Michiyama}, \&
  {Ando}}]{Iono2016}
{Iono}, D., {Yun}, M.~S., {Aretxaga}, I., {et~al.} 2016, \apjl, 829, L10,
  \dodoi{10.3847/2041-8205/829/1/L10}

\bibitem[{{Ivison} {et~al.}(2007){Ivison}, {Greve}, {Dunlop}, {Peacock},
  {Egami}, {Smail}, {Ibar}, {van Kampen}, {Aretxaga}, {Babbedge}, {Biggs},
  {Blain}, {Chapman}, {Clements}, {Coppin}, {Farrah}, {Halpern}, {Hughes},
  {Jarvis}, {Jenness}, {Jones}, {Mortier}, {Oliver}, {Papovich},
  {P{\'e}rez-Gonz{\'a}lez}, {Pope}, {Rawlings}, {Rieke}, {Rowan-Robinson},
  {Savage}, {Scott}, {Seigar}, {Serjeant}, {Simpson}, {Stevens}, {Vaccari},
  {Wagg}, \& {Willott}}]{Ivison2007b}
{Ivison}, R.~J., {Greve}, T.~R., {Dunlop}, J.~S., {et~al.} 2007, \mnras, 380,
  199, \dodoi{10.1111/j.1365-2966.2007.12044.x}

\bibitem[{{Kaasinen} {et~al.}(2019){Kaasinen}, {Scoville}, {Walter}, {Da
  Cunha}, {Popping}, {Pavesi}, {Darvish}, {Casey}, {Riechers}, \&
  {Glover}}]{Kaasinen2019}
{Kaasinen}, M., {Scoville}, N., {Walter}, F., {et~al.} 2019, \apj, 880, 15,
  \dodoi{10.3847/1538-4357/ab253b}

\bibitem[{{Karim} {et~al.}(2011){Karim}, {Schinnerer},
  {Mart{\'\i}nez-Sansigre}, {Sargent}, {van der Wel}, {Rix}, {Ilbert},
  {Smol{\v{c}}i{\'c}}, {Carilli}, {Pannella}, {Koekemoer}, {Bell}, \&
  {Salvato}}]{Karim2011}
{Karim}, A., {Schinnerer}, E., {Mart{\'\i}nez-Sansigre}, A., {et~al.} 2011,
  \apj, 730, 61, \dodoi{10.1088/0004-637X/730/2/61}

\bibitem[{{Karim} {et~al.}(2013){Karim}, {Swinbank}, {Hodge}, {Smail},
  {Walter}, {Biggs}, {Simpson}, {Danielson}, {Alexander}, {Bertoldi}, {de
  Breuck}, {Chapman}, {Coppin}, {Dannerbauer}, {Edge}, {Greve}, {Ivison},
  {Knudsen}, {Menten}, {Schinnerer}, {Wardlow}, {Wei{\ss}}, \& {van der
  Werf}}]{Karim2013}
{Karim}, A., {Swinbank}, A.~M., {Hodge}, J.~A., {et~al.} 2013, \mnras, 432, 2,
  \dodoi{10.1093/mnras/stt196}

\bibitem[{{Kartaltepe} {et~al.}(2012){Kartaltepe}, {Dickinson}, {Alexander},
  {Bell}, {Dahlen}, {Elbaz}, {Faber}, {Lotz}, {McIntosh}, {Wiklind}, {Altieri},
  {Aussel}, {Bethermin}, {Bournaud}, {Charmandaris}, {Conselice}, {Cooray},
  {Dannerbauer}, {Dav{\'e}}, {Dunlop}, {Dekel}, {Ferguson}, {Grogin}, {Hwang},
  {Ivison}, {Kocevski}, {Koekemoer}, {Koo}, {Lai}, {Leiton}, {Lucas}, {Lutz},
  {Magdis}, {Magnelli}, {Morrison}, {Mozena}, {Mullaney}, {Newman}, {Pope},
  {Popesso}, {van der Wel}, {Weiner}, \& {Wuyts}}]{Kartaltepe2012}
{Kartaltepe}, J.~S., {Dickinson}, M., {Alexander}, D.~M., {et~al.} 2012, \apj,
  757, 23, \dodoi{10.1088/0004-637X/757/1/23}

\bibitem[{{Kartaltepe} {et~al.}(2015){Kartaltepe}, {Sanders}, {Silverman},
  {Kashino}, {Chu}, {Zahid}, {Hasinger}, {Kewley}, {Matsuoka}, {Nagao},
  {Riguccini}, {Salvato}, {Schawinski}, {Taniguchi}, {Treister}, {Capak},
  {Daddi}, \& {Ohta}}]{Kartaltepe2015}
{Kartaltepe}, J.~S., {Sanders}, D.~B., {Silverman}, J.~D., {et~al.} 2015,
  \apjl, 806, L35, \dodoi{10.1088/2041-8205/806/2/L35}

\bibitem[{{Kashino} {et~al.}(2019){Kashino}, {Silverman}, {Sanders},
  {Kartaltepe}, {Daddi}, {Renzini}, {Rodighiero}, {Puglisi}, {Valentino},
  {Juneau}, {Arimoto}, {Nagao}, {Ilbert}, {Le F{\`e}vre}, \&
  {Koekemoer}}]{Kashino2019}
{Kashino}, D., {Silverman}, J.~D., {Sanders}, D., {et~al.} 2019, \apjs, 241,
  10, \dodoi{10.3847/1538-4365/ab06c4}

\bibitem[{{Koekemoer} {et~al.}(2011){Koekemoer}, {Faber}, {Ferguson}, {Grogin},
  {Kocevski}, {Koo}, {Lai}, {Lotz}, {Lucas}, {McGrath}, {Ogaz}, {Rajan},
  {Riess}, {Rodney}, {Strolger}, {Casertano}, {Castellano}, {Dahlen},
  {Dickinson}, {Dolch}, {Fontana}, {Giavalisco}, {Grazian}, {Guo}, {Hathi},
  {Huang}, {van der Wel}, {Yan}, {Acquaviva}, {Alexander}, {Almaini}, {Ashby},
  {Barden}, {Bell}, {Bournaud}, {Brown}, {Caputi}, {Cassata}, {Challis},
  {Chary}, {Cheung}, {Cirasuolo}, {Conselice}, {Roshan Cooray}, {Croton},
  {Daddi}, {Dav{\'e}}, {de Mello}, {de Ravel}, {Dekel}, {Donley}, {Dunlop},
  {Dutton}, {Elbaz}, {Fazio}, {Filippenko}, {Finkelstein}, {Frazer}, {Gardner},
  {Garnavich}, {Gawiser}, {Gruetzbauch}, {Hartley}, {H{\"a}ussler},
  {Herrington}, {Hopkins}, {Huang}, {Jha}, {Johnson}, {Kartaltepe},
  {Khostovan}, {Kirshner}, {Lani}, {Lee}, {Li}, {Madau}, {McCarthy},
  {McIntosh}, {McLure}, {McPartland}, {Mobasher}, {Moreira}, {Mortlock},
  {Moustakas}, {Mozena}, {Nandra}, {Newman}, {Nielsen}, {Niemi}, {Noeske},
  {Papovich}, {Pentericci}, {Pope}, {Primack}, {Ravindranath}, {Reddy},
  {Renzini}, {Rix}, {Robaina}, {Rosario}, {Rosati}, {Salimbeni}, {Scarlata},
  {Siana}, {Simard}, {Smidt}, {Snyder}, {Somerville}, {Spinrad}, {Straughn},
  {Telford}, {Teplitz}, {Trump}, {Vargas}, {Villforth}, {Wagner}, {Wandro},
  {Wechsler}, {Weiner}, {Wiklind}, {Wild}, {Wilson}, {Wuyts}, \&
  {Yun}}]{Koekemoer2011}
{Koekemoer}, A.~M., {Faber}, S.~M., {Ferguson}, H.~C., {et~al.} 2011, \apjs,
  197, 36, \dodoi{10.1088/0067-0049/197/2/36}

\bibitem[{{Kriek} {et~al.}(2015){Kriek}, {Shapley}, {Reddy}, {Siana}, {Coil},
  {Mobasher}, {Freeman}, {de Groot}, {Price}, {Sanders}, {Shivaei}, {Brammer},
  {Momcheva}, {Skelton}, {van Dokkum}, {Whitaker}, {Aird}, {Azadi}, {Kassis},
  {Bullock}, {Conroy}, {Dav{\'e}}, {Kere{\v{s}}}, \& {Krumholz}}]{Kriek2015}
{Kriek}, M., {Shapley}, A.~E., {Reddy}, N.~A., {et~al.} 2015, \apjs, 218, 15,
  \dodoi{10.1088/0067-0049/218/2/15}

\bibitem[{{Krogager} {et~al.}(2014){Krogager}, {Zirm}, {Toft}, {Man}, \&
  {Brammer}}]{Krogager2014}
{Krogager}, J.~K., {Zirm}, A.~W., {Toft}, S., {Man}, A., \& {Brammer}, G. 2014,
  \apj, 797, 17, \dodoi{10.1088/0004-637X/797/1/17}

\bibitem[{{Laigle} {et~al.}(2016){Laigle}, {McCracken}, {Ilbert}, {Hsieh},
  {Davidzon}, {Capak}, {Hasinger}, {Silverman}, {Pichon}, {Coupon}, {Aussel},
  {Le Borgne}, {Caputi}, {Cassata}, {Chang}, {Civano}, {Dunlop}, {Fynbo},
  {Kartaltepe}, {Koekemoer}, {Le F{\`e}vre}, {Le Floc'h}, {Leauthaud}, {Lilly},
  {Lin}, {Marchesi}, {Milvang-Jensen}, {Salvato}, {Sanders}, {Scoville},
  {Smolcic}, {Stockmann}, {Taniguchi}, {Tasca}, {Toft}, {Vaccari}, \&
  {Zabl}}]{Laigle2016}
{Laigle}, C., {McCracken}, H.~J., {Ilbert}, O., {et~al.} 2016, \apjs, 224, 24,
  \dodoi{10.3847/0067-0049/224/2/24}

\bibitem[{{Lang} {et~al.}(2019){Lang}, {Schinnerer}, {Smail},
  {Dudzevi{\v{c}}i{\={u}}t{\.{e}}}, {Swinbank}, {Liu}, {Leslie}, {Almaini},
  {An}, {Bertoldi}, {Blain}, {Chapman}, {Chen}, {Conselice}, {Cooke}, {Coppin},
  {Dunlop}, {Farrah}, {Fudamoto}, {Geach}, {Gullberg}, {Harrington}, {Hodge},
  {Ivison}, {Jim{\'e}nez-Andrade}, {Magnelli}, {Micha{\l}owski}, {Oesch},
  {Scott}, {Simpson}, {Smol{\v{c}}i{\'c}}, {Stach}, {Thomson}, {Toft},
  {Vardoulaki}, {Wardlow}, {Weiss}, \& {van der Werf}}]{Lang2019}
{Lang}, P., {Schinnerer}, E., {Smail}, I., {et~al.} 2019, \apj, 879, 54,
  \dodoi{10.3847/1538-4357/ab1f77}

\bibitem[{{Le Floc'h} {et~al.}(2005){Le Floc'h}, {Papovich}, {Dole}, {Bell},
  {Lagache}, {Rieke}, {Egami}, {P{\'e}rez-Gonz{\'a}lez}, {Alonso-Herrero},
  {Rieke}, {Blaylock}, {Engelbracht}, {Gordon}, {Hines}, {Misselt}, {Morrison},
  \& {Mould}}]{LeFloch2005}
{Le Floc'h}, E., {Papovich}, C., {Dole}, H., {et~al.} 2005, \apj, 632, 169,
  \dodoi{10.1086/432789}

\bibitem[{{Lee} {et~al.}(2016){Lee}, {Hennawi}, {White}, {Prochaska},
  {Font-Ribera}, {Schlegel}, {Rich}, {Suzuki}, {Stark}, {Le F{\`e}vre},
  {Nugent}, {Salvato}, \& {Zamorani}}]{Lee2016}
{Lee}, K.-G., {Hennawi}, J.~F., {White}, M., {et~al.} 2016, \apj, 817, 160,
  \dodoi{10.3847/0004-637X/817/2/160}

\bibitem[{{Lilly} {et~al.}(2007){Lilly}, {Le F{\`e}vre}, {Renzini}, {Zamorani},
  {Scodeggio}, {Contini}, {Carollo}, {Hasinger}, {Kneib}, {Iovino}, {Le Brun},
  {Maier}, {Mainieri}, {Mignoli}, {Silverman}, {Tasca}, {Bolzonella},
  {Bongiorno}, {Bottini}, {Capak}, {Caputi}, {Cimatti}, {Cucciati}, {Daddi},
  {Feldmann}, {Franzetti}, {Garilli}, {Guzzo}, {Ilbert}, {Kampczyk}, {Kovac},
  {Lamareille}, {Leauthaud}, {Le Borgne}, {McCracken}, {Marinoni}, {Pello},
  {Ricciardelli}, {Scarlata}, {Vergani}, {Sanders}, {Schinnerer}, {Scoville},
  {Taniguchi}, {Arnouts}, {Aussel}, {Bardelli}, {Brusa}, {Cappi}, {Ciliegi},
  {Finoguenov}, {Foucaud}, {Franceschini}, {Halliday}, {Impey}, {Knobel},
  {Koekemoer}, {Kurk}, {Maccagni}, {Maddox}, {Marano}, {Marconi}, {Meneux},
  {Mobasher}, {Moreau}, {Peacock}, {Porciani}, {Pozzetti}, {Scaramella},
  {Schiminovich}, {Shopbell}, {Smail}, {Thompson}, {Tresse}, {Vettolani},
  {Zanichelli}, \& {Zucca}}]{Lilly2007}
{Lilly}, S.~J., {Le F{\`e}vre}, O., {Renzini}, A., {et~al.} 2007, \apjs, 172,
  70, \dodoi{10.1086/516589}

\bibitem[{{Lilly} {et~al.}(2009){Lilly}, {Le Brun}, {Maier}, {Mainieri},
  {Mignoli}, {Scodeggio}, {Zamorani}, {Carollo}, {Contini}, {Kneib}, {Le
  F{\`e}vre}, {Renzini}, {Bardelli}, {Bolzonella}, {Bongiorno}, {Caputi},
  {Coppa}, {Cucciati}, {de la Torre}, {de Ravel}, {Franzetti}, {Garilli},
  {Iovino}, {Kampczyk}, {Kovac}, {Knobel}, {Lamareille}, {Le Borgne}, {Pello},
  {Peng}, {P{\'e}rez-Montero}, {Ricciardelli}, {Silverman}, {Tanaka}, {Tasca},
  {Tresse}, {Vergani}, {Zucca}, {Ilbert}, {Salvato}, {Oesch}, {Abbas},
  {Bottini}, {Capak}, {Cappi}, {Cassata}, {Cimatti}, {Elvis}, {Fumana},
  {Guzzo}, {Hasinger}, {Koekemoer}, {Leauthaud}, {Maccagni}, {Marinoni},
  {McCracken}, {Memeo}, {Meneux}, {Porciani}, {Pozzetti}, {Sanders},
  {Scaramella}, {Scarlata}, {Scoville}, {Shopbell}, \& {Taniguchi}}]{Lilly2009}
{Lilly}, S.~J., {Le Brun}, V., {Maier}, C., {et~al.} 2009, \apjs, 184, 218,
  \dodoi{10.1088/0067-0049/184/2/218}

\bibitem[{{Liu} {et~al.}(2019){Liu}, {Lang}, {Magnelli}, {Schinnerer},
  {Leslie}, {Fudamoto}, {Bondi}, {Groves}, {Jim{\'e}nez-Andrade}, {Harrington},
  {Karim}, {Oesch}, {Sargent}, {Vardoulaki}, {B{\v{a}}descu}, {Moser},
  {Bertoldi}, {Battisti}, {da Cunha}, {Zavala}, {Vaccari}, {Davidzon},
  {Riechers}, \& {Aravena}}]{Liu2019a}
{Liu}, D., {Lang}, P., {Magnelli}, B., {et~al.} 2019, \apjs, 244, 40,
  \dodoi{10.3847/1538-4365/ab42da}

\bibitem[{{Lonsdale} {et~al.}(2006){Lonsdale}, {Farrah}, \&
  {Smith}}]{Lonsdale2006}
{Lonsdale}, C.~J., {Farrah}, D., \& {Smith}, H.~E. 2006, in Astrophysics Update
  2, ed. J.~W. {Mason}, 285

\bibitem[{{Lotz} {et~al.}(2004){Lotz}, {Primack}, \& {Madau}}]{Lotz2004}
{Lotz}, J.~M., {Primack}, J., \& {Madau}, P. 2004, \aj, 128, 163,
  \dodoi{10.1086/421849}

\bibitem[{{Lotz} {et~al.}(2008){Lotz}, {Davis}, {Faber}, {Guhathakurta},
  {Gwyn}, {Huang}, {Koo}, {Le Floc'h}, {Lin}, {Newman}, {Noeske}, {Papovich},
  {Willmer}, {Coil}, {Conselice}, {Cooper}, {Hopkins}, {Metevier}, {Primack},
  {Rieke}, \& {Weiner}}]{Lotz2008}
{Lotz}, J.~M., {Davis}, M., {Faber}, S.~M., {et~al.} 2008, \apj, 672, 177,
  \dodoi{10.1086/523659}

\bibitem[{{Lutz} {et~al.}(2011){Lutz}, {Poglitsch}, {Altieri}, {Andreani},
  {Aussel}, {Berta}, {Bongiovanni}, {Brisbin}, {Cava}, {Cepa}, {Cimatti},
  {Daddi}, {Dominguez-Sanchez}, {Elbaz}, {F{\"o}rster Schreiber}, {Genzel},
  {Grazian}, {Gruppioni}, {Harwit}, {Le Floc'h}, {Magdis}, {Magnelli},
  {Maiolino}, {Nordon}, {P{\'e}rez Garc{\'\i}a}, {Popesso}, {Pozzi},
  {Riguccini}, {Rodighiero}, {Saintonge}, {Sanchez Portal}, {Santini}, {Shao},
  {Sturm}, {Tacconi}, {Valtchanov}, {Wetzstein}, \& {Wieprecht}}]{Lutz2011}
{Lutz}, D., {Poglitsch}, A., {Altieri}, B., {et~al.} 2011, \aap, 532, A90,
  \dodoi{10.1051/0004-6361/201117107}

\bibitem[{{Ma} \& {Yan}(2015)}]{Ma2015}
{Ma}, Z., \& {Yan}, H. 2015, \apj, 811, 58, \dodoi{10.1088/0004-637X/811/1/58}

\bibitem[{{Magnelli} {et~al.}(2009){Magnelli}, {Elbaz}, {Chary}, {Dickinson},
  {Le Borgne}, {Frayer}, \& {Willmer}}]{Magnelli2009}
{Magnelli}, B., {Elbaz}, D., {Chary}, R.~R., {et~al.} 2009, \aap, 496, 57,
  \dodoi{10.1051/0004-6361:200811443}

\bibitem[{{Magnelli} {et~al.}(2013){Magnelli}, {Popesso}, {Berta}, {Pozzi},
  {Elbaz}, {Lutz}, {Dickinson}, {Altieri}, {Andreani}, {Aussel},
  {B{\'e}thermin}, {Bongiovanni}, {Cepa}, {Charmandaris}, {Chary}, {Cimatti},
  {Daddi}, {F{\"o}rster Schreiber}, {Genzel}, {Gruppioni}, {Harwit}, {Hwang},
  {Ivison}, {Magdis}, {Maiolino}, {Murphy}, {Nordon}, {Pannella}, {P{\'e}rez
  Garc{\'\i}a}, {Poglitsch}, {Rosario}, {Sanchez-Portal}, {Santini}, {Scott},
  {Sturm}, {Tacconi}, \& {Valtchanov}}]{Magnelli2013}
{Magnelli}, B., {Popesso}, P., {Berta}, S., {et~al.} 2013, \aap, 553, A132,
  \dodoi{10.1051/0004-6361/201321371}

\bibitem[{{Marchesi} {et~al.}(2016){Marchesi}, {Civano}, {Elvis}, {Salvato},
  {Brusa}, {Comastri}, {Gilli}, {Hasinger}, {Lanzuisi}, {Miyaji}, {Treister},
  {Urry}, {Vignali}, {Zamorani}, {Allevato}, {Cappelluti}, {Cardamone},
  {Finoguenov}, {Griffiths}, {Karim}, {Laigle}, {LaMassa}, {Jahnke}, {Ranalli},
  {Schawinski}, {Schinnerer}, {Silverman}, {Smolcic}, {Suh}, \&
  {Trakhtenbrot}}]{Marchesi2016}
{Marchesi}, S., {Civano}, F., {Elvis}, M., {et~al.} 2016, \apj, 817, 34,
  \dodoi{10.3847/0004-637X/817/1/34}

\bibitem[{{Marsan} {et~al.}(2017){Marsan}, {Marchesini}, {Brammer}, {Geier},
  {Kado-Fong}, {Labb{\'e}}, {Muzzin}, \& {Stefanon}}]{Marsan2017}
{Marsan}, Z.~C., {Marchesini}, D., {Brammer}, G.~B., {et~al.} 2017, \apj, 842,
  21, \dodoi{10.3847/1538-4357/aa7206}

\bibitem[{{Masters} {et~al.}(2017){Masters}, {Stern}, {Cohen}, {Capak},
  {Rhodes}, {Castander}, \& {Paltani}}]{Masters2017}
{Masters}, D.~C., {Stern}, D.~K., {Cohen}, J.~G., {et~al.} 2017, \apj, 841,
  111, \dodoi{10.3847/1538-4357/aa6f08}

\bibitem[{{McCracken} {et~al.}(2012){McCracken}, {Milvang-Jensen}, {Dunlop},
  {Franx}, {Fynbo}, {Le F{\`e}vre}, {Holt}, {Caputi}, {Goranova}, {Buitrago},
  {Emerson}, {Freudling}, {Hudelot}, {L{\'o}pez-Sanjuan}, {Magnard}, {Mellier},
  {M{\o}ller}, {Nilsson}, {Sutherland}, {Tasca}, \& {Zabl}}]{McCracken2012}
{McCracken}, H.~J., {Milvang-Jensen}, B., {Dunlop}, J., {et~al.} 2012, \aap,
  544, A156, \dodoi{10.1051/0004-6361/201219507}

\bibitem[{{Momcheva} {et~al.}(2017){Momcheva}, {van Dokkum}, {van der Wel},
  {Brammer}, {MacKenty}, {Nelson}, {Leja}, {Muzzin}, \& {Franx}}]{Momcheva2017}
{Momcheva}, I.~G., {van Dokkum}, P.~G., {van der Wel}, A., {et~al.} 2017,
  \pasp, 129, 015004, \dodoi{10.1088/1538-3873/129/971/015004}

\bibitem[{{Mowla} {et~al.}(2019){Mowla}, {van Dokkum}, {Brammer}, {Momcheva},
  {van der Wel}, {Whitaker}, {Nelson}, {Bezanson}, {Muzzin}, {Franx},
  {MacKenty}, {Leja}, {Kriek}, \& {Marchesini}}]{Mowla2019}
{Mowla}, L.~A., {van Dokkum}, P., {Brammer}, G.~B., {et~al.} 2019, \apj, 880,
  57, \dodoi{10.3847/1538-4357/ab290a}

\bibitem[{{Muzzin} {et~al.}(2013){Muzzin}, {Marchesini}, {Stefanon}, {Franx},
  {Milvang-Jensen}, {Dunlop}, {Fynbo}, {Brammer}, {Labb{\'e}}, \& {van
  Dokkum}}]{Muzzin2013}
{Muzzin}, A., {Marchesini}, D., {Stefanon}, M., {et~al.} 2013, \apjs, 206, 8,
  \dodoi{10.1088/0067-0049/206/1/8}

\bibitem[{{Nanayakkara} {et~al.}(2016){Nanayakkara}, {Glazebrook}, {Kacprzak},
  {Yuan}, {Tran}, {Spitler}, {Kewley}, {Straatman}, {Cowley}, {Fisher},
  {Labbe}, {Tomczak}, {Allen}, \& {Alcorn}}]{Nanayakkara2016}
{Nanayakkara}, T., {Glazebrook}, K., {Kacprzak}, G.~G., {et~al.} 2016, \apj,
  828, 21, \dodoi{10.3847/0004-637X/828/1/21}

\bibitem[{{Oliver} {et~al.}(2012){Oliver}, {Bock}, {Altieri}, {Amblard},
  {Arumugam}, {Aussel}, {Babbedge}, {Beelen}, {B{\'e}thermin}, {Blain},
  {Boselli}, {Bridge}, {Brisbin}, {Buat}, {Burgarella},
  {Castro-Rodr{\'\i}guez}, {Cava}, {Chanial}, {Cirasuolo}, {Clements},
  {Conley}, {Conversi}, {Cooray}, {Dowell}, {Dubois}, {Dwek}, {Dye}, {Eales},
  {Elbaz}, {Farrah}, {Feltre}, {Ferrero}, {Fiolet}, {Fox}, {Franceschini},
  {Gear}, {Giovannoli}, {Glenn}, {Gong}, {Gonz{\'a}lez Solares}, {Griffin},
  {Halpern}, {Harwit}, {Hatziminaoglou}, {Heinis}, {Hurley}, {Hwang}, {Hyde},
  {Ibar}, {Ilbert}, {Isaak}, {Ivison}, {Lagache}, {Le Floc'h}, {Levenson},
  {Faro}, {Lu}, {Madden}, {Maffei}, {Magdis}, {Mainetti}, {Marchetti},
  {Marsden}, {Marshall}, {Mortier}, {Nguyen}, {O'Halloran}, {Omont}, {Page},
  {Panuzzo}, {Papageorgiou}, {Patel}, {Pearson}, {P{\'e}rez-Fournon}, {Pohlen},
  {Rawlings}, {Raymond}, {Rigopoulou}, {Riguccini}, {Rizzo}, {Rodighiero},
  {Roseboom}, {Rowan-Robinson}, {S{\'a}nchez Portal}, {Schulz}, {Scott},
  {Seymour}, {Shupe}, {Smith}, {Stevens}, {Symeonidis}, {Trichas}, {Tugwell},
  {Vaccari}, {Valtchanov}, {Vieira}, {Viero}, {Vigroux}, {Wang}, {Ward},
  {Wardlow}, {Wright}, {Xu}, \& {Zemcov}}]{Oliver2012}
{Oliver}, S.~J., {Bock}, J., {Altieri}, B., {et~al.} 2012, \mnras, 424, 1614,
  \dodoi{10.1111/j.1365-2966.2012.20912.x}

\bibitem[{{Pavesi} {et~al.}(2018){Pavesi}, {Riechers}, {Sharon},
  {Smol{\v{c}}i{\'c}}, {Faisst}, {Schinnerer}, {Carilli}, {Capak}, {Scoville},
  \& {Stacey}}]{Pavesi2018}
{Pavesi}, R., {Riechers}, D.~A., {Sharon}, C.~E., {et~al.} 2018, \apj, 861, 43,
  \dodoi{10.3847/1538-4357/aac6b6}

\bibitem[{{Peng} {et~al.}(2002){Peng}, {Ho}, {Impey}, \& {Rix}}]{Peng2002}
{Peng}, C.~Y., {Ho}, L.~C., {Impey}, C.~D., \& {Rix}, H.-W. 2002, \aj, 124,
  266, \dodoi{10.1086/340952}

\bibitem[{{Peng} {et~al.}(2010){Peng}, {Ho}, {Impey}, \& {Rix}}]{Peng2010}
---. 2010, \aj, 139, 2097, \dodoi{10.1088/0004-6256/139/6/2097}

\bibitem[{{Peth} {et~al.}(2016){Peth}, {Lotz}, {Freeman}, {McPartland},
  {Mortazavi}, {Snyder}, {Barro}, {Grogin}, {Guo}, {Hemmati}, {Kartaltepe},
  {Kocevski}, {Koekemoer}, {McIntosh}, {Nayyeri}, {Papovich}, {Primack}, \&
  {Simons}}]{Peth2016}
{Peth}, M.~A., {Lotz}, J.~M., {Freeman}, P.~E., {et~al.} 2016, \mnras, 458,
  963, \dodoi{10.1093/mnras/stw252}

\bibitem[{{Rieke} {et~al.}(2004){Rieke}, {Young}, {Engelbracht}, {Kelly},
  {Low}, {Haller}, {Beeman}, {Gordon}, {Stansberry}, {Misselt}, {Cadien},
  {Morrison}, {Rivlis}, {Latter}, {Noriega-Crespo}, {Padgett}, {Stapelfeldt},
  {Hines}, {Egami}, {Muzerolle}, {Alonso-Herrero}, {Blaylock}, {Dole}, {Hinz},
  {Le Floc'h}, {Papovich}, {P{\'e}rez-Gonz{\'a}lez}, {Smith}, {Su}, {Bennett},
  {Frayer}, {Henderson}, {Lu}, {Masci}, {Pesenson}, {Rebull}, {Rho}, {Keene},
  {Stolovy}, {Wachter}, {Wheaton}, {Werner}, \& {Richards}}]{Rieke2004}
{Rieke}, G.~H., {Young}, E.~T., {Engelbracht}, C.~W., {et~al.} 2004, \apjs,
  154, 25, \dodoi{10.1086/422717}

\bibitem[{{Rodriguez-Gomez} {et~al.}(2019){Rodriguez-Gomez}, {Snyder}, {Lotz},
  {Nelson}, {Pillepich}, {Springel}, {Genel}, {Weinberger}, {Tacchella},
  {Pakmor}, {Torrey}, {Marinacci}, {Vogelsberger}, {Hernquist}, \&
  {Thilker}}]{Rodriguez-Gomez2019}
{Rodriguez-Gomez}, V., {Snyder}, G.~F., {Lotz}, J.~M., {et~al.} 2019, \mnras,
  483, 4140, \dodoi{10.1093/mnras/sty3345}

\bibitem[{{Roseboom} {et~al.}(2012){Roseboom}, {Bunker}, {Sumiyoshi}, {Wang},
  {Dalton}, {Akiyama}, {Bock}, {Bonfield}, {Buat}, {Casey}, {Chapin},
  {Clements}, {Conley}, {Curtis-Lake}, {Cooray}, {Dunlop}, {Farrah}, {Ham},
  {Ibar}, {Iwamuro}, {Kimura}, {Lewis}, {Macaulay}, {Magdis}, {Maihara},
  {Marsden}, {Mauch}, {Moritani}, {Ohta}, {Oliver}, {Page}, {Schulz}, {Scott},
  {Symeonidis}, {Takato}, {Tamura}, {Totani}, {Yabe}, \&
  {Zemcov}}]{Roseboom2012}
{Roseboom}, I.~G., {Bunker}, A., {Sumiyoshi}, M., {et~al.} 2012, \mnras, 426,
  1782, \dodoi{10.1111/j.1365-2966.2012.21777.x}

\bibitem[{{Runge} \& {Yan}(2018)}]{Runge2018}
{Runge}, J., \& {Yan}, H. 2018, \apj, 853, 47, \dodoi{10.3847/1538-4357/aaa020}

\bibitem[{{Salvato} {et~al.}(2011){Salvato}, {Ilbert}, {Hasinger}, {Rau},
  {Civano}, {Zamorani}, {Brusa}, {Elvis}, {Vignali}, {Aussel}, {Comastri},
  {Fiore}, {Le Floc'h}, {Mainieri}, {Bardelli}, {Bolzonella}, {Bongiorno},
  {Capak}, {Caputi}, {Cappelluti}, {Carollo}, {Contini}, {Garilli}, {Iovino},
  {Fotopoulou}, {Fruscione}, {Gilli}, {Halliday}, {Kneib}, {Kakazu},
  {Kartaltepe}, {Koekemoer}, {Kovac}, {Ideue}, {Ikeda}, {Impey}, {Le Fevre},
  {Lamareille}, {Lanzuisi}, {Le Borgne}, {Le Brun}, {Lilly}, {Maier},
  {Manohar}, {Masters}, {McCracken}, {Messias}, {Mignoli}, {Mobasher}, {Nagao},
  {Pello}, {Puccetti}, {Perez-Montero}, {Renzini}, {Sargent}, {Sanders},
  {Scodeggio}, {Scoville}, {Shopbell}, {Silvermann}, {Taniguchi}, {Tasca},
  {Tresse}, {Trump}, \& {Zucca}}]{Salvato2011}
{Salvato}, M., {Ilbert}, O., {Hasinger}, G., {et~al.} 2011, \apj, 742, 61,
  \dodoi{10.1088/0004-637X/742/2/61}

\bibitem[{{Sanders} {et~al.}(2007){Sanders}, {Salvato}, {Aussel}, {Ilbert},
  {Scoville}, {Surace}, {Frayer}, {Sheth}, {Helou}, {Brooke}, {Bhattacharya},
  {Yan}, {Kartaltepe}, {Barnes}, {Blain}, {Calzetti}, {Capak}, {Carilli},
  {Carollo}, {Comastri}, {Daddi}, {Ellis}, {Elvis}, {Fall}, {Franceschini},
  {Giavalisco}, {Hasinger}, {Impey}, {Koekemoer}, {Le F{\`e}vre}, {Lilly},
  {Liu}, {McCracken}, {Mobasher}, {Renzini}, {Rich}, {Schinnerer}, {Shopbell},
  {Taniguchi}, {Thompson}, {Urry}, \& {Williams}}]{Sanders2007}
{Sanders}, D.~B., {Salvato}, M., {Aussel}, H., {et~al.} 2007, \apjs, 172, 86,
  \dodoi{10.1086/517885}

\bibitem[{{Schreiber} {et~al.}(2018){Schreiber}, {Glazebrook}, {Nanayakkara},
  {Kacprzak}, {Labb{\'e}}, {Oesch}, {Yuan}, {Tran}, {Papovich}, {Spitler}, \&
  {Straatman}}]{Schreiber2018}
{Schreiber}, C., {Glazebrook}, K., {Nanayakkara}, T., {et~al.} 2018, \aap, 618,
  A85, \dodoi{10.1051/0004-6361/201833070}

\bibitem[{{Scoville} {et~al.}(2007){Scoville}, {Aussel}, {Brusa}, {Capak},
  {Carollo}, {Elvis}, {Giavalisco}, {Guzzo}, {Hasinger}, {Impey}, {Kneib},
  {LeFevre}, {Lilly}, {Mobasher}, {Renzini}, {Rich}, {Sanders}, {Schinnerer},
  {Schminovich}, {Shopbell}, {Taniguchi}, \& {Tyson}}]{Scoville2007}
{Scoville}, N., {Aussel}, H., {Brusa}, M., {et~al.} 2007, \apjs, 172, 1,
  \dodoi{10.1086/516585}

\bibitem[{{S{\'e}rsic}(1963)}]{Sersic1963}
{S{\'e}rsic}, J.~L. 1963, Boletin de la Asociacion Argentina de Astronomia La
  Plata Argentina, 6, 41

\bibitem[{{Siebenmorgen} \& {Kr{\"u}gel}(2007)}]{Siebenmorgen2007}
{Siebenmorgen}, R., \& {Kr{\"u}gel}, E. 2007, \aap, 461, 445,
  \dodoi{10.1051/0004-6361:20065700}

\bibitem[{{Silverman} {et~al.}(2015){Silverman}, {Kashino}, {Sanders},
  {Kartaltepe}, {Arimoto}, {Renzini}, {Rodighiero}, {Daddi}, {Zahid}, {Nagao},
  {Kewley}, {Lilly}, {Sugiyama}, {Baronchelli}, {Capak}, {Carollo}, {Chu},
  {Hasinger}, {Ilbert}, {Juneau}, {Kajisawa}, {Koekemoer}, {Kovac}, {Le
  F{\`e}vre}, {Masters}, {McCracken}, {Onodera}, {Schulze}, {Scoville},
  {Strazzullo}, \& {Taniguchi}}]{Silverman2015a}
{Silverman}, J.~D., {Kashino}, D., {Sanders}, D., {et~al.} 2015, \apjs, 220,
  12, \dodoi{10.1088/0067-0049/220/1/12}

\bibitem[{{Simpson} {et~al.}(2015){Simpson}, {Smail}, {Swinbank}, {Chapman},
  {Geach}, {Ivison}, {Thomson}, {Aretxaga}, {Blain}, {Cowley}, {Chen},
  {Coppin}, {Dunlop}, {Edge}, {Farrah}, {Ibar}, {Karim}, {Knudsen},
  {Meijerink}, {Micha{\l}owski}, {Scott}, {Spaans}, \& {van der
  Werf}}]{Simpson2015a}
{Simpson}, J.~M., {Smail}, I., {Swinbank}, A.~M., {et~al.} 2015, \apj, 807,
  128, \dodoi{10.1088/0004-637X/807/2/128}

\bibitem[{{Simpson} {et~al.}(2017){Simpson}, {Smail}, {Swinbank}, {Ivison},
  {Dunlop}, {Geach}, {Almaini}, {Arumugam}, {Bremer}, {Chen}, {Conselice},
  {Coppin}, {Farrah}, {Ibar}, {Hartley}, {Ma}, {Micha{\l}owski}, {Scott},
  {Spaans}, {Thomson}, \& {van der Werf}}]{Simpson2017}
---. 2017, \apj, 839, 58, \dodoi{10.3847/1538-4357/aa65d0}

\bibitem[{{Simpson} {et~al.}(2019){Simpson}, {Smail}, {Swinbank}, {Chapman},
  {Chen}, {Geach}, {Matsuda}, {Wang}, {Wang}, {Yang}, {Ao}, {Asquith},
  {Bourne}, {Coogan}, {Coppin}, {Gullberg}, {Hine}, {Ho}, {Hwang}, {Ivison},
  {Kato}, {Lacaille}, {Lewis}, {Liu}, {Micha{\l}owski}, {Oteo}, {Sawicki},
  {Scholtz}, {Smith}, {Thomson}, \& {Wardlow}}]{Simpson2019}
---. 2019, \apj, 880, 43, \dodoi{10.3847/1538-4357/ab23ff}

\bibitem[{{Simpson} {et~al.}(2020){Simpson}, {Smail},
  {Dudzevi{\v{c}}i{\={u}}t{\.{e}}}, {Matsuda}, {Hsieh}, {Wang}, {Swinbank},
  {Stach}, {An}, {Birkin}, {Ao}, {Bunker}, {Chapman}, {Chen}, {Coppin},
  {Ikarashi}, {Ivison}, {Mitsuhashi}, {Saito}, {Umehata}, {Wang}, \&
  {Zhao}}]{Simpson2020}
{Simpson}, J.~M., {Smail}, I., {Dudzevi{\v{c}}i{\={u}}t{\.{e}}}, U., {et~al.}
  2020, \mnras, 495, 3409, \dodoi{10.1093/mnras/staa1345}

\bibitem[{{Smail} {et~al.}(2004){Smail}, {Chapman}, {Blain}, \&
  {Ivison}}]{Smail2004}
{Smail}, I., {Chapman}, S.~C., {Blain}, A.~W., \& {Ivison}, R.~J. 2004, \apj,
  616, 71, \dodoi{10.1086/424896}

\bibitem[{{Smol{\v{c}}i{\'c}} {et~al.}(2017{\natexlab{a}}){Smol{\v{c}}i{\'c}},
  {Novak}, {Bondi}, {Ciliegi}, {Mooley}, {Schinnerer}, {Zamorani}, {Navarrete},
  {Bourke}, {Karim}, {Vardoulaki}, {Leslie}, {Delhaize}, {Carilli}, {Myers},
  {Baran}, {Delvecchio}, {Miettinen}, {Banfield}, {Balokovi{\'c}}, {Bertoldi},
  {Capak}, {Frail}, {Hallinan}, {Hao}, {Herrera Ruiz}, {Horesh}, {Ilbert},
  {Intema}, {Jeli{\'c}}, {Kl{\"o}ckner}, {Krpan}, {Kulkarni}, {McCracken},
  {Laigle}, {Middleberg}, {Murphy}, {Sargent}, {Scoville}, \&
  {Sheth}}]{Smolcic2017a}
{Smol{\v{c}}i{\'c}}, V., {Novak}, M., {Bondi}, M., {et~al.} 2017{\natexlab{a}},
  \aap, 602, A1, \dodoi{10.1051/0004-6361/201628704}

\bibitem[{{Smol{\v{c}}i{\'c}} {et~al.}(2017{\natexlab{b}}){Smol{\v{c}}i{\'c}},
  {Delvecchio}, {Zamorani}, {Baran}, {Novak}, {Delhaize}, {Schinnerer},
  {Berta}, {Bondi}, {Ciliegi}, {Capak}, {Civano}, {Karim}, {Le Fevre},
  {Ilbert}, {Laigle}, {Marchesi}, {McCracken}, {Tasca}, {Salvato}, \&
  {Vardoulaki}}]{Smolcic2017b}
{Smol{\v{c}}i{\'c}}, V., {Delvecchio}, I., {Zamorani}, G., {et~al.}
  2017{\natexlab{b}}, \aap, 602, A2, \dodoi{10.1051/0004-6361/201630223}

\bibitem[{{Soifer} \& {Neugebauer}(1991)}]{Soifer1991}
{Soifer}, B.~T., \& {Neugebauer}, G. 1991, \aj, 101, 354,
  \dodoi{10.1086/115691}

\bibitem[{{Stach} {et~al.}(2018){Stach}, {Smail}, {Swinbank}, {Simpson},
  {Geach}, {An}, {Almaini}, {Arumugam}, {Blain}, {Chapman}, {Chen},
  {Conselice}, {Cooke}, {Coppin}, {Dunlop}, {Farrah}, {Gullberg}, {Hartley},
  {Ivison}, {Maltby}, {Micha{\l}owski}, {Scott}, {Simpson}, {Thomson},
  {Wardlow}, \& {van der Werf}}]{Stach2018}
{Stach}, S.~M., {Smail}, I., {Swinbank}, A.~M., {et~al.} 2018, \apj, 860, 161,
  \dodoi{10.3847/1538-4357/aac5e5}

\bibitem[{{Stach} {et~al.}(2019){Stach}, {Dudzevi{\v{c}}i{\={u}}t{\.{e}}},
  {Smail}, {Swinbank}, {Geach}, {Simpson}, {An}, {Almaini}, {Arumugam},
  {Blain}, {Chapman}, {Chen}, {Conselice}, {Cooke}, {Coppin}, {da Cunha},
  {Dunlop}, {Farrah}, {Gullberg}, {Hodge}, {Ivison}, {Kocevski},
  {Micha{\l}owski}, {Miyaji}, {Scott}, {Thomson}, {Wardlow}, {Weiss}, \& {van
  der Werf}}]{Stach2019}
{Stach}, S.~M., {Dudzevi{\v{c}}i{\={u}}t{\.{e}}}, U., {Smail}, I., {et~al.}
  2019, \mnras, 487, 4648, \dodoi{10.1093/mnras/stz1536}

\bibitem[{{Sutherland} \& {Saunders}(1992)}]{Sutherland1992}
{Sutherland}, W., \& {Saunders}, W. 1992, \mnras, 259, 413,
  \dodoi{10.1093/mnras/259.3.413}

\bibitem[{{Swinbank} {et~al.}(2010){Swinbank}, {Smail}, {Chapman}, {Borys},
  {Alexander}, {Blain}, {Conselice}, {Hainline}, \& {Ivison}}]{Swinbank2010}
{Swinbank}, A.~M., {Smail}, I., {Chapman}, S.~C., {et~al.} 2010, \mnras, 405,
  234, \dodoi{10.1111/j.1365-2966.2010.16485.x}

\bibitem[{{Targett} {et~al.}(2013){Targett}, {Dunlop}, {Cirasuolo}, {McLure},
  {Bruce}, {Fontana}, {Galametz}, {Paris}, {Dav{\'e}}, {Dekel}, {Faber},
  {Ferguson}, {Grogin}, {Kartaltepe}, {Kocevski}, {Koekemoer}, {Kurczynski},
  {Lai}, \& {Lotz}}]{Targett2013}
{Targett}, T.~A., {Dunlop}, J.~S., {Cirasuolo}, M., {et~al.} 2013, \mnras, 432,
  2012, \dodoi{10.1093/mnras/stt482}

\bibitem[{{Trump} {et~al.}(2007){Trump}, {Impey}, {McCarthy}, {Elvis},
  {Huchra}, {Brusa}, {Hasinger}, {Schinnerer}, {Capak}, {Lilly}, \&
  {Scoville}}]{Trump2007}
{Trump}, J.~R., {Impey}, C.~D., {McCarthy}, P.~J., {et~al.} 2007, \apjs, 172,
  383, \dodoi{10.1086/516578}

\bibitem[{{Trump} {et~al.}(2009){Trump}, {Impey}, {Kelly}, {Elvis}, {Merloni},
  {Bongiorno}, {Gabor}, {Hao}, {McCarthy}, {Huchra}, {Brusa}, {Cappelluti},
  {Koekemoer}, {Nagao}, {Salvato}, \& {Scoville}}]{Trump2009}
{Trump}, J.~R., {Impey}, C.~D., {Kelly}, B.~C., {et~al.} 2009, \apj, 700, 49,
  \dodoi{10.1088/0004-637X/700/1/49}

\bibitem[{{van der Wel} {et~al.}(2016){van der Wel}, {Noeske}, {Bezanson},
  {Pacifici}, {Gallazzi}, {Franx}, {Mu{\~n}oz-Mateos}, {Bell}, {Brammer},
  {Charlot}, {Chauk{\'e}}, {Labb{\'e}}, {Maseda}, {Muzzin}, {Rix}, {Sobral},
  {van de Sande}, {van Dokkum}, {Wild}, \& {Wolf}}]{vanderWel2016}
{van der Wel}, A., {Noeske}, K., {Bezanson}, R., {et~al.} 2016, \apjs, 223, 29,
  \dodoi{10.3847/0067-0049/223/2/29}

\bibitem[{{Veilleux} {et~al.}(2002){Veilleux}, {Kim}, \&
  {Sanders}}]{Veilleux2002}
{Veilleux}, S., {Kim}, D.~C., \& {Sanders}, D.~B. 2002, \apjs, 143, 315,
  \dodoi{10.1086/343844}

\bibitem[{{Wang} {et~al.}(2014){Wang}, {Viero}, {Clarke}, {Bock}, {Buat},
  {Conley}, {Farrah}, {Guo}, {Heinis}, {Magdis}, {Marchetti}, {Marsden},
  {Norberg}, {Oliver}, {Page}, {Roehlly}, {Roseboom}, {Schulz}, {Smith},
  {Vaccari}, \& {Zemcov}}]{Wang2014}
{Wang}, L., {Viero}, M., {Clarke}, C., {et~al.} 2014, \mnras, 444, 2870,
  \dodoi{10.1093/mnras/stu1569}

\bibitem[{{Wang} {et~al.}(2016){Wang}, {Elbaz}, {Daddi}, {Finoguenov}, {Liu},
  {Schreiber}, {Mart{\'\i}n}, {Strazzullo}, {Valentino}, {van der Burg},
  {Zanella}, {Ciesla}, {Gobat}, {Le Brun}, {Pannella}, {Sargent}, {Shu}, {Tan},
  {Cappelluti}, \& {Li}}]{Wang2016}
{Wang}, T., {Elbaz}, D., {Daddi}, E., {et~al.} 2016, \apj, 828, 56,
  \dodoi{10.3847/0004-637X/828/1/56}

\bibitem[{{Wiklind} {et~al.}(2014){Wiklind}, {Conselice}, {Dahlen},
  {Dickinson}, {Ferguson}, {Grogin}, {Guo}, {Koekemoer}, {Mobasher},
  {Mortlock}, {Fontana}, {Dav{\'e}}, {Yan}, {Acquaviva}, {Ashby}, {Barro},
  {Caputi}, {Castellano}, {Dekel}, {Donley}, {Fazio}, {Giavalisco}, {Grazian},
  {Hathi}, {Kurczynski}, {Lu}, {McGrath}, {de Mello}, {Peth}, {Safarzadeh},
  {Stefanon}, \& {Targett}}]{Wiklind2014}
{Wiklind}, T., {Conselice}, C.~J., {Dahlen}, T., {et~al.} 2014, The Messenger,
  156, 40

\bibitem[{{Yan} \& {Ma}(2016)}]{Yan2016}
{Yan}, H., \& {Ma}, Z. 2016, \apjl, 820, L16,
  \dodoi{10.3847/2041-8205/820/1/L16}

\bibitem[{{Yan} {et~al.}(2020){Yan}, {Ma}, {Huang}, \& {Fan}}]{Yan2020}
{Yan}, H., {Ma}, Z., {Huang}, J.-S., \& {Fan}, L. 2020, \apjs, 249, 1,
  \dodoi{10.3847/1538-4365/ab964a}

\bibitem[{{Yan} {et~al.}(2014){Yan}, {Stefanon}, {Ma}, {Willner}, {Somerville},
  {Ashby}, {Dav{\'e}}, {P{\'e}rez-Gonz{\'a}lez}, {Cava}, {Wiklind}, {Kocevski},
  {Rafelski}, {Kartaltepe}, {Cooray}, {Koekemoer}, \& {Grogin}}]{Yan2014}
{Yan}, H., {Stefanon}, M., {Ma}, Z., {et~al.} 2014, \apjs, 213, 2,
  \dodoi{10.1088/0067-0049/213/1/2}

\bibitem[{{Yun} {et~al.}(2015){Yun}, {Aretxaga}, {Gurwell}, {Hughes},
  {Monta{\~n}a}, {Narayanan}, {Rosa-Gonz{\'a}lez}, {S{\'a}nchez-Arg{\"u}elles},
  {Schloerb}, {Snell}, {Vega}, {Wilson}, {Zeballos}, {Chavez}, {Cybulski},
  {D{\'\i}az-Santos}, {De La Luz}, {Erickson}, {Ferrusca}, {Gim}, {Heyer},
  {Iono}, {Pope}, {Rogstad}, {Scott}, {Souccar}, {Terlevich}, {Terlevich},
  {Wilner}, \& {Zavala}}]{Yun2015}
{Yun}, M.~S., {Aretxaga}, I., {Gurwell}, M.~A., {et~al.} 2015, \mnras, 454,
  3485, \dodoi{10.1093/mnras/stv1963}

\bibitem[{{Zavala} {et~al.}(2018){Zavala}, {Aretxaga}, {Dunlop},
  {Micha{\l}owski}, {Hughes}, {Bourne}, {Chapin}, {Cowley}, {Farrah}, {Lacey},
  {Targett}, \& {van der Werf}}]{Zavala2018}
{Zavala}, J.~A., {Aretxaga}, I., {Dunlop}, J.~S., {et~al.} 2018, \mnras, 475,
  5585, \dodoi{10.1093/mnras/sty217}

\end{thebibliography}
	
\end{document}